\documentclass[12pt]{article}

\usepackage{latexsym}
\usepackage{amsmath}
\usepackage{amssymb}
\catcode`\@=11
\newcount\hour
\newcount\minute
\newtoks\amorpm \hour=\time\divide\hour by 60\minute
=\time{\multiply\hour by 60 \global\advance\minute by-\hour}
\edef\standardtime{{\ifnum\hour<12 \global\amorpm={am}%
        \else\global\amorpm={pm}\advance\hour by-12 \fi
        \ifnum\hour=0 \hour=12 \fi
        \number\hour:\ifnum\minute<10
        0\fi\number\minute\the\amorpm}}
\edef\militarytime{\number\hour:\ifnum\minute<10
0\fi\number\minute}
\def\draftlabel#1{{\@bsphack\if@filesw {\let\thepage\relax
   \xdef\@gtempa{\write\@auxout{\string
      \newlabel{#1}{{\@currentlabel}{\thepage}}}}}\@gtempa
   \if@nobreak \ifvmode\nobreak\fi\fi\fi\@esphack}
        \gdef\@eqnlabel{#1}}
\def\@eqnlabel{}
\def\@vacuum{}
\def\marginnote#1{}
\def\draftmarginnote#1{\marginpar{\raggedright\scriptsize\tt#1}}
\def\draft{
        \pagestyle{plain}
        \overfullrule=2pt
        \oddsidemargin -.1truein
        \def\@oddhead{\sl \phantom{\today\quad\militarytime} \hfil
        \smash{\Large\sl DRAFT} \hfil \today\quad\militarytime}
        \let\@evenhead\@oddhead
        \let\label=\draftlabel
        \let\marginnote=\draftmarginnote
        \def\ps@empty{\let\@mkboth\@gobbletwo
        \def\@oddfoot{\hfil \smash{\Large\sl DRAFT} \hfil}
        \let\@evenfoot\@oddhead}
        \def\@eqnnum{(\theequation)\rlap{\kern\marginparsep\tt\@eqnlabel}%
        \global\let\@eqnlabel\@vacuum}  }

\renewcommand{\theequation}{\thesection.\arabic{equation}}
\renewcommand{\thefootnote}{\fnsymbol{footnote}}
\newcommand{\newsection}{    
\setcounter{equation}{0}\section}
\def\appendix#1{\addtocounter{section}{1}\setcounter{equation}{0}
\renewcommand{\thesection}{\Alph{section}}
\section*{Appendix \thesection\protect\indent \parbox[t]{11.15cm}{#1}}
\addcontentsline{toc}{section}{Appendix \thesection\ \ \ #1}}

\def \lc {{light-cone}}

\def \bi{\bibitem}
\def \la {\label}

\def \b {\beta}
\def \Om {\Omega}

\def \s{\sigma}
\def \d {\partial}

\def\be{\begin{equation}}
\def\ee{\end{equation}}

\def\nat {{\natural}}
 \def \lc {light-cone\ }

\catcode`\@=11
\def \lc {light cone\ }

\def\bea{\begin{eqnarray}}
\def\eea{\end{eqnarray}}
\def\beann{\begin{eqnarray*}}
\def\eeann{\end{eqnarray*}}
\def\beq{\begin{equation}}
\def\eeq{\end{equation}}
\def\ba{\begin{array}}
\def\ea{\end{array}}
\def\ben{\begin{enumerate}}
\def\een{\end{enumerate}}
 \def \l {\lambda}
 \def\m {\mu}
\def\s {\sigma }
 \def \la {\label}
 \def\be{\begin{equation}}
\def\ee{\end{equation}}

\def \la {\label}

\def \r {\rho}

\font\mybb=msbm10 at 11pt

\def\bb#1{\hbox{\mybb#1}}

\def\bR {\bb{R}}

\def\bC {\bb{C}}

\def\e  {\epsilon}

 \def\ep {\epsilon}

\def \ee {\epsilon}
\def \te {\tilde \epsilon}

\def \g {\gamma}
\def \bi{\bibitem}
\def\a{\alpha }

\def \ep {\epsilon}

\def \s {\sigma}

\def \r {\rho}
\def \d {\delta}

\def \l {\lambda}
\def \m {\mu}
\def \g {\gamma}

\def \b {\beta}

\def\lc{\lrcorner}

\newcommand{\smallfrac}{\tfrac}

\def\be{\begin{equation}}
\def\ee{\end{equation}}

\def \bi {\bibitem}
\def \la{\label}

\def \nn {\nonumber}

\newcommand{\cont}[1]{{}_{#1}{}^{#1}}

\begin{document}
\date{November 2002}
\begin{titlepage}
\begin{center}
\hfill hep-th/0503046\\
\vspace{4.0cm}
{\Large \bf Systematics of M-theory spinorial geometry}\\[.2cm]

\vspace{1.5cm}
 {\large  U. Gran,   G. Papadopoulos and D. Roest }

 \vspace{0.5cm}
Department of Mathematics\\
King's College London\\
Strand\\
London WC2R 2LS
\end{center}

\vskip 1.5 cm
\begin{abstract}
We reduce the classification of all supersymmetric backgrounds in
eleven dimensions to the evaluation of the supercovariant derivative
and of an integrability condition, which contains the field
equations, on six types of spinors.  We determine the expression of
the supercovariant derivative on all six types of spinors and give
in each case the field equations that do not arise as the
integrability conditions of Killing spinor equations. The Killing
spinor equations of a background become a linear system for the
fluxes, geometry and spacetime derivatives of the functions that
determine the spinors. The solution of the linear system expresses
the fluxes in terms of the geometry and specifies the restrictions
on the geometry of spacetime for all supersymmetric backgrounds. We
also show that the minimum number of field equations that is needed
for  a supersymmetric configuration  to be a solution of
eleven-dimensional supergravity can be found by solving a linear
system. The linear systems of the Killing spinor equations and their
integrability conditions are given in both a timelike and a null
spinor basis. We illustrate the construction with examples.

\end{abstract}
\end{titlepage}
\newpage
\setcounter{page}{1}
\renewcommand{\thefootnote}{\arabic{footnote}}
\setcounter{footnote}{0}

\setcounter{section}{0}
\setcounter{subsection}{0}
\newsection{Introduction}

The last ten years, the supersymmetric solutions of ten- and
eleven- dimensional supergravities have given new insights into
understanding of string theory and M-theory, see e.g. \cite{townsend, oz}. Most of the
solutions have been found using ans\"atze adapted to the
requirements of  physical problems. Recently, the realization
that there are new maximally supersymmetric solutions \cite{georgea} and the
rediscovery of some old ones \cite{kowalski, georgeb}  has led to a more systematic
exploration of supersymmetric solutions in supergravity theories.
The maximally supersymmetric solutions of ten and eleven
dimensional supergravities have been classified in \cite{jffgp}
using the integrability conditions of the Killing spinor equations
which leads to the vanishing of the supercovariant curvature. A
method\footnote{For a refinement see \cite{oisin}.} based on the Killing spinor bi-linear forms has also been
used to solve the Killing spinor equations of eleven-dimensional
supergravity for backgrounds with one Killing spinor \cite{pakis,
gutowski}. However this method has not been applied to
eleven-dimensional backgrounds with more than one supersymmetry.

In \cite{joe}, a new method to investigate the Killing spinor equations
of supergravities has been proposed. It is based on  a description of spinors
in terms of forms, the gauge symmetry of Killing spinor equations and
an oscillator basis in the space of spinors \cite{joe}.
This has been applied to systematically
explore the supersymmetric solutions of eleven-dimensional
supergravity with one, two, three and four supersymmetries
and to solve the Killing spinor equations of IIB supergravity
for one Killing spinor \cite{jan}.

In this paper, we shall show that the method of \cite{joe} can be
extended further to investigate all supersymmetric eleven-dimensional
 backgrounds\footnote{This includes backgrounds with both $SU(5)$ and $(Spin(7)\ltimes \bR^8)\times \bR$
  invariant spinors.}. For this, we use the linearity of the Killing spinor
equations to show that the supercovariant derivative ${\cal D}$ of
eleven-dimensional supergravity acting on any spinor $\epsilon$ can
be decomposed into a linear combination of six ``irreducible"
components. These six irreducible components are given by the action
of the supercovariant derivative, ${\cal D}\sigma_I$, on six types
of spinors
 \bea
  1~,~~~e_{i}~,~~~e_{ij}~,~~~e_{ijk}~,~~~e_{ijkl}~,~~~e_{12345}~,
 \eea
which are collectively denoted by $\sigma_I = e_{i_1 \cdots i_I}$
with $I=0,\dots 5$. These spinors can also be labeled by the
irreducible representations of $U(5)$ on the space
$\Lambda^*(\bC^5)$ of forms. We compute ${\cal D}\sigma_I$. As a
result, one can compute ${\cal D}\epsilon$ for any number of spinors
$\epsilon$ and then use the basis in the space of spinors
\cite{joe}, see also  appendix A, to express the Killing spinor
equations as a linear system for the geometry, fluxes and spacetime
derivatives of the functions that determine the Killing spinors
$\epsilon$. Therefore, we show that the Killing spinor equations for
any number of Killing spinors reduce to a linear system and we give
all the coefficients and all the unknowns of the system. The
solution of this system expresses the fluxes in terms of the
geometry and gives the restrictions of the geometry required by
supersymmetry.

It has been known for some time that the integrability conditions of the Killing
spinor equations imply some of the field equations of supergravity theories, e.g.
 in maximal supersymmetric backgrounds the integrability
conditions of Killing spinor equations imply all the field equations
\cite{jffgp}. The first integrability condition of the Killing
spinor equations is ${\cal R}_{AB}\epsilon=[{\cal D}_A, {\cal
D}_B]\epsilon=0$, where ${\cal R}$ is the curvature of the
supercovariant connection. This integrability condition has various
components one of which, ${\cal I}_A\epsilon=\Gamma^B{\cal
R}_{AB}\epsilon=0$, contains
 the field equations of eleven-dimensional supergravity \cite{pakis}.
Since the integrability conditions ${\cal R}\epsilon=0$ and ${\cal
I} \epsilon=0$ of the Killing spinor equations are linear algebraic
equations for the Killing spinor $\epsilon$, they again can be
decomposed in terms of the ${\cal R}\sigma_I$ and ${\cal I}
\sigma_I$. We give all the expressions for ${\cal I} \sigma_I$.
Since the integrability conditions of any number of Killing spinors
can be written in terms of ${\cal I} \sigma_I$, one can use the
basis of \cite{joe} to find which components of the field equations
are implied as integrability conditions of the Killing spinor
equations. In particular, one finds a linear system with the
components of the field equations as unknowns and the functions that
determine the Killing spinors as coefficients. The components of the
field equations that are {\it not}  determined as solutions of this
linear system are those that have to be imposed as additional
conditions to the Killing spinor equations for a configuration with
any number of supersymmetries to be a solution of the theory. We
remark that such an analysis can be done for ${\cal R}\epsilon=0$.
This would be an extension of the method used in \cite{jffgp} to solve the
Killing spinor equations for maximally supersymmetric
spacetimes\footnote {One may have to consider higher order
integrability conditions \cite{liu}.}.

The main aim of this paper is to be used as a manual for
systematically constructing all supersymmetric solutions of
eleven-dimensional supergravity. Because of this, we first present
the general formulae for ${\cal D}\sigma_I$ and ${\cal I}\sigma_I$.
However, these are rather involved when expressed in terms of the
oscillator basis in the space of spinors,
 see \cite{joe} and appendix A.
Because of this, we state the results in tables which have been put
in appendices. The construction of the linear systems associated
with ${\cal D}_A \epsilon_h=0$ and ${\cal I}\epsilon_h=0$ for any
number of Killing spinors $h=1,\dots,N$ can be read off from these
tables.

As we have explained, the construction of the linear systems
associated with the Killing spinor equations and  with the
integrability conditions can be done in the the basis of \cite{joe}
for any number of Killing spinors. However, if one or more  Killing
spinors are null, i.e. they are representatives of the orbit of
$Spin(10,1)$ with stability subgroup $(Spin(7)\ltimes \bR^8)\times
\bR$, it may be convenient to use another basis in the space of spinors.
Such a basis adapted to null spinors has been constructed in
\cite{jan} in the context of IIB supergravity and it is extended to
eleven dimensions in appendix A. We shall refer to the spinor basis
presented in \cite{joe} as ``timelike'' and that constructed from
IIB supergravity as ``null'' basis. In the null basis, one can find
simple expressions for the representatives of the $(Spin(7)\ltimes
\bR^8)\times \bR$ orbit. So one expects that the linear systems
associated with null Killing spinors will be easier to solve in the
null basis than the analogous linear systems derived in the timelike
basis. We show that the linear systems associated with the Killing
spinor equations and  of the integrability conditions in the null
basis can be derived from those we have constructed in the timelike
basis after a suitable replacement the time direction with the tenth
spatial direction and taking into account the way that $\Gamma^0$
and $\Gamma^\nat$ act\footnote{We denote the tenth direction with
$\nat$.} on the spinor basis. The rules of relating the linear
systems in the null basis and in timelike basis are given in detail
in section five. It turns out that these rules are very simple. Because of this
we shall focus on the timelike case and will only discuss the null case in
section five. Some partial results on the construction of the
linear system for the Killing spinor equations in another null basis
have been obtained in \cite{oisinb}.

To illustrate our construction we solve the Killing spinor equations
for backgrounds with two supersymmetries and
the {\sl most  general} $SU(4)$ invariant Killing spinors. Special cases have
 already been investigated in \cite{joe}.
Then we find for several configurations with one, two, three
and four supersymmetries
  the minimal set of field equations that in addition should
  be imposed in order
to be  solutions of eleven-dimensional supergravity.
In the process, we explain how the tables in the appendices can be used.

Our analysis is in the context of eleven-dimensional supergravity. But it can
be extended to the effective theory of M-theory which includes higher order corrections, e.g.
see \cite{minasian}. For example, our conclusion about the six types of spinors is not altered
by the inclusion of higher order corrections.

This paper is organized as follows: In section two, we summarize the
Killing spinor equations ${\cal D}\epsilon=0$ and give the
integrability conditions ${\cal I}\epsilon=0$ of eleven-dimensional
supergravity. In section three, we show how a general spinor is
related to the six types of spinors $\sigma_I$, and
  express the
Killing spinor equations ${\cal D}\epsilon=0$ and associated integrability
conditions ${\cal I}\epsilon=0$ in terms of  ${\cal D}\sigma_I$ and
${\cal I}\sigma_I$, respectively. In section four, we derive some  general
formulae that give ${\cal D}\sigma_I$ and ${\cal I}\sigma_I$ in terms
of the timelike canonical basis (\ref{hermbasis}). In section five, we
 give ${\cal D}\sigma_I$ and ${\cal I}\sigma_I$ in terms
of the null canonical basis. In section six, we summarize
the conditions on the geometry and fluxes for the most general background with two
supersymmetries and $SU(4)$ invariant spinors and
analyze the geometry of spacetime. In section seven, we analyze the field equations
of some backgrounds with one, two and four supersymmetries. In section eight,
we solve both the Killing spinor and field equations of a background with four
supersymmetries and $SU(4)$ invariant spinors and in section nine, we give our conclusions.
In appendix A, we summarize the properties of $Spin(10,1)$  spinors.
In appendix B, we give the conditions on the geometry and the
expressions for the fluxes of backgrounds which admit one $SU(5)$
invariant Killing spinor. These results can be found in \cite{joe}
but are summarized here for convenience. In appendix C, we give the tables
with the expressions for ${\cal D}\sigma_I$ expanded in the basis
(\ref{hermbasis}). In appendix D, we give the tables
with the expressions for ${\cal I}\sigma_I$ expanded in the basis
(\ref{hermbasis}). In appendix E, we solve the Killing spinor equations
for backgrounds which admit two Killing spinors which are invariant
under the $SU(4)$ subgroup of $Spin(10,1)$.

\newsection{Killing spinor equations and integrability conditions}

\subsection{Killing spinor equations}

The Killing spinor equations of eleven-dimensional supergravity
\cite{julia} are the vanishing of the gravitino supersymmetry
transformation restricted on the bosonic fields of the theory. The
bosonic fields are the metric $g$ and a four-form field strength
${\cal F}$. The Killing spinors of eleven-dimensional supergravity
are in the Majorana representation $\Delta_{32}$ of $Spin(10,1)$.
The supercovariant connection of eleven-dimensional supergravity is
\be {\cal D}_A \epsilon= \nabla_A \epsilon+ \Sigma_A \epsilon
\label{supcon} \ee where \be
\nabla_A\epsilon=\partial_A\epsilon+\tfrac{1}{4}\Omega_{A,BC}
\Gamma^{BC}\epsilon~, \ee i.e.~$\nabla_A$ is the spin covariant
derivative induced from the Levi-Civita connection, \be \Sigma_A =
-\tfrac{1}{288} (\Gamma_A{}^{B_1 \cdots B_4}- 8 \delta_{A}^{B_1}
\Gamma^{B_2 \cdots B_4 }) {\cal F}_{B_1 \cdots B_4} ~,
\label{supconb} \ee and ${\cal F}$ is the four-form field strength
(or flux), $A,B,\ldots=0, \dots, 9, 10$ are frame indices. The
supercovariant connection is a covariant derivative on the spinor
bundle of eleven-dimensional spacetime associated with the Majorana
representation of $Spin(10,1)$. However, ${\cal D}$ is not induced
from the tangent bundle because of the term (\ref{supconb}) which
depends on the flux ${\cal F}$.

As has been explained in \cite{joe}, the supercovariant connection
has gauge symmetry $Spin(10,1)$ and this can be used to bring the Killing spinors
into a canonical or normal form up to an induced Lorentz transformation
on the spacetime frame and fluxes ${\cal F}$. In this way, one can simplify the conditions
imposed by supersymmetry of the fluxes and geometry of a background by choosing the
Killing spinors to lie at a particular directions.
This simplification is possible for backgrounds with one and two supersymmetries.
It turns out that the stability subgroup of two generic spinors in $Spin(10,1)$
is the identity. Therefore, one does not expect that there will be a simplification
in the form of a third spinor in a generic background with three supersymmetries.
This is unless
the conditions on the geometry and on the fluxes imposed by the first two spinors
necessitate the vanishing of many components of $\Omega$ and ${\cal F}$ and so
the equations for the third Killing spinor are not involved.
In any case there are several special backgrounds with more than two supersymmetries
that admit spinors which have a non-trivial stability subgroup in $Spin(10,1)$.

Since in the basis of gamma matrices we have adopted, the frame time
direction is distinguished from the rest, it is convenient to
decompose the frame indices as $A=(0, i)$, where $i=1,\dots, 10$.
Then we introduce an orthonormal frame $\{e^A: A=0, \dots, 10\}$ and
write the spacetime metric as \be ds^2= -(e^0)^2+\sum_{i=1}^{10}
(e^i)^2~. \ee In this frame, the four-form field strength ${\cal F}$
can be expanded in electric and magnetic parts as \be {\cal
F}=\tfrac{1}{3!} e^0\wedge G_{ijk} e^i\wedge e^j\wedge e^k+
\tfrac{1}{4!} F_{ijkl} e^i\wedge e^j\wedge e^k\wedge e^l~. \ee The
spin (Levi-Civita) connection has non-vanishing components \be
\Omega_{0, ij}~,~~~~\Omega_{0,
0j}~,~~~~\Omega_{i,0j}~,~~~~\Omega_{i,jk}~. \ee The Killing spinor
equations decomposes as
\bea
\partial_0 \epsilon+\tfrac{1}{4} \Omega_{0, ij}\Gamma^{ij}\epsilon-\tfrac{1}{2}\Omega_{0,0i}
\Gamma_0\Gamma^{i}\epsilon-\tfrac{1}{288}\bigl(\Gamma_0
\Gamma^{ijkl} F_{ijkl}-8 G_{ijk} \Gamma^{ijk}\bigr)\epsilon&=&0\,,
\nonumber \\
\partial_i\epsilon+ \tfrac{1}{4} \Omega_{i,jk} \Gamma^{jk}\epsilon-
\tfrac{1}{2} \Omega_{i,0j} \Gamma_0 \Gamma^j\epsilon -\tfrac{1}{288}
\bigl (\Gamma_i{}^{jklm} F_{jklm}&& \nonumber \\ +4 \Gamma_0
\Gamma_i{}^{jkl} G_{jkl} -24 \Gamma_0 G_{ijk} \Gamma^{jk}- 8
F_{ijkl} \Gamma^{jkl}\bigr) \epsilon&=&0\,.
\la{deckse}
\eea This is the form of
the Killing spinor equations that we shall use later to derive our
results.

\subsection{Integrability conditions and field equations}

The integrability conditions of the Killing spinor equations ${\cal
D}\epsilon=0$ are \bea [{\cal D}_A, {\cal D}_B]\epsilon={\cal
R}_{AB}\epsilon=0~, \eea where ${\cal R}$ is the supercovariant
curvature which has been computed in \cite{nicolai, jffgp}. It has
been observed in \cite{pakis} that, using the Bianchi identity of
the Riemann curvature of spacetime, $\Gamma^B {\cal R}_{AB}
\epsilon=0$ can be written as
\begin{eqnarray}
 {\cal I}_A \epsilon & = & [ E_{AB} \Gamma^{B}
  + L_{C_1C_2C_3} (\Gamma_A{}^{C_1C_2C_3}
  - 6 \delta_A^{C_1} \Gamma^{C_2C_3}) + \nonumber \\
&& \;\; +B_{C_1\dots C_5} (\Gamma_A{}^{C_1 \cdots C_5}
 - 10 \delta_A^{C_1} \Gamma^{C_2 \cdots C_5}) ]\epsilon=0 \,,
 \label{intcond}
 \end{eqnarray}
 where
 \bea
 E_{AB}&:=& R_{AB} -\tfrac{1}{12} {\cal F}_{AC_1C_2C_3}{\cal F}_B{}^{C_1C_2C_3}
 +\tfrac{1}{144} g_{AB}{\cal F}_{C_1\cdots C_4} {\cal F}^{C_1\cdots
 C_4}\,, \nonumber \\
 L_{ABC}&:=&-\tfrac{1}{36} *(d*{\cal F}-\tfrac{1}{2} {\cal F}\wedge {\cal
 F})_{ABC}\,, \nonumber \\
 B_{A_1\dots A_5}& := & \tfrac{1}{6!} (d{\cal F})_{A_1\dots A_5}~.
 \eea
 The above integrability conditions can be written in terms of the frame $(e^0, e^i)$.
 This computation is similar to the one for the Killing spinor equations
 and we shall not repeat it here.
  It is clear that some of the components of the field equations
 (and Bianchi identity) are satisfied as integrability conditions
 of the Killing spinor equations\footnote{For an alternative approach see \cite{Bellorin:2005hy}.}. Sometimes it is customary to impose enough
 conditions on the field equations  and on Bianchi identity ${\cal F}$ such that
 all Einstein equations are satisfied. This is because the field equations
 and Bianchi identity of  ${\cal F}$ are easier to solve.

\newsection{The six types of spinors}

A direct consequence of the construction of the $Spin(10,1)$
 Majorana spinor representation in appendix A is that
 any Majorana Killing spinor can
be written in terms of forms as
\bea
\epsilon&=& f (1+e_{12345})+ i
g (1- e_{12345}) +\sqrt{2} u^i (e_i+
\tfrac{1}{4!}\epsilon_i{}^{jklm} e_{jklm}) + i \sqrt{2} v^i
(e_i-\tfrac{1}{4!}\epsilon_i{}^{jklm} e_{jklm}) \nonumber \\ &&+
\tfrac{1}{2} w^{ij} (e_{ij}-\tfrac{1}{3!} \epsilon_{ij}{}^{klm}
e_{klm}) + \tfrac{i}{2} z^{ij}(e_{ij}+\tfrac{1}{3!}
\epsilon_{ij}{}^{klm} e_{klm})~, \la{genspinor}
\eea
where $f,g,
u^i, v^i, w^{ij}$ and $z^{ij}$ are real spacetime functions. The six
types of spinors $e_{i_1 \cdots i_I}$ with $i=0,\ldots,5$ correspond
to the irreducible representation of $U(5)$ on $\Lambda^*(\bC^5)$
and are denoted by $\sigma_I$.

The Killing spinor equations for $\epsilon$ are
\bea {\cal
D}_A\epsilon=\partial_A f (1+e_{12345})+ i
\partial_A g (1- e_{12345})+ \sqrt{2}\partial_Au^i (e_i+
\tfrac{1}{4!}\epsilon_i{}^{jklm} e_{jklm}) \nonumber \\ +i
\sqrt{2}\partial_Av^i (e_i- \tfrac{1}{4!}\epsilon_i{}^{jklm}
e_{jklm}) + \tfrac{1}{2}
\partial_A w^{ij} (e_{ij}-\tfrac{1}{3!} \epsilon_{ij}{}^{klm} e_{klm})
\nonumber \\ +\tfrac{i}{2}  \partial_A z^{ij}(e_{ij}+\tfrac{1}{3!}
\epsilon_{ij}{}^{klm} e_{klm}) +f{\cal D}_A (1+e_{12345})+ i g{\cal
D}_A (1+e_{12345}) \nonumber \\ +\sqrt{2} u^i {\cal D}_A(e_i+
\tfrac{1}{4!}\epsilon_i{}^{jklm} e_{jklm}) +i\sqrt{2} v^i {\cal
D}_A(e_i- \tfrac{1}{4!}\epsilon_i{}^{jklm} e_{jklm})
\nonumber \\ +
\tfrac{1}{2}w^{ij} {\cal D}_A(e_{ij}-\tfrac{1}{3!}
\epsilon_{ij}{}^{klm} e_{klm}) +\tfrac{i}{2}   z^{ij} {\cal
D}_A(e_{ij}+\tfrac{1}{3!} \epsilon_{ij}{}^{klm} e_{klm})=0~.
\la{genspcov}
\eea
 Thus the Killing spinor equations reduce to the
evaluation of the supercovariant derivative on the spinors
$\sigma_I$. So it remains to compute the
\bea
{\cal D}_A1~,~~~{\cal D}_Ae_i~,~~~ {\cal D}_Ae_{ij}~,~~~{\cal
D}_Ae_{ijk}~,~~~{\cal D}_Ae_{ijkl}~,~~~{\cal D}_Ae_{12345}~,
\la{suptypes}
\eea
and express
the result in the basis (\ref{hermbasis}). Note that in some cases it is possible
to put some spinors in a canonical or normal form using the $Spin(10,1)$ gauge
symmetry of the supercovariant connection ${\cal D}$, see \cite{joe}. As a result
the spinors depend on less functions than those indicated in (\ref{genspinor}). In such
cases, it is helpful to consider the orbits of $Spin(10,1)$ in the space of spinors
\cite{bryant, jose}.

The same analysis can be done for the integrability condition
${\cal I}\epsilon=0$ of a Killing spinor $\epsilon$. Since this
condition is linear, we have
\bea
{\cal I}\epsilon&=& f {\cal
I}(1+e_{12345})+ i g {\cal I}(1- e_{12345}) +\sqrt{2} u^i {\cal
I}(e_i+ \tfrac{1}{4!}\epsilon_i{}^{jklm} e_{jklm})
\nonumber
\\
&& + i \sqrt{2} v^i
{\cal I}(e_i-\tfrac{1}{4!}\epsilon_i{}^{jklm} e_{jklm})
+ \tfrac{1}{2} w^{ij} {\cal I}(e_{ij}-\tfrac{1}{3!}
\epsilon_{ij}{}^{klm} e_{klm})
\nonumber
\\
&& + \tfrac{i}{2} z^{ij}{\cal
I}(e_{ij}+\tfrac{1}{3!} \epsilon_{ij}{}^{klm} e_{klm})~.
\la{ingenspinor}
\eea
Therefore to find which field equations are
determined by the Killing spinor equations, it suffices to compute
\bea
{\cal I}_A1~,~~~{\cal I}_A e_i~,~~~{\cal I}_A e_{ij}~,~~~ {\cal
I}_A e_{klm}~,~~~ {\cal I}_A e_{jklm}~,~~~{\cal I}_A e_{12345}\,,
\eea
and then solve the linear system.

\newsection{Linear systems in a timelike basis}

\subsection{The linear system of Killing spinor equations}

We would like to explain how one evaluates the supercovariant
derivative on an arbitrary basis element
 \be
  e_{i_1 \cdots i_I} = \frac{1}{2^{I / 2}} \Gamma^{\bar i_1}
  \cdots \Gamma^{\bar i_I} 1 \,.
 \label{basis-element}
 \ee
where the indices $i_1, \ldots, i_I$ pick out $I$ holomorphic
indices (with $0 \leq I \leq 5$) from the range $\a = 1, \ldots, 5$.
It will be convenient to distinguish between the indices that do
appear in the basis element (\ref{basis-element}) and those that do
not: we split the holomorphic indices $\a$ into the
indices\footnote{The $i_1,\ldots,i_I$ should not be thought of as
indices in this context, but rather as fixed labels for a particular
spinor.} $a = (i_1, \ldots, i_I)$ and the remaining $5-I$ indices
$p$, and similarly for the anti-holomorphic indices $\bar \a$. Note
that $\Gamma^{\bar a}$ and $\Gamma^p$ annihilate the spinor $e_{i_1
\cdots i_I}$ while $\Gamma^{a}$ and $\Gamma^{\bar p}$ act as
creation operators. For this reason it is useful to define the new
indices $\rho, \sigma, \tau$ consisting of the combination
 \be
  \rho = (\bar a_1, \ldots, \bar a_I, p_1, \ldots,
  p_{5-I}) \,, \qquad  \bar{\rho} = (a_1, \ldots, a_I, \bar p_1, \ldots, \bar p_{5-I})
 \,,
 \ee
where $\Gamma^\rho$ and $\Gamma^{\bar \rho}$ are the annihilation
and creation operators, respectively, for the spinor  $e_{i_1 \cdots
i_I}$. Note that the indices $\a$ and $\rho$ are
 identical for $I=0$,
i.e.~for the spinor $1$. For $I > 0$, i.e.~for any other basis
element, these indices differ.

In terms of the basis\footnote{Note that in this basis $e_{i_1
\cdots i_I}$ is the Clifford algebra vacuum.}
 \be
  \{ e_{i_1 \cdots i_I}, \Gamma^{\bar \sigma_1} e_{i_1 \cdots i_I}, \ldots,
  \Gamma^{\bar \sigma_1 \cdots \bar \sigma_5} e_{i_1 \cdots i_I}
  \} \,, \label{element-basis}
 \ee
the supercovariant derivative with $A =0$ can be expanded in the
following contributions:
 \bea
 {{\cal D}}_0 e_{i_1 \cdots i_I} &=& [\tfrac{1}{2}\Omega_{0, \tau}{}^{ \tau}
 + (-1)^{I+1} \tfrac{i}{24} F_{\tau_1}{}^{ \tau_1}{}_{ \tau_2} {}^{
 \tau_2}] e_{i_1 \cdots i_I}
 + [(-1)^I \tfrac{i}{2} \Omega_{0,0 \bar \sigma}
  + \tfrac{1}{6} G_{\bar \sigma \tau}{}^{\tau}] \Gamma^{\bar \sigma} e_{i_1 \cdots i_I}
   \nonumber \\
  &+& [\tfrac{1}{4} \Omega_{0,\bar \sigma_1 \bar \sigma_2}
  + (-1)^{I+1} \tfrac{i}{24} F_{\bar \sigma_1 \bar \sigma_2 \tau}{}^{\tau}]
  \Gamma^{\bar \sigma_1 \bar \sigma_2} e_{i_1 \cdots i_I}
  + [\tfrac{1}{36} G_{\bar \sigma_1 \bar \sigma_2 \bar \sigma_3}]
\Gamma^{\bar \sigma_1 \bar \sigma_2 \bar \sigma_3} e_{i_1 \cdots
i_I} \nonumber \\ &+& [(-1)^{I+1} \tfrac{i}{288} F_{\bar \sigma_1
\cdots \bar \sigma_4}] \Gamma^{\bar \sigma_1 \cdots \bar \sigma_4}
e_{i_1 \cdots i_I}~. \la{genzcom}
 \eea
Observe that the component $\Gamma^{\bar \sigma_1 \cdots \bar
\sigma_5} e_{i_1 \cdots i_I}$ vanishes. Similarly, the expression
for $A = \rho$ read
 \bea
 {{\cal D}}_{\rho} e_{i_1
\cdots i_I}  &=& [\tfrac{1}{2}\Omega_{\rho, \sigma}{}^{ \sigma}
 + (-1)^{I} \tfrac{i}{4} G_{\rho \sigma} {}^{\sigma}] e_{i_1 \cdots i_I}
 + [(-1)^I \tfrac{i}{2} \Omega_{\rho,0 \bar \sigma}
  + \tfrac{1}{4} F_{\rho \bar \sigma \tau}{}^{\tau}
  - \tfrac{1}{24} g_{\rho \bar \sigma} F \cont{\tau_1} \cont{\tau_2}]
  \Gamma^{\bar \sigma} e_{i_1 \cdots i_I}
   \nonumber \\
  &+& [\tfrac{1}{4} \Omega_{\rho,\bar \sigma_1 \bar \sigma_2}
  + (-1)^I \tfrac{i}{8} G_{\rho \bar \sigma_1 \bar \sigma_2} + [
    (-1)^{I+1} \tfrac{i}{12} g_{\rho [ \bar \sigma_1} G_{\bar \sigma_2 ]}] \cont{\tau}]
    \Gamma^{\bar \sigma_1 \bar \sigma_2} e_{i_1 \cdots i_I}
     \nonumber \\
&+& [\tfrac{1}{24} F_{\rho \bar \sigma_1 \bar \sigma_2 \bar
\sigma_3} -
  \tfrac{1}{24} g_{\rho [ \bar \sigma_1} F_{\bar \sigma_2 \bar \sigma_3 ]} \cont{\tau}]
  \Gamma^{\bar \sigma_1 \bar \sigma_2 \bar \sigma_3} e_{i_1 \cdots i_I}
  \nonumber \\
   &+& [(-1)^{I+1} \tfrac{i}{72} g_{\rho [ \bar \sigma_1}
G_{\bar \sigma_2 \bar \sigma_3 \bar \sigma_4 ]}] \Gamma^{\bar
\sigma_1 \cdots \bar \sigma_4} e_{i_1 \cdots i_I} + [
 -\tfrac{1}{288} g_{\rho [ \bar \sigma_1} F_{\bar \sigma_2 \cdots \bar \sigma_5
 ]}]
 \Gamma^{\bar \sigma_1 \cdots \bar \sigma_5} e_{i_1 \cdots i_I}
 \la{genrcom}
 \eea
Finally, for $A =\bar \rho$ we find
 \bea
 {{\cal D}}_{{\bar \rho}} e_{i_1 \cdots
i_I}&=& [\tfrac{1}{2}\Omega_{{\bar \rho},\sigma}{}^{\sigma}
 + (-1)^{I} \tfrac{i}{12} G_{{\bar
\rho}  \sigma} {}^{ \sigma}] e_{i_1 \cdots i_I} + [(-1)^I
\tfrac{i}{2} \Omega_{{\bar \rho},0 \bar \sigma}
  + \tfrac{1}{12} F_{{\bar
\rho} \bar \sigma \tau}{}^{\tau}]\Gamma^{\bar \sigma} e_{i_1 \cdots
i_I}
 \nonumber \\
 &+& [\tfrac{1}{4} \Omega_{{\bar
\rho},\bar \sigma_1 \bar \sigma_2}
  + (-1)^I \tfrac{i}{24} G_{{\bar
\rho} \bar \sigma_1 \bar \sigma_2}]  \Gamma^{\bar \sigma_1 \bar \sigma_2}
 e_{i_1 \cdots i_I}
 + [\tfrac{1}{72} F_{{\bar
\rho} \bar \sigma_1 \bar \sigma_2 \bar \sigma_3}]
 \Gamma^{\bar \sigma_1 \bar \sigma_2 \bar \sigma_3} e_{i_1 \cdots i_I}~.
 \la{genrbcom}
 \eea
Observe that the components along $\Gamma^{\bar \sigma_1 \cdots \bar \sigma_4}
 e_{i_1 \cdots i_I}$
and $\Gamma^{\bar \sigma_1 \cdots \bar \sigma_5} e_{i_1 \cdots i_I}$
vanish.

It is convenient to convert the above expressions
 from  basis
(\ref{element-basis}) to the ``canonical'' basis
 \be
  \{ 1, \Gamma^{\bar \a} 1, \ldots, \Gamma^{\bar \a_1 \cdots \bar \a_5}
  1 \} \,. \label{canonical-basis}
 \ee
{}For this, we expand the products of $\Gamma^{\bar \rho}$ matrices,
which are creation operators on $e_{i_1 \cdots i_I}$, into a sum of
products of $\Gamma^{a}$ and $\Gamma^{\bar p}$ matrices, which are
annihilation and creation operators, respectively, on $1$. Then we
act on $e_{i_1 \cdots i_I}$ with the annihilation operators. In
particular, we have
 \bea
  {\cal D}_A e_{i_1 \cdots i_I}&=& \sum_k [{\cal D}_A e_{i_1 \cdots i_I}]_{\bar\rho_1\cdots \bar\rho_k}
  \Gamma^{\bar\rho_1\cdots\bar\rho_k} e_{i_1 \cdots i_I} \nonumber \\ &=& \sum_k \sum_{m+n=k}
 \frac{k!}{m! n!} [{\cal D}_{A} e_{i_1
 \cdots i_I}]_{a_1 \cdots a_m \bar p_1   \cdots \bar p_n}
 \Gamma^{a_1 \cdots a_m} \Gamma^{\bar p_1 \cdots
 \bar p_n} e_{i_1\dots i_I}
 \nonumber \\
 &=& \sum_k \sum_{m+n=k}
 \frac{k!}{m! n!}   \frac{(-1)^{[m/2]+nI}}{{2}^{{I / 2}-m} (I-m)!}
  \epsilon^{a_1 \cdots a_m}{}_{\bar a_{m+1} \cdots \bar a_I}
  \nonumber \\
  &&
  [{\cal D}_A e_{i_1 \cdots i_I}]_{a_1 \cdots a_m \bar p_1 \cdots \bar p_n}
   \Gamma^{\bar a_{m+1} \cdots \bar a_I \bar p_1 \cdots
  \bar p_n} 1\,,
  \la{genform}
 \eea
with the obvious restrictions $m \leq I$ and $n \leq 5-I$ and the
convention that $\epsilon_{\bar i_1 \cdots \bar i_I} = 1$. Using the
expressions  (\ref{genzcom}), (\ref{genrcom}) and (\ref{genrbcom})
for the components of ${\cal D}_A e_{i_1 \cdots i_I}$ in the basis
(\ref{element-basis}) which appear in square brackets in
(\ref{genform}), one can easily compute the components of ${\cal
D}_A e_{i_1 \cdots i_I}$ in the canonical basis
(\ref{canonical-basis}). For convenience we give the explicit
expressions for the different basis elements in
appendix~\ref{Killing-list}.

{}From the expression (\ref{genform}) one can also derive a relation
between the Killing spinors equations from $e_{i_1 \cdots i_I}$ and
$e_{i_{I+1} \cdots i_{5}}$, whose labels satisfy $\e_{i_1 \cdots i_I
i_{I+1} \cdots i_{5}} = 1$. The key observation is that, in the
basis (\ref{basis-element}),
 \be
  ({\cal D}_{A} e_{i_1 \cdots i_I})_{{\bar \sigma}_1 \cdots {\bar
\sigma}_i} =
  ({{\cal D}}_{A} e_{i_{I+1} \cdots i_5})^*_{{\bar \sigma}_1 \cdots
{\bar \sigma}_i} \,,
\la{complexconj}
 \ee
where the notation, i.e.~the division of $\a$ and $\bar \a$ into
$\s$ and $\bar \s$, is based on $e_{i_1 \cdots i_I}$ and not on
$e_{i_{I+1} \cdots i_5}$ (as it will be in the remainder of this
section). Converting both expressions to the canonical basis using
(\ref{genform}), one finds that the previous relation translates
into
 \bea
  ({{\cal D}}_{A} e_{i_1 \cdots i_I})_{{\bar a}_{1} \cdots {\bar a}_m \bar
p_1 \cdots {\bar p}_n} = \frac{2^{2-m-n} (-)^{[(m+n)/2] + [I/2]}
(5-m-n)!}{(m+n)!(I-m)!(5-I-n)!} \cdot && \nonumber
\\
  \cdot \te_{{\bar a}_{1} \cdots {\bar a}_m \bar p_1 \cdots {\bar p}_n}{}^{a_{m+1} \cdots a_I
  { p}_{n+1} \cdots { p}_{5-I}}
  ({{\cal D}}_{A} e_{i_{I+1} \cdots i_5})^*_{a_{m+1} \cdots a_I { p}_{n+1}
\cdots { p}_{5-I}} \,. &&
 \eea
After the addition of the complex conjugated and dualised version of
this expression to its original, one finds that the components of
the combination $e_{i_1 \cdots i_I} + (-1)^{[I/2]} e_{i_{I+1} \cdots
i_5}$ are related to each other:
 \bea
  ({{\cal D}}_{A} e_{i_1 \cdots i_I} + (-1)^{[I/2]} {{\cal D}}_{A} e_{i_{I+1}
\cdots i_5})_{{\bar a}_{1} \cdots {\bar a}_m \bar p_1 \cdots {\bar
p}_n} = \frac{2^{2-m-n} (-)^{[(m+n)/2]}
(5-m-n)!}{(m+n)!(I-m)!(5-I-n)!} \cdot && \nonumber
\\
  \cdot \te_{{\bar a}_{1} \cdots {\bar a}_m \bar p_1 \cdots {\bar p}_n}{}^{a_{m+1} \cdots a_I
  { p}_{n+1} \cdots { p}_{5-I}}
  ({{\cal D}}_{A} e_{i_1 \cdots i_I} + (-1)^{[I/2]} {{\cal D}}_{A} e_{i_{I+1}
\cdots i_5})^*_{a_{m+1} \cdots a_I { p}_{n+1} \cdots { p}_{5-I}} \,.
&& \label{dual-relation}
 \eea
A similar expression holds for the components of $i{{\cal D}}_{A} e_{i_1 \cdots i_I} -i (-1)^{[I/2]} {{\cal D}}_{A} e_{i_{I+1}
\cdots i_5}$.
This relates the Killing spinor equations of any real Majorana
spinor (\ref{genspinor}). For this reason one only has to consider
half of all equations; in appendix~\ref{Killing-list} we give all
$A=\bar \a$ equations plus the $A=0$ equations coming with less than
three $\Gamma^{\bar \a}$-matrices.

\subsection{The linear system of integrability conditions}

As we have explained the integrability condition (\ref{intcond}) on
any Killing spinor, ${\cal I}\epsilon$,  can be expressed in terms
of ${\cal I}\sigma_I$. In turn ${\cal I}\sigma_I$ can be expanded in
the basis (\ref{element-basis}). For this, one inserts
  $e_{i_1 \cdots i_I}$ in (\ref{intcond}), expands the resulting
  equation in  (\ref{element-basis}) and sets $A = 0$  to find that
 \bea
 {{{\cal I}}}_0 e_{i_1 \cdots i_I} &=&  [(-1)^{I+1} i E_{00} - 12 L_{0  \sigma}{}^{ \sigma} - 120 B_{0
   \sigma}{}^{ \sigma}{}_{\tau}{}^{\tau}]  e_{i_1 \cdots i_I}
\nonumber \\
  &+& [E_{0 \bar \sigma} +
  (-1)^{I+1} 6 i L_{\bar \sigma \tau}{}^{\tau} +
  (-1)^{I+1} 60 i B_{\bar \sigma \tau_1}{}^{\tau_1}{}_{\tau_2}{}^{\tau_2}]
  \Gamma^{\bar \sigma} e_{i_1 \cdots i_I}
   \nonumber \\
  &+& [- 6 L_{0 \bar \sigma_1 \bar \sigma_2} - 120 B_{0 \bar \sigma_1 \bar \sigma_2}
  \cont{\tau}] \Gamma^{\bar \sigma_1 \bar \sigma_2} e_{i_1 \cdots i_I}
   \nonumber \\
 &+& [(-1)^{I+1} i L_{\bar \sigma_1 \bar \sigma_2 \bar \sigma_3}
 + (-)^{I+1} 20 i B_{\bar \sigma_1 \bar \sigma_2 \bar \sigma_3} \cont{\tau}]
 \Gamma^{\bar \sigma_1 \bar \sigma_2 \bar \sigma_3} e_{i_1 \cdots i_I}
  \nonumber \\
&+& [-10 B_{0 \bar \sigma_1 \cdots \bar \sigma_4} \Gamma^{\bar
\sigma_1 \cdots \bar \sigma_4} e_{i_1 \cdots i_I} + (-1)^{I+1} i
B_{\bar \sigma_1 \cdots \bar \sigma_5} ]\Gamma^{\bar \sigma_1 \cdots
\bar \sigma_5} e_{i_1 \cdots i_I}
 \eea
Similarly for $A = \rho$,  one finds
 \bea
{{{\cal I}}}_{\rho} e_{\a_1 \cdots \a_I}  &=& [(-1)^{I+1} i E_{0
\rho} - 18 L_{\rho} \cont{ \sigma}- 180
  B_{\rho} \cont{ \tau_1} \cont{\tau_2}] e_{i_1 \cdots i_I}
   \nonumber \\
  &+&[ E_{\rho \bar \sigma} + (-1)^{I+1} 6 i g_{\rho \bar \sigma} L_{0}
  \cont{\tau} + (-1)^I 18 i L_{0 \rho \bar \sigma}
  \nonumber \\
  && + (-1)^{I+1} 60 i g_{\rho \bar \sigma}
  B_{0} \cont{\tau_1} \cont{\tau_2}
  + (-1)^I 360 i B_{0 \rho \bar \sigma}
  \cont{\tau}] \Gamma^{\bar \sigma} e_{i_1 \cdots i_I}
  \nonumber \\
  &+& [6 g_{\rho [ \bar \sigma_1} L_{\bar \sigma_2 ]} \cont{\tau} - 9
  L_{\rho \bar \sigma_1 \bar \sigma_2} + 60 g_{\rho [ \bar \sigma_1} B_{\bar \sigma_2 ]}
  \cont{\tau_1} \cont{\tau_2} - 180 B_{\rho \bar \sigma_1 \bar \sigma_2}
  \cont{\tau}]\Gamma^{\bar \sigma_1 \bar \sigma_2} e_{i_1 \cdots i_I}
  \nonumber \\
   &+& [(-1)^{I+1} 3 i g_{\rho [ \bar \sigma_1} L_{0 \bar \sigma_2
\bar \sigma_3]} + (-1)^{I+1}  60 i
  g_{\rho [ \bar \sigma_1} B_{0 \bar \sigma_2 \bar \sigma_3 ]} \cont{\tau} + (-1)^I
  60 i  B_{0 \rho \bar \sigma_1 \bar \sigma_2 \bar \sigma_3}]
  \Gamma^{\bar \sigma_1 \bar \sigma_2 \bar \sigma_3} e_{i_1 \cdots i_I}
  \nonumber \\
&+&[ g_{\rho [ \bar \sigma_1} L_{\bar \sigma_2 \cdots \bar \sigma_4
]} + 20
  g_{\rho [ \bar \sigma_1} B_{\bar \sigma_2 \cdots \bar \sigma_4 ]}
  \cont{\tau} - 15 B_{\rho \bar \sigma_1 \cdots \bar \sigma_4}]
  \Gamma^{\bar \sigma_1 \cdots \bar \sigma_4} e_{i_1 \cdots i_I}
  \nonumber \\
   &+& [(-1)^{I+1} 5 i g_{\rho [ \bar \sigma_1} B_{0 \bar \sigma_2
\cdots \bar \sigma_5
  ] } ] \Gamma^{\bar \sigma_1 \cdots \bar \sigma_5} e_{i_1 \cdots i_I}
 \eea
Finally for $A =\bar \rho$, we find
 \bea
{{{\cal I}}}_{{\bar \rho}} e_{\a_1 \cdots \a_I}&=& [(-)^{I+1} i E_{0
\bar \rho} - 6 L_{\bar \rho} \cont{ \sigma} - 60
  B_{\bar \rho} \cont{ \sigma} \cont{\tau}] e_{i_1 \cdots i_I}
\nonumber \\
  &+& [E_{\bar \rho \bar \sigma} + (-)^I 6 i L_{0 \bar
\rho \bar \sigma} + (-)^I 120 i B_{0 \bar \rho \bar \sigma}
\cont{\tau}] \Gamma^{\bar \sigma} e_{i_1 \cdots i_I} \nonumber \\
 &+& [- 3 L_{\bar
\rho \bar \sigma_1 \bar \sigma_2} - 60 B_{\bar \rho \bar \sigma_1
\bar \sigma_2} \cont{\tau} ] \Gamma^{\bar \sigma_1 \bar \sigma_2}
e_{i_1 \cdots i_I} \nonumber \\ &+& [(-)^I 20 i B_{0 \bar \rho \bar
\sigma_1 \bar \sigma_2 \bar \sigma_2}] \Gamma^{\bar \sigma_1 \bar
\sigma_2 \bar \sigma_3} e_{i_1 \cdots i_I} + [-5 B_{\bar \rho \bar
\sigma_1 \cdots \bar \sigma_4}] \Gamma^{\bar \sigma_1 \cdots \bar
\sigma_4} e_{i_1 \cdots i_I}
\eea
Observe that the $\Gamma^{\bar
\sigma_1 \cdots \bar \sigma_5} e_{i_1 \cdots i_I}$ component of the
last integrability condition vanishes. It is straightforward to
convert the above expressions to the canonical basis
(\ref{canonical-basis}). This is completely similar to that for the
Killing spinor equations in (\ref{genform}) and we shall not repeat
the expression here. In addition, a relation similar to
(\ref{dual-relation}) holds for the integrability conditions. In
appendix~\ref{integrability-list} we give the explicit expressions
for ${\cal I}\sigma_I$ in the canonical basis.

\newsection{Linear systems in a null basis}

The construction of the linear systems in the previous section
applies to all Killing spinors, i.e.~to  spinors that represent the
orbit of $Spin(10,1)$ with stability subgroup $SU(5)$ and the
spinors that represent the orbit of $Spin(10,1)$ with stability
subgroup $(Spin(7)\ltimes \bR^8)\times \bR$. However, if it is known
that one of the Killing spinors represents the orbit with stability
subgroup $(Spin(7)\ltimes \bR^8)\times \bR$, it may be more
convenient to use the null basis of appendix A to construct the
linear systems for the Killing spinor equations and the associated
integrability conditions. This is because the gauge symmetry of the
supercovariant connection can be used to put that spinor along the
direction $1+e_{1234}$, see appendix A.

The timelike and null  bases in the space of spinors are oscillator
bases. Because of this, the linear system of the Killing spinor
equations and of the integrability conditions which we have
derived for the timelike basis in the previous section are easily
adapted  to the null basis. We shall demonstrate this for the linear
system of the Killing spinor equations. Since in the null basis the
tenth direction, which we denote by $\nat$, is separated from the
rest, see appendix A, we decompose the four-form field strength ${\cal
F}$  as
 \be
  {\cal F}=\tfrac{1}{3!} e^\natural \wedge G_{ijk} e^i\wedge e^j\wedge e^k+
  \tfrac{1}{4!} F_{ijkl} e^i\wedge e^j\wedge e^k\wedge e^l~,
 \ee
 where $i=0,1,2,\dots,9$. We have denoted the tenth component  and the remaining
 components of ${\cal F}$ as the electric and magnetic components, respectively, that appear in the
decomposition of ${\cal F}$ in the  timelike basis. The reason for this will become apparent.
The spin (Levi-Civita) connection has non-vanishing components
 \be
  \Omega_{\natural, ij}~,~~~~\Omega_{\natural,
  \natural j}~,~~~~\Omega_{i,\natural j}~,~~~~\Omega_{i,jk}~.
 \ee
The Killing spinor equations decomposes as
 \bea
  \partial_\natural \epsilon+\tfrac{1}{4} \Omega_{\natural, ij}\Gamma^{ij}\epsilon
  + \tfrac{1}{2}\Omega_{\natural,\natural i}
  \Gamma^\natural \Gamma^{i}\epsilon-\tfrac{1}{288}\bigl(\Gamma_\natural
  \Gamma^{ijkl} F_{ijkl}-8 G_{ijk} \Gamma^{ijk}\bigr)\epsilon&=&0\,,
  \nonumber
  \\
  \partial_i\epsilon+ \tfrac{1}{4} \Omega_{i,jk} \Gamma^{jk}\epsilon +
  \tfrac{1}{2} \Omega_{i,\natural j} \Gamma^\natural \Gamma^j\epsilon - \tfrac{1}{288}
  \bigl (\Gamma_i{}^{jklm} F_{jklm}&& \nonumber
  \\
  - 4 \Gamma^\natural
  \Gamma_i{}^{jkl} G_{jkl} + 24 \Gamma^\natural G_{ijk} \Gamma^{jk}- 8
  F_{ijkl} \Gamma^{jkl}\bigr) \epsilon&=&0\,.
 \eea
Observe that these formulae can be derived from those of the timelike basis in (\ref{deckse}) after the replacement
 $0 \rightarrow \nat$. It is clear from this that the linear system for the null basis associated
 with the Killing spinor equations can be derived from that of the timelike basis, we have derived, after taking into
 account the different way that $\Gamma_0$ and  $\Gamma_\nat$ act on the basis spinors.

Every spinor in the null basis can be written as a linear combination of six types of spinors. These
spinors are constructed by the creation operators of the null basis acting on the Clifford vacuum $1$,
see appendix A. In particular, we have that  $\Gamma_\natural = - \Gamma_0 \Gamma_1 \cdots \Gamma_9$ acts
 on our null basis of
spinors as
 \be
  \Gamma_\natural e_{i_1 \cdots i_I} = \Gamma^\natural e_{i_1 \cdots i_I} = (-1)^{I+1} e_{i_1 \cdots i_I} \,.
 \ee
This means that the difference between the timelike and the null case consists of replacing $\Gamma^0$ by
$\Gamma^\natural$ in most cases. This amounts to the replacement $i \rightarrow + 1$. The only exception is the
$F$-term in the ${\cal D}_\nat$ component of the supercovariant derivative,
where $\Gamma_0$ is replaced by $\Gamma_\natural$ and so there is an additional minus sign.

As in the timelike case, it will be convenient to distinguish between the indices that do
appear in the basis element $e_{i_1 \cdots i_I}$ and those that do
not. In particular, we split the indices $\a$ into the
indices $a = (i_1, \ldots, i_I)$ and the remaining $5-I$ indices
$p$, and similarly for the indices $\bar \a$.
Subsequently we define the new
indices $\rho, \sigma, \tau$ consisting of the combination\footnote{Note that
$\rho$ and $\bar \rho$ are no longer complex conjugate
due to the presence of the $(+,-)$ null indices.}
 \be
  \rho = (\bar a_1, \ldots, \bar a_I, p_1, \ldots,
  p_{5-I}) \,, \qquad  \bar{\rho} = (a_1, \ldots, a_I, \bar p_1, \ldots, \bar p_{5-I})
 \,,
 \ee
where $\Gamma^\rho$ and $\Gamma^{\bar \rho}$ are the annihilation
and creation operators, respectively, for the spinor  $e_{i_1 \cdots
i_I}$. Next consider the basis in the space of spinors associated
with the Clifford vacuum $e_{i_1 \cdots i_I}$, i.e.
 \be
  \{ e_{i_1 \cdots i_I}, \Gamma^{\bar \sigma_1} e_{i_1 \cdots i_I}, \ldots,
  \Gamma^{\bar \sigma_1 \cdots \bar \sigma_5} e_{i_1 \cdots i_I}
  \} \,, \label{element-basis-null}
 \ee
In this basis the supercovariant derivative with $A =\natural$ can be expanded as
 \bea
 {{\cal D}}_\natural e_{i_1 \cdots i_I} &=& [\tfrac{1}{2}\Omega_{\natural, \tau}{}^{ \tau}
 + (-1)^{I} \tfrac{1}{24} F_{\tau_1}{}^{ \tau_1}{}_{ \tau_2} {}^{
 \tau_2}] e_{i_1 \cdots i_I}
 + [(-1)^{I} \tfrac{1}{2} \Omega_{\natural,\natural \bar \sigma}
  + \tfrac{1}{6} G_{\bar \sigma \tau}{}^{\tau}] \Gamma^{\bar \sigma} e_{i_1 \cdots i_I}
   \nonumber \\
  &+& [\tfrac{1}{4} \Omega_{\natural,\bar \sigma_1 \bar \sigma_2}
  + (-1)^{I} \tfrac{1}{24} F_{\bar \sigma_1 \bar \sigma_2 \tau}{}^{\tau}]
  \Gamma^{\bar \sigma_1 \bar \sigma_2} e_{i_1 \cdots i_I}
  + [\tfrac{1}{36} G_{\bar \sigma_1 \bar \sigma_2 \bar \sigma_3}]
\Gamma^{\bar \sigma_1 \bar \sigma_2 \bar \sigma_3} e_{i_1 \cdots i_I} \nonumber \\ &+& [(-1)^{I} \tfrac{1}{288}
F_{\bar \sigma_1 \cdots \bar \sigma_4}] \Gamma^{\bar \sigma_1 \cdots \bar \sigma_4} e_{i_1 \cdots i_I}~.
\la{genzcom-null}
 \eea
Similarly, the expression for $A = \rho$ read
 \bea
 {{\cal D}}_{\rho} e_{\a_1
\cdots \a_I}  &=& [\tfrac{1}{2}\Omega_{\rho, \sigma}{}^{ \sigma}
 + (-1)^{I} \tfrac{1}{4} G_{\rho \sigma} {}^{\sigma}] e_{i_1 \cdots i_I}
 + [(-1)^{I} \tfrac{1}{2} \Omega_{\rho,\natural \bar \sigma}
  + \tfrac{1}{4} F_{\rho \bar \sigma \tau}{}^{\tau}
  - \tfrac{1}{24} g_{\rho \bar \sigma} F \cont{\tau_1} \cont{\tau_2}]
  \Gamma^{\bar \sigma} e_{i_1 \cdots i_I}
   \nonumber \\
  &+& [\tfrac{1}{4} \Omega_{\rho,\bar \sigma_1 \bar \sigma_2}
  + (-1)^{I} \tfrac{1}{8} G_{\rho \bar \sigma_1 \bar \sigma_2} + [
    (-1)^{I+1} \tfrac{1}{12} g_{\rho [ \bar \sigma_1} G_{\bar \sigma_2 ]}] \cont{\tau}]
    \Gamma^{\bar \sigma_1 \bar \sigma_2} e_{i_1 \cdots i_I}
     \nonumber \\
&+& [\tfrac{1}{24} F_{\rho \bar \sigma_1 \bar \sigma_2 \bar \sigma_3} -
  \tfrac{1}{24} g_{\rho [ \bar \sigma_1} F_{\bar \sigma_2 \bar \sigma_3 ]} \cont{\tau}]
  \Gamma^{\bar \sigma_1 \bar \sigma_2 \bar \sigma_3} e_{i_1 \cdots i_I}
  \nonumber \\
   &+& [(-1)^{I+1} \tfrac{1}{72} g_{\rho [ \bar \sigma_1}
G_{\bar \sigma_2 \bar \sigma_3 \bar \sigma_4 ]}] \Gamma^{\bar \sigma_1 \cdots \bar \sigma_4} e_{i_1 \cdots i_I} +
[
 -\tfrac{1}{288} g_{\rho [ \bar \sigma_1} F_{\bar \sigma_2 \cdots \bar \sigma_5
 ]}]
 \Gamma^{\bar \sigma_1 \cdots \bar \sigma_5} e_{i_1 \cdots i_I}\,.
 \la{genrcom-null}
 \eea
Finally, for $A =\bar \rho$ we find
 \bea
 {{\cal D}}_{{\bar \rho}} e_{\a_1 \cdots
\a_I}&=& [\tfrac{1}{2}\Omega_{{\bar \rho},\sigma}{}^{\sigma}
 + (-1)^{I} \tfrac{1}{12} G_{{\bar
\rho}  \sigma} {}^{ \sigma}] e_{i_1 \cdots i_I} + [(-1)^I \tfrac{1}{2} \Omega_{{\bar \rho},\natural \bar \sigma}
  + \tfrac{1}{12} F_{{\bar
\rho} \bar \sigma \tau}{}^{\tau}]\Gamma^{\bar \sigma} e_{i_1 \cdots i_I}
 \nonumber \\
 &+& [\tfrac{1}{4} \Omega_{{\bar
\rho},\bar \sigma_1 \bar \sigma_2}
  + (-1)^I \tfrac{1}{24} G_{{\bar
\rho} \bar \sigma_1 \bar \sigma_2}]  \Gamma^{\bar \sigma_1 \bar \sigma_2}
 e_{i_1 \cdots i_I}
 + [\tfrac{1}{72} F_{{\bar
\rho} \bar \sigma_1 \bar \sigma_2 \bar \sigma_3}]
 \Gamma^{\bar \sigma_1 \bar \sigma_2 \bar \sigma_3} e_{i_1 \cdots i_I}~.
 \la{genrbcom-null}
 \eea

To go from the basis (\ref{element-basis-null}) to the ``canonical''
basis which is associated with the Clifford vacuum $1$, one can use
the same expressions as for the timelike case.  So we shall not
repeat the formulae here.

The complex conjugation between the components of the supercovariant
derivative found in the timelike case does not extend to the null
basis in a straightforward way because  $\Gamma^+$ and $\Gamma^-$
are null instead of holomorphic. For this reason, it will be
convenient to treat the null indices separately. In what follows all
indices only take values in $1,\ldots, 4$. Instead of
(\ref{complexconj}), the following relations hold
 \bea
  ({\cal D}_{A} e_{i_1 \cdots i_I})_{{\bar \sigma}_1 \cdots {\bar
\sigma}_i} =
  ({{\cal D}}_{A} e_{i_{I+1} \cdots i_4})^*_{{\bar \sigma}_1 \cdots
{\bar \sigma}_i} \,, \quad
  ({\cal D}_{A} e_{i_1 \cdots i_I})_{{\bar \sigma}_1 \cdots {\bar
\sigma}_i +} =
  ({{\cal D}}_{A} e_{i_{I+1} \cdots i_4})^*_{{\bar \sigma}_1 \cdots
{\bar \sigma}_i +} \,, \notag \\
  ({\cal D}_{A} e_{i_1 \cdots i_I 5})_{{\bar \sigma}_1 \cdots {\bar
\sigma}_i} =
  ({{\cal D}}_{A} e_{i_{I+1} \cdots i_4 5})^*_{{\bar \sigma}_1 \cdots
{\bar \sigma}_i} \,, \quad
  ({\cal D}_{A} e_{i_1 \cdots i_I 5})_{{\bar \sigma}_1 \cdots {\bar
\sigma}_i +} =
  ({{\cal D}}_{A} e_{i_{I+1} \cdots i_4 5})^*_{{\bar \sigma}_1 \cdots
{\bar \sigma}_i +} \,,
 \eea
where $\epsilon_{i_1 \ldots i_I i_{I+1} \ldots i_4} = +1$. Using the relation to the ''canonical'' basis associated
with the Clifford vacuum $1$, these imply that
 \bea
  ({{\cal D}}_{A} e_{i_1 \cdots i_I})_{{\bar a}_{1} \cdots {\bar a}_m \bar
p_1 \cdots {\bar p}_n} = \frac{2^{2-m-n} (-)^{[(m+n)/2] + [I/2]}
(4-m-n)!}{(m+n)!(I-m)!(4-I-n)!} \cdot && \nonumber
\\
  \cdot \epsilon_{{\bar a}_{1} \cdots {\bar a}_m \bar p_1 \cdots {\bar p}_n}{}^{a_{m+1} \cdots a_I
  { p}_{n+1} \cdots { p}_{4-I}}
  ({{\cal D}}_{A} e_{i_{I+1} \cdots i_4})^*_{a_{m+1} \cdots a_I { p}_{n+1}
\cdots { p}_{4-I}} \,. &&
 \eea
Adding the complex conjugated and dualised version of
this expression to its original, one finds the following expression relating different components of
the combination $e_{i_1 \cdots i_I} + (-1)^{[I/2]} e_{i_{I+1} \cdots
i_4}$:
 \bea
  ({{\cal D}}_{A} e_{i_1 \cdots i_I} + (-1)^{[I/2]} {{\cal D}}_{A} e_{i_{I+1}
\cdots i_4})_{{\bar a}_{1} \cdots {\bar a}_m \bar p_1 \cdots {\bar
p}_n} = \frac{2^{2-m-n} (-)^{[(m+n)/2]}
(4-m-n)!}{(m+n)!(I-m)!(4-I-n)!} \cdot && \nonumber
\\
  \cdot \epsilon_{{\bar a}_{1} \cdots {\bar a}_m \bar p_1 \cdots {\bar p}_n}{}^{a_{m+1} \cdots a_I
  { p}_{n+1} \cdots { p}_{4-I}}
  ({{\cal D}}_{A} e_{i_1 \cdots i_I} + (-1)^{[I/2]} {{\cal D}}_{A} e_{i_{I+1}
\cdots i_4})^*_{a_{m+1} \cdots a_I { p}_{n+1} \cdots { p}_{4-I}} \,.
&& \label{dual-relation-null}
 \eea
The same expression holds with an extra $+$ index on the end of the list of components on both sides.
In addition, the same relation holds for the following combinations
 \be
  e_{i_1 \cdots i_I 5} + (-1)^{[I/2]} e_{i_{I+1} \cdots i_4 5} \,,
  \quad i e_{i_1 \cdots i_I} - i (-1)^{[I/2]} e_{i_{I+1} \cdots i_4} \,,
  \quad i e_{i_1 \cdots i_I 5} - i (-1)^{[I/2]} e_{i_{I+1} \cdots i_4 5} \,,
 \ee
 which together with $e_{i_1 \cdots i_I} + (-1)^{[I/2]}  e_{i_{I+1}
\cdots i_4}$ above span a basis in the space of Majorana spinors for the null case.
This concludes the investigation of the complex conjugation relations
of the  components of the Killing spinor equations  in the null basis.

The linear systems associated with the integrability conditions in the null and in the timelike
basis are related in a similar way to those of for the Killing spinor equations. One again
replaces in the linear system for the integrability conditions $0$ with $\nat$ and $i$ with $+1$.
 An additional sign appears in
the ${\cal I}_0$ component of the integrability conditions because
one replaces $\Gamma_0$ with $\Gamma_\nat$ as in the Killing spinor
equations case.  Because of the simplicity of the rules to derive the
linear systems associated with the null basis from those of the
timelike one, we shall not give further details for the former.

\newsection{N=2 backgrounds with $SU(4)$ invariant Killing spinors}

\subsection{The Killing spinor equations}

The most general $SU(4)$ invariant Killing spinors of a $N=2$
background  are
\bea
\eta_1&=&f (1+e_{12345})
 \nonumber \\
\eta_2&=&
g_1 (1+e_{12345})+g_2 i (1-e_{12345})+\sqrt{2} g_3 (e_5+ e_{1234})~.
\la{sufks}
\eea
where $f$, $g_1, g_2, g_3$ are real functions of the spacetime
which will be determined by the Killing spinor equations. We shall
assume that $g_3\not=0$ because otherwise the spinors are $SU(5)$
invariant and this case has already been investigated in \cite{joe}.
The Killing spinor equations of $\eta_1$ are as in the $N=1$ case. So
it remains to solve the Killing spinor equations for the second
spinor. Using the Killing spinor equations of $\eta_1$, the
 Killing spinor equations ${\cal D}_A \eta_2=0$ can be written as
\be
(\partial_Ag_1-g_1 \partial_A\log f) (1+e_{12345})+ i\partial_A
g_2 (1-e_{12345})+i g_2 {\cal D}_A (1-e_{12345})+\sqrt{2} {\cal
D}_A[g_3 (e_5+ e_{1234})]=0
\ee
Multiplying the above equation with
$g_3^{-1}$, we find that the Killing spinor equations for the second
spinor can be rewritten as
\bea
&&g_3^{-1}(\partial_Ag_1-g_1
\partial_A\log f+i\partial_A g_2)1 +g_3^{-1}(\partial_Ag_1-g_1
\partial_A\log f-i\partial_A g_2) e_{12345} \nonumber \\
&&+\sqrt{2}\partial_A\log g_3 (e_5+e_{1234})+
 ig_3^{-1} g_2 {\cal D}_A (1-e_{12345})+
 \sqrt{2} {\cal D}_A (e_5+ e_{1234})=0~.
 \la{sufkspin}
\eea

To proceed one can use the results in the appendix C to substitute
for ${\cal D}_A(e_5+e_{1234})$ and $i{\cal D}_A (1-e_{12345})$. The
resulting expressions have been given in
appendix E. It turns out that in solving the resulting linear systems
one has to distinguish between $g_2 = 0$ and $g_2 \neq 0$.
We will first consider the simplest case with $g_2=0$. This splits up in two subcases, depending on
whether $g_1$ vanishes or not. If $g_1=0$, the results have been given in \cite{joe}.
Here we shall summarize the $g_1\not=0$ case.  The conditions on the
function $g_3$ and $g_1$ are
 \be
  \partial_0 g_3 = 0 \,, \qquad \partial_\lambda \log g_3 = \partial_\lambda
  \log f \,, \qquad \partial_{\bar5} \log g_3 =  \partial_5
  \log f \,, \label{function-g3}
 \ee
 \be
  \partial_\lambda \log g_1 = \partial_\lambda \log f \,,
 \ee
and
 \be
  \partial_5 \log g_1 = \partial_5 \log f \,.
 \ee
We are left with the two equations (\ref{tfone-1}) and (\ref{tfone-2}), the first one of
which gives the time-dependence of the function $g_1$:
 \be
  g_3^{-1}\partial_0 g_1-i \Om_{0,05}+i \Om_{0,0\bar 5}=0 \,.
 \ee
The conditions on the $\Omega_{0,0i}$ components are \be
\Omega_{0,05}=-2\partial_5 \log f~,~~~~
\Omega_{0,0\lambda}=-2\partial_{\lambda} \log f~. \la{summary1} \ee
The conditions on the  $\Omega_{0,ij}$ components  are \be
\Omega_{0, 5\bar\lambda}=\Omega_{0, 5\lambda}= \Omega_{0,
5\bar5}=\Omega_{0,\sigma}{}^\sigma=0~,~~~~~ \Omega_{0,
\sigma_1\sigma_2}=\tfrac{i}{4} (\Omega_{5,\bar \rho_1\bar\rho_2}-
\Omega_{\bar 5,\bar
\rho_1\bar\rho_2})\te^{\bar\rho_1\bar\rho_2}{}_{\sigma_1\sigma_2}
\la{summary2} \ee and the traceless part of $\Omega_{0,
\lambda\bar\sigma}$ is not determined. The conditions on the
$\Omega_{\bar\lambda, ij}$ components are \bea
\Omega_{[\bar\sigma_1,\bar\sigma_2\bar\sigma_3]}&=&0~,~~~
\Omega_{\bar\lambda, \sigma_1\sigma_2}=-\Omega_{0,0[\sigma_1}
g_{\sigma_2]\bar\lambda}~,~~~ \Omega_{ \sigma,\bar\lambda}{}^\sigma=
-\tfrac{3}{2} \Omega_{0,0\bar\lambda}~, \cr
\Omega_{\lambda,\sigma}{}^\sigma&=&-\tfrac{1}{2}
(\Omega_{0,0\lambda}+2\Omega_{5,\lambda5})~. \la{summary3} \eea In
addition, we have \bea &&\Omega_{[\bar\sigma_1, \bar\sigma_2]\bar
5}=-\Omega_{\bar 5,\bar\sigma_1 \bar\sigma_2}~,~~~
\Omega_{[\bar\sigma_1, \bar\sigma_2] 5}=-\Omega_{5,\bar\sigma_1
\bar\sigma_2}~,~~~ \cr && \Omega_{(\bar\sigma_1, \bar\sigma_2)
5}=\Omega_{(\bar\sigma_1, \bar\sigma_2) \bar 5}~, ~~~
\Omega_{\bar\lambda, 5\bar5}=0~. \la{summary4} \eea The conditions
on the $\Omega_{\bar5, ij}$ components are \bea &&
 \Omega_{ 5, \bar\lambda 5}=\Omega_{\bar 5, \bar\lambda \bar5}~,~~~
\Omega_{ 5, \bar\lambda \bar5}=\Omega_{\bar 5, \bar\lambda 5}~, ~~~
\Omega_{\bar5, \bar\lambda \bar 5}-\Omega_{\bar5, \bar\lambda 5}=
-\Omega_{0,0\bar\lambda}~,~~~ \Omega_{\bar5, 5\bar
5}=-\Omega_{5,5\bar 5}~. \la{summary5}
 \eea
We also have the following relations
 \bea
  && \Omega_{0,05} + \Omega_{0,0 \bar 5} - \Omega_{5,5 \bar 5} +
  \Omega_{\bar 5, 5 \bar 5} + 2 \Omega_{5,} \cont{\lambda} - 2\Omega_{\bar 5,}
  \cont{\lambda} \,, \cr
  && 2 \Omega_{(\bar \lambda, \sigma) \bar 5} + \tfrac{1}{3}
  g_{\bar \lambda \sigma} ( - \tfrac{1}{2} \Omega_{0,05}
  - \tfrac{1}{2} \Omega_{0,0 \bar 5} - \Omega_{5,5 \bar 5} +
  \Omega_{\bar 5, 5 \bar 5} - \Omega_{5,} \cont{\rho} + \Omega_{\bar 5,}
  \cont{\rho}) = 0 \,,
  \label{extra-conditions}
 \eea
and
 \bea
  \Omega_{5,} \cont{\lambda} = \Omega_{\bar 5,} \cont{\lambda} \,,
  ~~~
  \Omega_{5,5 \bar 5} = \tfrac{1}{2} (\Omega_{0,05} + \Omega_{0,0
  \bar 5}) \,, ~~~
  \Omega_{(\bar \lambda,  \sigma) \bar 5} = \tfrac{1}{4} g_{\bar \lambda
  \sigma} (\Omega_{0,05} + \Omega_{0,0
  \bar 5})\,.
 \eea
All fluxes are expressed in terms of the geometry via
the relations summarized in appendix B. In addition, we find that
 \be
  F_{\lambda\bar\sigma 5\bar5}=-2i
\Omega_{0, \lambda\bar\sigma}~,~~~  F_{\bar \lambda\bar5
\sigma_1\sigma_2}=\tfrac{1}{2}
\Omega_{\bar\lambda,\bar\rho_1\bar\rho_2}
\te^{\bar\rho_1\bar\rho_2}{}_{\sigma_1\sigma_2} \,.
 \ee
This concludes the analysis of the $N=2$
$SU(4)$ case with $g_2 = 0$.

The Killing spinor equations for the case with $g_2 \neq 0$ are rather different
from those with $g_2 = 0$. The solution of this linear system is described in section E.2.
Here, we summarize the conditions on functions that
determine the spinors, the geometry and the fluxes.

The conditions on the functions $f, g_1,g_2$ and $g_3$ are
\bea
&& \partial_0 g_3=0 \,, \quad
g_3^{-1} \partial_0 g_2-(\Omega_{0,05}+\Omega_{0,0\bar5})=0 \,, \quad
g_3^{-1} \partial_0 g_1-i(\Omega_{0,05}-\Omega_{0,0\bar5})=0 \,, \nonumber \\
&& \partial_{\bar\r}\log(g_1/f)=0 \,, \quad
\partial_{\bar\r}\log(g_2/f)-2g_3
g_2^{-1}\Om_{0,\bar\r\bar5}=0 \,, \quad
\partial_{\bar\r}\log
g_3+\Om_{5,\bar\r\bar5}+\Om_{\bar5,\bar5\bar\r}-\tfrac{1}{2}\Om_{0,0\bar\r}=0 \,, \nn \\
&& \partial_{\bar 5}\log (g_3 f)=0 \,, \quad
\partial_{\bar 5}\log(g_2 f^{-1})=\partial_{\bar 5}\log(g_1 f^{-1})=0 ~.
\la{sumfun}
\eea

The conditions on the geometry are
\bea
&&g_3^{-1} g_2
[\Omega_{0,05}+2\Omega_{5,\r}{}^\r]+2\Omega_{0,\r}{}^\r=0 \,,\qquad
\Omega_{0,5\bar 5}=0 \,,\qquad \Omega_{\bar\r,\s
5}+\Omega_{\s,\bar\r5}=0 \nonumber \\
&&\Omega_{5,5\bar5}=0 \,,\qquad
\Omega_{(\bar\r,\bar\s)5}=\Omega_{(\bar\r,\bar\s)\bar5}=0 \,,\qquad
\Om_{\l,\s}{}^\s+\Om_{\bar5,\l\bar5}+\tfrac{1}{2}\Om_{0,0\l}=0\,,\nonumber\\
&& -4i g_3^{-1} g_2 \Omega_{\bar5, \bar\r\bar\s}-4i
\Omega_{0,\bar\r\bar\s} -\Omega_{5,\l_1\l_2}
\te^{\l_1\l_2}{}_{\bar\r\bar\s}+ \Omega_{\bar 5,\l_1\l_2}
\te^{\l_1\l_2}{}_{\bar\r\bar\s}=0 \nonumber \\
&&\Omega_{[\bar\r,\bar\s]5}+\Omega_{5,\bar\r\bar\s}=0 \,,\qquad
\Omega_{\bar\r, 5\bar5}=0 \,,\qquad
\Om_{\bar5,\bar\r5}-\Om_{5,\bar\r\bar5}=0\,,\qquad
\Om_{5,\r\s}+\Om_{[\r,\s]5}=0 \nonumber \\
&&\Om_{\bar\l_1,\bar\l_2\bar\l_3}\te^{\bar\l_1,\bar\l_2\bar\l_3}{}_
\r+2i \Om_{0,\r\bar5}=0\,,\qquad
-\Om_{\bar\rho,\l}{}^{\bar\rho}-\Om_{\bar5,\l5}-\Om_{\l,\tau}{}^\tau
-\Om_{0,0\l}=0\nonumber\\
&&\Om_{5,5\l}=-\Om_{\bar5,\l \bar5}\,,\qquad
g_3^{-1}g_2 \Om_{\bar5,\l\bar5}=-\Om_{0,\l\bar5}\,,\qquad
\Om_{0,\s\bar5}=\Om_{0\s5}\nonumber\\
&&\Om_{\bar\rho,\l_1\l_2}+\tfrac{2}{3}(\Om_{5,5[\l_1}-\Om_{\bar5,5[\l_1}+\tfrac{1}{2}\Om_{0,0[\l_1})g_{\l_2]\bar\rho}
-\tfrac{i}{6}g_3^{-1}g_2(\Om_{\bar\rho,\bar\s_1\bar\s_2}-\Om_{\bar\s_1,\bar\s_2
\bar\r}) \te^{\bar\s_1\bar\s_2}{}_{\l_1\l_2}=0\nonumber\\
&&-g_3^{-1}g_2 \Om_{0,\r\bar5}-\Om_{\bar5,\r
5}-\Om_{5,5\r}+\Om_{0,0\r}=0~.
\la{sumgeom}
\eea

The conditions on the fluxes that arise from the requirement of
$N=1$ supersymmetry have been summarized in appendix B. The
additional conditions that arise for two supersymmetries are
\bea
F_{\r\bar\s5\bar5}&=&-2i\Omega_{0,\r\bar\s} \nonumber \\
F_{\bar\rho\bar5\l_1\l_2} & = & \tfrac{8i}{3}\Om_{0,\bar5[\l_1}g_{\l_2]\bar\r}
+ \tfrac{1}{2}(\Om_{\bar\rho,\bar\s_1\bar\s_2}+\Om_{[\bar\r,\bar\s_1\bar\s_2]})\te^{\bar\s_1\bar\s_2}{}_{\l_1\l_2}~.
\eea

\subsection{The geometry of spacetime}

Using the results of \cite{joe}, it is straightforward to compute the
spacetime form bilinears associated with the Killing spinors (\ref{sufks})
for both $g_2=0$ and $g_2\not=0$.
These are a zero form
\bea
\alpha(\eta_1, \eta_2)=-2 fg_2~,
\eea
three one-forms
\bea
\kappa(\eta_1,\eta_1)&=&-2 f^2 e^0~,
\cr
\kappa(\eta_1,\eta_2)&=&-2 f g_1 e^0+ 2\sqrt{2} f g_3 e^{10}~,
\cr
\kappa(\eta_2, \eta_2)&=&-2(g_1^2+g_2^2+2g_3^2) e^0+4 \sqrt{2} g_1g_3 e^{10} +
4 \sqrt{2} g_2g_3 e^{5}~,
\eea
three two forms,
\bea
\omega(\eta_1,\eta_2)&=& 2 f^2 \omega~,
\cr
\omega(\eta_2,\eta_2)&=& 2 (g_1^2+ g_2^2) \omega+ 4 g_3^2 \hat\omega- 4\sqrt{2} g_1 g_3
 e^0\wedge e^5+ 4\sqrt{2} g_2 g_3  e^0\wedge e^{10}~,
 \cr
\omega(\eta_1, \eta_2)&=& 2 f g_1 \omega - 2 \sqrt{2} f g_3 e^0\wedge e^5~,
\eea
one three form
\bea
\xi(\eta_1,\eta_2)=- 2 \sqrt{2} fg_3\, \omega^{SU(4)}\wedge e^5~,
\eea
one four-form
\bea
\zeta(\eta_1,\eta_2)={f g_2\over \sqrt{2}} \omega\wedge\omega + 2 \sqrt{2} f g_3
[{\rm Im}\,\epsilon- e^0\wedge \omega^{SU(4)}\wedge e^{10}]~,
\eea
and three five-forms
\bea
\tau(\eta_1, \eta_1)&=&2 f^2[{\rm Im}\,\epsilon+{1\over2} e^0\wedge\omega\wedge\omega]~,
\cr
\tau(\eta_1, \eta_2)&=&2 f g_1 [{\rm Im}\,\epsilon+{1\over2} e^0\wedge\omega\wedge\omega]
\cr
&+& 2 f g_2 {\rm Re}\,\epsilon- 2 \sqrt{2} f g_3 [ e^0\wedge {\rm Re}\,\epsilon^{SU(4)}
+{1\over2} \omega^{SU(4)}\wedge \omega^{SU(4)}\wedge e^{10}]~,
\cr
\tau(\eta_2, \eta_2)&=& 2 g_1^2 [{\rm Im}\,\epsilon+{1\over2} e^0\wedge\omega\wedge\omega]
+2 g_2^2 [-{\rm Im}\,\epsilon+{1\over2} e^0\wedge\omega\wedge\omega]
\cr
&+& 4 g_3^2 [{\rm Im}\, \hat\epsilon+{1\over2}\hat\omega\wedge \hat\omega\wedge e^0]
+ 4 g_1 g_2 {\rm Re}\,\epsilon
\cr
&-& 4 \sqrt{2} g_1 g_3 [e^0\wedge {\rm Re}\,\epsilon^{SU(4)}
+{1\over2} \omega^{SU(4)}\wedge \omega^{SU(4)}\wedge e^{10}]
\cr
&+&4 \sqrt{2} g_2 g_3 [ e^0\wedge {\rm Im}\,\epsilon^{SU(4)}-{1\over2} \omega^{SU(4)}
\wedge \omega^{SU(4)}\wedge e^5]~,
\eea
where
\bea
\omega&=&-e^1\wedge e^6-e^2\wedge e^7-e^3\wedge e^8-e^4\wedge e^9- e^5\wedge e^{10}~,
\cr
\hat \omega&=&e^1\wedge e^6+e^2\wedge e^7+e^3\wedge e^8+e^4\wedge e^9- e^5\wedge e^{10}~,
\cr
\omega^{SU(4)}&=&e^1\wedge e^6+e^2\wedge e^7+e^3\wedge e^8+e^4\wedge e^9~,
\cr
\epsilon&=& (e^1+i e^6)\wedge\dots\wedge (e^5+i e^{10})~,
\cr
\epsilon^{SU(4)}&=&(e^1+i e^6)\wedge\dots\wedge (e^4+i e^{9})~,
\cr
\hat\epsilon&=&(e^1+i e^6)\wedge\dots\wedge(e^4+i e^9)\wedge (-e^5+i e^{10})~.
\eea

All the above forms specify the geometry of spacetime. Instead of investigating
the properties of all spacetime form bilinears, we shall mostly focus on the
properties of the three  one-form  bilinears. It is convenient to rescale them with a
factor of $1/2$ and rewrite them in the Hermitian frame basis as
\bea
\kappa(\eta_1,\eta_1)&=&-f^2 e^0~,
\cr
\kappa(\eta_1,\eta_2)&=&- f g_1 e^0-i f g_3 e^5+
i f g_3 e^{\bar 5} ~,
\cr
\kappa(\eta_2, \eta_2)
&=& -(g_1^2+g_2^2+2g_3^2) e^0+ 2 g_3 (g_2-i g_1) e^5+ 2 g_3
(g_2+i g_1) e^{\bar 5}~.
\eea
The associated vector fields $X,Y$ and $Z$, respectively, are Killing. This can be easily
verified using the conditions summarized in (\ref{sumfun}) and (\ref{sumgeom}).
In addition it turns out that $X,Y$ and $Z$ mutually commute, i.e. $[X,Y]=0$ and similarly
for the rest of the pairs.
In addition, we have that
\bea
g(X,X)&=&- f^4~,
\cr
g(Y,Y)&=&-f^2 g_1^2+ 2 f^2 g_3^2~,
\cr
g(Z,Z)&=&-[g_1^2+g_2^2-2 g_3^2]^2~,
\cr
g(X,Y)&=&-f^3 g_1~,
\cr
g(X,Z)&=&- [g_1^2+g_2^2+2 g_3^2] f^2~,
\cr
g(Y,Z)&=&-f g_1^3+ 4 f g_1 g_3^2~.
\eea
The vector field $X$ is timelike while as one expects  $Z$ is  timelike or null.

The Killing vector fields do not commute. So in general one cannot adapt coordinates to all three
Killing vectors. The form of the metric can be written by adapting coordinates to one of the Killing vector
fields say $X$.

\newsection{Solutions to the integrability conditions}

\subsection{$N=1$ backgrounds with $SU(5)$ invariant spinors}

The Killing spinor is
$\eta = f(1+e_{12345})$. The integrability condition on this
spinor implies the vanishing of the combination
\begin{eqnarray}
  ({\cal I}_A 1)_{\bar{\a}_1 \cdots \bar{\a}_i }+ ({\cal I}_A e_{12345})_{\bar{\a}_1 \cdots \bar{\a}_i} & = 0 \,,
 \end{eqnarray}
for $i=0,\ldots,5$. These integrability conditions guarantee the
vanishing of the Bianchi components $B_{0 \bar \alpha \bar \beta
\bar{\g} \bar{\d}}$ and $B_{0 \alpha \bar \beta \bar \g \bar \d}$.
The remaining field equations are subject to the relations
 \begin{eqnarray}
  0 & = & E_{00} - 12 i L_{0} \cont \a - 120 i B_{0} \cont \a \cont \b + 4 i B_{\a_1 \cdots \a_5} \te^{\a_1 \cdots \a_5} \,, \\
  0 & = & E_{0 \bar \a} - 180 i B_{\bar \a} \cont \b \cont \g \,,\\
  0 & = & E_{\alpha \bar{\b}} - 6 i g_{\alpha \bar{\b}} L_{0} \cont \g + 18 i L_{0 \a \bar \b} - 60 i
  g_{\alpha \bar{\b}} B_{0} \cont \g \cont \d + 360 i B_{0 \a \bar \b} \cont \g
  - 10 B_{\a \g_1 \cdots \g_4} \te_{\bar \b}{}^{\g_1 \cdots \g_4} \,,
  \cr
  0 & = & E_{\bar \a \bar \b} - 18 i L_{0 \bar \a \bar \b} + (80 g_{\bar \a \g_1} B_{\g_2 \cdots \g_4} \cont{\d}
  - 30 B_{\bar \a \g_1 \cdots \g_4}) \te_{\bar \b}{}^{\g_1 \cdots \g_4} \,, \\
  0 & = & L_{\a \bar \b \bar \g} - 20 g_{\a [ \bar \b} B_{\bar \g]} \cont \d \cont \e + 20 B_{\a \bar \b \bar \g} \cont \d \,, \\
  0 & = & L_{\bar{\a} \bar{\b} \bar \g} + 20 B_{\bar \a \bar{\b} \bar \g} \cont \d
  + \tfrac{1}{2} i L_{0 \d \e} \te_{\bar \a \bar \b \bar \g}{}^{\d \e} \,.
 \end{eqnarray}
These can be solved by explicitly imposing the components
 \be
  \{ L_{0 \alpha \bar{\b}}, L_{0 \bar{\a} \bar{\b}}, B_{0 \alpha \beta \bar{\g} \bar{\d}}, B_{\bar \alpha \bar \beta \bar{\g} \bar{\d} \bar{\e}}, B_{\alpha \bar \beta \bar{\g} \bar{\d} \bar{\e}}, B_{\alpha \beta \bar{\g} \bar{\d} \bar{\e}} \} \,,
 \ee
Therefore, in the $N=1$ $SU(5)$ case, one still needs to impose the above components of the Bianchi identity plus the electric part of the gauge field equation\footnote{The fact that the magnetic part of the gauge field equation is implied by $N=1$ $SU(5)$ supersymmetry and the Bianchi identity can also be derived from the bilinear formalism of \cite{pakis}.} to satisfy all field equations.

\subsection{$N=2$ backgrounds with $SU(5)$ invariant spinors}

The Killing spinors are
 \be
  \eta_1 = f_1 \eta^{SU(5)} \,, \qquad\eta_2 = f_2 \eta^{SU(5)} + f_3 \theta^{SU(5)} \,,
 \ee
with $f_1$ and $f_3$ non-vanishing.
Independent of the functions $f_1, f_2, f_3$, the integrability conditions arising from these spinors are
 \bea
  ({\cal I}_A 1)_{\bar{\a}_1 \cdots \bar{\a}_i} = ({\cal I}_A e_{12345})_{\bar{\a}_1 \cdots \bar{\a}_i} = 0 \,,
 \eea
for $i=0,\ldots,5$. From these conditions one can derive that the
field equations do not automatically vanish are
 \be
  \{ E_{00}, E_{0 \bar{\a}}, E_{\alpha \bar{\b}}, L_{0 \alpha \bar{\b}}, L_{\alpha \bar{\b} \bar{\g}},
  \tilde{B}_{0 \alpha \beta \bar{\g} \bar{\d}}, B_{\alpha \beta \bar{\g} \bar{\d} \bar{\e}}
  \},
 \ee
where the tilde means traceless part, subject to the relations
 \begin{eqnarray}
  0 & = & E_{00} - 12 i L_{0} \cont{\alpha} \,, \\
  0 & = & E_{0 \bar{\a}} - 180 i B_{\bar{\a}} \cont{\beta} \cont{\gamma} \,, \\
  0 & = & E_{\alpha \bar{\b}} - 6 i g_{\alpha \bar{\b}} L_{0} \cont{\gamma} + 18 i L_{0 \alpha \bar{\b}} \,, \\
  0 & = & L_{\alpha \bar{\b} \bar{\g}} - 20 g_{a [ \bar{\b}} B_{\bar{\g} ]} \cont{\delta} \cont{\epsilon}
  + 20 B_{\alpha \bar{\b} \bar{\g}} \cont{\delta} \,.
 \end{eqnarray}
One can solve these equations by explicitly checking
 \be
  \{ L_{0 \alpha \bar{\b}}, \tilde{B}_{0 \alpha \beta \bar{\g} \bar{\d}}, B_{\alpha \beta \bar{\g} \bar{\d} \bar{\e}} \} \,,
 \label{N=2fe}
 \ee
after which all other field equations are implied.

\subsection{$N=4$ backgrounds with $SU(4)$ invariant spinors} \label{N=4-intcond}

The  Killing spinors are
 \bea
  \eta_1 &=& f_1 \eta^{SU(5)} \,, \\
  \eta_2 &=& f_2 \eta^{SU(5)} + f_3 \theta^{SU(5)} \,, \\
  \eta_3 &=& f_4 \eta^{SU(5)} + f_5 \theta^{SU(5)} + f_6 \eta^{SU(4)} \,, \\
  \eta_4 &=& f_7 \eta^{SU(5)} + f_8 \theta^{SU(5)} + f_9 \eta^{SU(4)} + f_{10} \theta^{SU(4)} \,,
 \label{N=4-spinors}
 \eea
with $f_1, f_3, f_6$ and $f_{10}$ non-vanishing. In this case, independent of the ten space-time functions, the integrability conditions (\ref{intcond}) of the four Killing spinors correspond to the conditions
 \begin{eqnarray}
  ({\cal I}_A 1)_{\bar{\lambda}_1 \cdots \bar{\lambda}_i} = ({\cal I}_A 1)_{\bar{\lambda}_1 \cdots \bar{\lambda}_i \bar 5} =
  ({\cal I}_A e_{12345})_{\bar{\lambda}_1 \cdots \bar{\lambda}_i} = ({\cal I}_A e_{12345})_{\bar{\lambda}_1 \cdots \bar{\lambda}_i \bar 5} = 0 \,, \\
  ({\cal I}_A e_5)_{\bar{\lambda}_1 \cdots \bar{\lambda}_i} = ({\cal I}_A e_5)_{\bar{\lambda}_1 \cdots \bar{\lambda}_i \bar 5} =
  ({\cal I}_A e_{1234})_{\bar{\lambda}_1 \cdots \bar{\lambda}_i} = ({\cal I}_A e_{1234})_{\bar{\lambda}_1 \cdots \bar{\lambda}_i \bar 5} = 0 \,,
 \end{eqnarray}
for $i=0,\ldots,4$. These imply all but the following field equations:
 \be
  \{ E_{00}, E_{\lambda \bar{\mu}}, E_{5 \bar{5}}, L_{0 5 \bar{5}}, \tilde{L}_{\lambda \bar{\mu} \bar{\nu}},
  \tilde{B}_{0 \lambda \mu \bar{\nu} \bar{\rho}}, B_{\lambda \mu \bar{\nu} \bar{\rho} \bar{\sigma}},
  \tilde{B}_{\lambda \mu \bar{\nu} \bar{\rho} \bar 5},
  \}, \label{N=4fe}
 \ee
(where the tilde means traceless part) subject to the relations
 \begin{eqnarray}
  0 & = & E_{00} - 12 i L_{0 5 \bar{5}} \,, \\
  0 & = & E_{\lambda \bar{\mu}} - 6 i g_{\lambda \bar{\mu}} L_{0 5 \bar{5}} \,, \\
  0 & = & E_{5 \bar{5}} + 12 i L_{0 5 \bar{5}} \,, \\
  0 & = & L_{\lambda \bar{\mu} \bar{\nu}} + 20 B_{\lambda \bar{\mu} \bar{\nu}}
  \cont{\rho} \,.
 \end{eqnarray}
These can be solved by requiring the components of the
Bianchi identity in (\ref{N=4fe}) to vanish and by imposing the field equation $L_{0 5
\bar{5}}=0$.

\newsection{Resolved  membranes}

In this section we will consider the class of solutions which admit Killing spinors as in
(\ref{N=4-spinors}) with the restrictions
 \be
  f_2=f_4=f_5=f_7=f_8=f_9=0 \,.
 \label{N=4-restricted-spinors}
 \ee
as analyzed in \cite{joe}. We shall show here that the most general solution is
a resolved rotating membrane wrapped on a two-torus $T^2$.

We will start by summarizing the conditions for $SU(4)$
backgrounds to admit the four Killing spinors. Firstly, this
background has three commuting Killing vectors, one of which is timelike and the other two are spacelike.
Because of this, we introduce three coordinates $x^i$, $i=0,1,2$, adapted to these three Killing vector fields
which are thought of as the worldvolume coordinates of a membrane.
  We define the frames\footnote{The Hermitian frame directions $e^5$ and $e^{\bar 5}$  are related to $e^1$ and $e^2$ as
 \be
  e^5 = (e^1 + i e^2) / \sqrt{2} \,, \qquad
  e^{\bar 5} = (e^1 - i e^2) / \sqrt{2}  \,.
 \ee}
 \bea
  e^i = f^2 (dx^i + \alpha^i) \,,
 \eea
where the $\alpha^i$ are independent of the world-volume coordinates and only take values in the 8D transverse space.
Since the Killing vector fields are orthogonal and of the same length, the metric in this frame reads
 \be
  ds^2 = f^4 g_{ij} (dx^i + \alpha^i) (dx^j + \alpha^j)  + 2 g_{\lambda \bar \mu} e^\lambda e^{\bar \mu} \,,
 \ee
where $g_{ij} = {\rm diag}(-1,1,1)$,  $g_{\lambda \bar \mu} = \delta_{\lambda \bar \mu}$ and $e^\lambda$, $\lambda=1,\dots,8$,
is a Hermitian frame for the metric on the space of orbits $B$ of the three isometries.
The eight-dimensional space $B$ is complex and  is identified
with the transverse space of the membrane.  The Killing spinor equations imply \cite{joe} that the
 four-form field strength can be written as
 \be
  {\cal F} = - d (e^0 \wedge e^1 \wedge e^2) + \tilde{F}^{(2,2)} \,,
 \ee
 where $\tilde F^{(2,2)}$ is a traceless (2,2) form on $B$ and so self-dual. In addition,
the components  $\Omega_{i,AB}$ of the
spin connection satisfy the conditions
 \bea
  \Omega_{i, jk} = 0 \,, \qquad & & \Omega_{i, j \lambda} = 2 g_{ij} \partial_\lambda \log f \,, \\
  \Omega_{i, \lambda \mu} = 0 \,,  \qquad & & \Omega_{i, \lambda}{}^\lambda = 0 \,, \qquad
  \Omega_{i, \lambda \bar \mu} = - \tfrac{1}{2} f^2 (d \alpha_i)_{\lambda \bar \mu} \,,
 \eea
while the  $\Omega_{\lambda,AB}$ components read
 \bea
  \Omega_{\lambda, ij} = 0 \,, \qquad && \Omega_{\lambda, i \mu} = 0 \,, \\
  \Omega_{\lambda, i \bar \mu} = - \tfrac{1}{2} f^2 (d \alpha_i)_{\lambda \bar \mu} \,, \qquad &&
  \Omega_{\lambda, \mu \nu} = 0 \,, \\
  \Omega_{\lambda, \bar \mu \bar \nu} = - 2 g_{\lambda [ \bar \mu} \partial_{\bar \nu ]} \log f \,, \qquad &&
  \Omega_{\lambda, \mu}{}^\mu = \partial_\lambda \log f \,,
  \eea
where $d\alpha^i_{\lambda\bar\mu}=\partial_\lambda\alpha^i_{\bar\mu}- \partial_{\bar\mu} \alpha^i_\lambda$.

We now turn to the field equations.
As explained in section (\ref{N=4-intcond}), the integrability conditions for the $N=4$ $SU(4)$-invariant Killing spinors imply
that one only needs to impose the vanishing of a number of components of the Bianchi equation and one component of the ${\cal F}$ field equation.
Specifically, one has to impose the vanishing of
 \be
  \{ L_{012}, \tilde{B}_{i \lambda \mu \bar{\nu} \bar{\rho}}, B_{\lambda \mu \bar{\nu} \bar{\rho} \bar{\sigma}} \}, \label{N=4fe-M2}
 \ee
where the $\tilde{}$ denotes traceless part of the associated quantity.  Let us first consider the Bianchi identity.
The components with a world-volume index imply independence of $\tilde F^{(2,2)}$ of the world-volume Killing directions:
 \be
  \partial_i \tilde F^{(2,2)} = 0 \,.
 \ee
The remaining $(2,3)$ component of the Bianchi equation implies $F^{(2,2)}$ to be a closed form on $B$, i.e.
 \be
  d_8\tilde  F^{(2,2)} = 0 \,,
 \ee
where $d_8 = e^\lambda\partial_\lambda  + e^{\bar \lambda} \partial_{\bar \lambda} $.
Since $\tilde F{(2,2)}$  is  self-dual, $\tilde F{(2,2)}$ is also co-closed and so  a harmonic $(2,2)$ form on $B$. The only component of the
field equation that one needs to impose is $L_{012}$, which implies
 \be
  D^I\partial_I \log f = {f^4\over 6} g_{ij}\, d\alpha^i\cdot d\alpha^j+ \tfrac{1}{12} F^{(2,2)} \cdot F^{(2,2)}~,~~~I=\lambda, \bar\lambda
 \ee
where the $D_I$ is the Levi-Civita connection of $ds^2=2\gamma_{\lambda\bar\mu} e^\lambda e^{\bar\mu}$ of
the transverse space $B$ and the inner product  of a k-form $\phi$ is $\phi\cdot \phi={1\over k!} \phi_{I_1\dots I_k}
\phi^{I_1\dots I_k}$.

This solution can be written in a more familiar form by rescaling the $e^\lambda$ as  $e^\lambda=f^{-1}\hat{e}^\lambda $
 and identifying $H = f^{-6}$. The metric and flux then become
 \bea
  ds^2 & = & H^{-2/3} g_{ij} (dx^i + \alpha^i) (dx^j + \alpha^j)  + 2 H^{1/3} g_{\lambda \bar \mu}
  \hat{e}^\lambda \hat{e}^{\bar \mu}\,,
  \cr
  {\cal F} & = & - d ( e^0 \wedge e^1 \wedge e^2) + \tilde{F}^{(2,2)}
   \,, \label{resolvedM2}
 \eea
 where $e^i=H^{-{1\over3}} (dx^i+\alpha^i)$.
The components of the spin connection of the rescaled metric $d\hat s^2=g_{\lambda \bar \mu}
  \hat{e}^\lambda \hat{e}^{\bar \mu}$ and frame satisfy
 \be
  \hat{\Omega}_{\lambda,  \mu  \nu} = 0 \,, \qquad
  \hat{\Omega}_{\lambda, \bar \mu \bar \nu} = 0 \,, \qquad
  \hat{\Omega}_{\lambda,  \mu}{}^\mu = 0 \,,
 \ee
and hence $d\hat s^2$  is a
Calabi-Yau metric. The Laplacian equation for $f$  in terms of $H$ becomes
 \be
  -\hat D^I \partial_I H = g_{ij}\, d\alpha^i\cdot d\alpha^j+ \tfrac{1}{2} \tilde F^{(2,2)} \cdot \tilde F^{(2,2)} \,,
  \la{feqrm}
 \ee
where $\hat D_I$ is the Levi-Civita connection of $d\hat s^2$ and inner products have been taken with respect to $d\hat s^2$.
The equation (\ref{feqrm}) for $\alpha^i=0$ has been explored before in the context of resolved membranes, see e.g.~\cite{Becker:1996gj, Duff:1997hf,
Cvetic:2000mh, Cvetic:2000db}. A case  $\alpha^0 \neq 0$ has been considered
 in \cite{pakis}, corresponding to a rotating resolved M2-brane but $d\alpha$ was taken to be (2,0) and (0,2) instead of (1,1)
 and traceless that we have here. The
 case $\alpha^i \neq 0$, $i=0,1,2$ has been considered in \cite{Chen:2004kj} and solutions were found
 with specific transverse spaces. The interpretation of these solutions are resolved rotating membranes wrapped
 on a two-torus $T^2$ which fibres over the transverse space $B$.
Here we have shown that this is the most general supersymmetric configuration with four supersymmetries
for the $SU(4)$-invariant Killing spinors (\ref{N=4-spinors}) subject to (\ref{N=4-restricted-spinors}).

It is worth pointing out that in the right-hand-side of (\ref{feqrm}) the contribution of the rotation
has a different sign from those of the wrapping of the membrane on $T^2$. We can use this
to give necessary and sufficient conditions for the existence of {\it non-singular} solutions on any compact, connected without boundary,
Calabi-Yau manifold $B$. First observe that $\alpha^i$ can be thought of as the connection of a line bundle $L^{(i)}$ over the
Calabi-Yau manifold $B$. Since the curvature $\beta^i=d\alpha^i$ is (1,1), this line bundle is holomorphic.
In addition a necessary and sufficient condition
for $L^{(i)}$ to admit a connection $\alpha^i$ such that the curvature $\beta^i$ is traceless, i.e. to satisfy the
Donaldson condition, is
\bea
\int_B \beta^i\wedge \hat\omega\wedge \hat\omega\wedge \hat\omega=0
\la{conna}
\eea
where $\hat\omega$ is the K\"ahler form of the Calabi-Yau metric $d\hat s^2$, see e.g. \cite{green}.
Next turn to (\ref{feqrm}). Since the left-hand-side of (\ref{feqrm}) is a Laplacian on $H$, one can use Hodge theory
to invert the Laplace operator and solve for  $H$. A necessary and sufficient condition for the existence of a (well-defined)
solution on $B$ is that the right-hand-side expression in (\ref{feqrm}) is orthogonal to the harmonic zero-forms.
This translates to the condition
\bea
\int_B d{\rm vol}~(g_{ij}\, \beta^i\cdot \beta^j+ \tfrac{1}{2} \tilde F^{(2,2)} \cdot \tilde F^{(2,2)})=0~,
\eea
where $d{\rm vol}$ is the volume form of the Calabi-Yau metric $d\hat s^2$. This relation can be
rewritten using the traceless condition of $\beta^i$ and the self-duality condition of $\tilde F^{(2,2)}$ as
\bea
\int_B~\big( - \tfrac12 g_{ij}\, \beta^i\wedge
\beta^j\wedge\hat\omega\wedge\hat\omega+\tfrac12 \tilde F^{(2,2)}\wedge
\tilde F^{(2,2)}\big)=0~.
\la{connb}
\eea
Observe that the above relation  depends on the cohomology class of
$\hat\omega$, $\beta^i$ and  $\tilde F^{(2,2)}$ and it may be
interpreted as a cancelation of membrane, rotation and wrapping
fluxes when integrated over the compact Calabi-Yau manifold $B$. The
above condition is the sum of squares and therefore  if there is no
rotation, i.e. $\alpha^0=0$, then $\beta^i=0$, $i=1,2$ and $\tilde
F^{(2,2)}=0$ and so there is only a trivial solution, i.e. $H={\rm
const}$. However if  $\alpha^0\not=0$, then there are solutions of
(\ref{connb}) for non-trivial $\beta^i$ and $\tilde F^{(2,2)}$
provided that (\ref{conna}) and (\ref{connb}) are satisfied.
Furthermore observe that  $H$ is determined up to a constant in
(\ref{feqrm}), and it is bounded because $B$ is compact. Therefore,
it is always
 possible to choose $H$ to be positive,
$H>0$.
In such cases, one can find a {\it non-singular} solution of eleven-dimensional supergravity preserving
four supersymmetries with metric and flux given in (\ref{resolvedM2}).
In the context of M-theory, there are corrections to the flux field equation. In particular, one has \cite{vafa, duff}
\bea
d*\mathcal F- \tfrac{1}{2} \mathcal F \wedge \mathcal F =\kappa
X_8~,
\eea
where $\kappa$ are some units,  $X_8={1\over 192} (p_1^2-4p_2^2)$,  and $p_1$ and $p_2$ are the Pontryagin classes of spacetime.
It is clear that in this case the condition (\ref{connb}) is modified as
\bea
\int_B~\big(- \tfrac12 g_{ij}\, \beta^i\wedge
\beta^j\wedge\hat\omega\wedge\hat\omega+ \tfrac12 \tilde F^{(2,2)} \wedge
\tilde F^{(2,2)}+\kappa X_8\big)=0~,
\la{connbc}
\eea
and this new condition is required for the existence of non-singular solutions. To summarize, the conditions
(\ref{conna}) and (\ref{connb}) are necessary
and sufficient  for the existence of a resolved (non-singular), rotating and wrapped membrane solution
of eleven-dimensional supergravity with transverse space a
compact, connected without boundary Calabi-Yau manifold. Incidentally, these type of solutions resemble those found in the
context of flux tubes in \cite{mateos} and it may be worth exploring this further.

\newsection{Concluding remarks}

The Killing spinor equations of any background of eleven-dimensional supergravity
theory have been reduced to the evaluation of the supercovariant derivative
${\cal D}\sigma_I$ on six types
of spinors $\sigma_I$. The expressions for all ${\cal D}\sigma_I$ have been given
in both a timelike and a null spinor basis.
In addition the integrability conditions of the Killing spinor equations which
encode the field equations of the theory
have been investigated. It is shown that these integrability
conditions can be expressed as a linear combination of the six types of spinors
 ${\cal I}\sigma_I$. We give the expressions of all ${\cal I}\sigma_I$ again in
 both a timelike and a null spinor basis.
 As a result, one can determine the field equations of the theory which arise
 as integrability conditions of the Killing spinor equations. In this way,
 one can specify
 the minimal set of additional field equations required for a supersymmetric
 configuration to be a solution of the supergravity field equations.
 We have also presented some examples to illustrate our construction. In particular,
 we have given a class of resolved, rotating, wrapped membranes with transverse
 space a Calabi-Yau manifold preserving four supersymmetries. We have also shown
 that these are the most general  supersymmetric solutions for a class of
 $SU(4)$ invariant Killing spinors.

This paper has given the systematics of how to classify all supersymmetric
solutions in eleven dimensions.
The above construction can be used to reduce the Killing spinor equations
 to a linear system for the fluxes, geometry and spacetimes derivatives
of the functions
that determine the Killing spinors. This system is of increasing complexity
with the number of Killing spinors that a background admits. Nevertheless, we
have determined
all the coefficients and unknowns of this linear system for all supersymmetric
backgrounds. A similar conclusion applies for the linear system that arises
in the integrability conditions which determines the minimal set of field equations
which should be satisfied. Therefore, the classification of supersymmetric backgrounds
is associated  with two linear systems, one is related to the Killing spinor equations
and the other to the field equations.

The two linear systems systems can always be solved.
A question arises whether they are tractable
for all supersymmetric backgrounds. In the general situation, they will be rather involved.
However, some simplifications
may occur. The Killing spinors can be simplified by using the
gauge symmetry $Spin(10,1)$ of the supercovariant connection to put them
at particular directions in space of spinors, i.e. to put them in a canonical or normal form.
This typically reduces the number of functions that the spinors depend on.
Further simplifications occur  whenever the spinors have some residual symmetry,
i.e. some non-trivial
stability subgroup in $Spin(10,1)$. This happens in many supersymmetric backgrounds
of interest and in particular in those that appear in compactifications with fluxes.
A detailed discussion of this has appeared in \cite{joe}. However, it is known that
there are backgrounds for which the Killing spinors have the identity  in $Spin(10,1)$
as stability subgroup. This phenomenon occurs even for backgrounds
with two supersymmetries. For such backgrounds there is no apparent simplification.
Nevertheless, it may be  possible in practice to solve these linear systems in general
in many cases. For example, since the systems are linear this can be done with the
help of computers.

\section*{Acknowledgements}

The work of U.G. is funded by the The Swedish Research Council
and in addition the research of both U.G. and D.R. is funded
by the PPARC grant PPA/G/O/2002/00475.

\setcounter{section}{0}

\renewcommand{\thesection}{\Alph{section}}

\newsection{Spinors from forms} \label{Spinors}

\subsection{A timelike basis}

The realization of Majorana spinors of $Spin(10,1)$ in terms of
forms has been described in \cite{joe}, see also   \cite{wang,
lawson, harvey}. Here we shall summarize some of the features of the
construction. For a detailed account of the construction see
\cite{joe}.

Let $e_1, \dots, e_{10}$ be an orthonormal basis in $V=\bR^{10}$.
Next consider the subspace $U=\bR^5$ in $V$ generated by $e_1,
\dots, e_5$. The Euclidean inner product on $V$ can be extended to a
Hermitian inner product in $V_{\bC}=V\otimes \bC$ and then
restricted in $U_{\bC}=U\otimes \bC$ denoted by $<,>$, i.e. on
$U_{\bC}$ is
\be
<z^i e_i, w^j e_j>=\sum_{i=1}^{5}  (z^*)^i w^i~,
\ee
where $(z*)^i$ is the standard complex conjugate of $z^i$.
The space of  $Spin(10)$ Dirac spinors  is $\Delta_c=\Lambda^*(U_{\bC})$.
The above inner product can be easily extended to $\Delta_c$ and it is called
the Dirac inner product on the space of $Spin(10)$ spinors.
The gamma matrices  act on $\Delta_c$ as
\bea
\Gamma_i\eta&=& e_i\wedge\eta+e_i\lc \eta~,~~~~ i\leq 5
\nonumber \\
\Gamma_{5+i}\eta&=&ie_i\wedge\eta-ie_i\lc \eta~,~~~~ i\leq 5
\eea
where $e_i\lc$ is the adjoint of $e_i\wedge$ with respect to $<,>$.
Moreover we have that the Weyl representations of $Spin(10)$ are
 $\Delta_{16}^+=\Lambda^{\rm even}U_{\bC}$ and
$\Delta_{16}^-=\Lambda^{\rm odd}U_{\bC}$.
 Clearly $\Gamma_i: \Delta_{16}^\pm\rightarrow \Delta_{16}^\mp$.
The linear maps $\Gamma_i$ are Hermitian with respect to the inner
product $<,>$, $<\Gamma_i \eta, \theta>=<\eta, \Gamma_i\theta>$, and
satisfy the Clifford algebra relations $\Gamma_i\Gamma_j+\Gamma_j
\Gamma_i=2 \delta_{ij}$. The Majorana Pin(10) invariant inner
product on $\Delta_c$ is
\be
B(\eta, \theta)=<B(\eta^*), \theta>~,
\la{inmajor}
\ee
where the linear map denoted with the same symbol as the inner product
is $B=\Gamma_6\dots\Gamma_{\nat}$ and\footnote{From here on, we shall adopt
 the notation to denote
the tenth direction with $\nat=10$.}
$\Gamma_\nat=\Gamma_{10}$. $B$ is skew-symmetric.

The spinor representations of $Spin(10,1)$ are constructed by first
setting $\Gamma_0=\Gamma_1\dots \Gamma_{\nat}$. It is easy to see
that $\Gamma^2_0=-1$ as expected and that $\Gamma_0$ anticommutes
with $\Gamma_i$. The Dirac representation of $Spin(10)$ is the same
as that of $Spin(10,1)$. The Dirac inner product on  $Spin(10,1)$
representation,  $\Delta_c$, is \bea D(\eta, \theta)= <\Gamma_0\eta,
\theta> \eea and the $Pin(10)$ Majorana inner product
(\ref{inmajor}) extends to the Majorana inner product of
$Spin(10,1)$. It remains to impose the Majorana condition on the
$Spin(10,1)$ representation, $\Delta_c$. This is \be \eta^*=
\Gamma_0 B(\eta)~,~~~~~~\eta\in \Delta_c \la{maj} \ee The
$Spin(10,1)$ Majorana spinors $\Delta_{32}=\{\eta\in \Delta_c~{\rm
s.t.}\, \eta^*= \Gamma_0 B(\eta)$. For completeness, the spacetime
form bilinears associated with the Majorana spinors $\eta, \theta$
are
\be
\alpha(\eta, \theta)={1\over k!} B(\eta,\Gamma_{A_1\dots
A_k} \theta) e^{A_1}\wedge\dots\wedge e^{A_k}~,~~~~~~~k=0,\dots,
9,\nat~. \la{forms}
\ee

Another ingredient in solving the Killing spinor equations and their
integrability conditions is the construction of a basis
in the space of $Spin(10,1)$ Dirac
 spinors $\Delta_c$. It turns out that
 \be
\Delta_c= \sum_{k=0}^5\Lambda^{0,k} \cdot 1~,
\ee
 where $\cdot$
denotes Clifford multiplication. Therefore \be \Gamma^{\bar
\alpha_1\dots \bar\alpha_k}\cdot 1~,~~~~~k=0, \dots, 5
\la{hermbasis} \ee is a {\it basis} in the space of spinors
$\Delta_c$, where
\bea
\Gamma_{\bar\a}&=&\tfrac{1}{\sqrt{2}}
(\Gamma_\a+i \Gamma_{\a+5})~,~~~ \Gamma^\a=g^{\a\bar\b}
\Gamma_{\bar\b}~,~~~\a=1,\dots,5 \nonumber \\
\Gamma_{\a}&=&\tfrac{1}{\sqrt{2}} (\Gamma_\a-i \Gamma_{\a+5})~,~~~
\Gamma^{\bar\a}=g^{\bar\a\b} \Gamma_{\b} ~,~~~\a=1,\dots,5~,
\eea
 and $g_{\a\bar\b}=\delta_{\a\bar\b}$.
The Clifford algebra relations in this basis are
$\Gamma_\a\Gamma_{\bar \b}+\Gamma_{\bar \b} \Gamma_\a=2g_{\a\bar\b}$,
$\Gamma_\a \Gamma_\b+\Gamma_\b\Gamma_\a= \Gamma_{\bar\a} \Gamma_{\bar\b}
+\Gamma_{\bar\b}\Gamma_{\bar\a}=0$. Observe that
$(\Gamma_j+i \Gamma_{j+5}) 1=0$
and similarly $(\Gamma_j-i \Gamma_{j+5}) e_1\wedge\dots\wedge e_5=0$.
 In particular,
\be
e_{12345}={1\over8\cdot 5!} \te_{\bar\a_1\dots\bar\a_5}
\Gamma^{\bar\a_1\dots\bar\a_5}\, 1~,
\ee
where $\te_{\bar 1\bar
2\bar 3\bar 4 \bar 5}=\sqrt{2}$. We shall extensively use this basis
for spinors  to analyze the Killing spinor equations and their
integrability conditions. As in the above equation, throughout the
paper we suppress the sign of the Clifford multiplication, e.g.
instead of $\Gamma^{\bar\a}\cdot 1$ we write $\Gamma^{\bar\a}\, 1$.

\subsection{A null basis}

An alternative way to construct the Majorana representation of $Spin(10,1)$ is to begin
from the $Spin(9,1)$ spinor representations. The realization  of the spinor representations
of $Spin(9,1)$ in terms of forms has been presented in \cite{jan}. We shall repeat
this construction and then we shall explain the application to $Spin(10,1)$.
Let  $U=\bR<e_1,\dots,e_5>$ be a vector space spanned by $e_1,\dots,e_5$ orthonormal vectors.
The space of Dirac $Spin(9,1)$ spinors is
$\Delta_c=\Lambda^*(U\otimes \bC)$.
The gamma matrices are represented on $\Delta_c$ as
\bea
\Gamma_0\eta&=& -e_5\wedge\eta +e_5\lc\eta~,~~~~
\Gamma_5\eta= e_5\wedge\eta+e_5\lc \eta
\cr
\Gamma_i\eta&=& e_i\wedge \eta+ e_i\lc \eta~,~~~~~~i=1,\dots,4
\cr
\Gamma_{5+i}\eta&=& i e_i\wedge\eta-ie_i\lc\eta~,
\eea
where $\lc$ is the adjoint of $\wedge$ with respect to the (auxiliary) inner
product
\be
 <z^a e_a, w^b e_b>=\sum_{a=1}^{5}  (z^a)^* w^a~,~~~~
\ee
on $U\otimes \bC$  and then extended
to $\Delta_c$.   $(z^a)^*$ is the standard complex conjugate of $z^a$.
The gamma matrices have been chosen such that
  $\{\Gamma_i; i=1,\dots, 9\}$  are Hermitian and $\Gamma_0$ is anti-Hermitian with respect to the (auxiliary)
inner product $<,>$.

 The above gamma matrices satisfy the
Clifford algebra relations $\Gamma_A\Gamma_B+\Gamma_B \Gamma_A=2 \eta_{AB}$ with respect to the Lorentzian inner
product as expected. The Dirac inner product on the space of spinors $\Delta_c$ is $D(\eta,\theta)=<\Gamma_0\eta,
\theta>$ and the $Pin (9,1)$ invariant (Majorana)  inner product is
\be
B(\eta,\theta)= <B(\eta^*), \theta>~,~~~~~~~~
\ee
where  $B=\Gamma_{06789}$. Observe that $B(\eta, \theta)=-B(\theta,\eta)$.

The Dirac representation of $Spin(10,1)$ is identified with the Dirac representation of $Spin(9,1)$. The tenth
gamma matrix is chosen as
\bea
\Gamma_\nat=-\Gamma_{0123456789}~.
\eea
One can verify that $\Gamma_\nat^2=1$ and that anticommutes with the rest of gamma matrices.
The Dirac inner product is  $D(\eta,\theta)=<\Gamma_0\eta,
\theta>$, i.e. the same as that of the $Spin(9,1)$. In addition, since $B$ is
a $Pin(9,1)$ invariant inner product, it extends to a $Spin(10,1)$ invariant Majorana inner product.
It remains to construct the Majorana representation of $Spin(10,1)$. For this, we impose the condition
that the Dirac conjugate is equal to the Majorana conjugate.
 It turns out that it is convenient to chose as a
reality condition
\be
\eta=-\Gamma_0 B(\eta^*)~,
\ee
or equivalently
\be
\eta^*=\Gamma_{6789}\eta~.
\la{rcon}
\ee
The map $C=\Gamma_{6789}$ is also called charge conjugation matrix. In this basis the
$(Spin(7)\ltimes \bR^8)\times \bR$-invariant
Majorana spinor is $1+e_{1234}$. The simplicity of this representative of the $Spin(10,1)$ orbit
with stability subgroup $(Spin(7)\ltimes \bR^8)\times \bR$ suggests that if one of the Killing spinors is null,
then it may be simpler to use
this basis
to solve the Killing spinor equations.

To solve the Killing spinor equations of eleven-dimensional supergravity, it is convenient to use an oscillator
basis  in the space of spinors $\Delta_c$.  For this  write
\be
\Gamma_{\bar\a}= {1\over \sqrt {2}}(\Gamma_\a+i \Gamma_{\a+5})~,~~~~~~~~~
\Gamma_\pm={1\over \sqrt{2}} (\Gamma_5\pm\Gamma_0)
~,~~~~~~~~~\Gamma_{\a}= {1\over \sqrt {2}}(\Gamma_\a-i \Gamma_{\a+5})~.
\la{hbasis}
\ee
Observe that the Clifford algebra relations in the above basis are
$\Gamma_A\Gamma_B+\Gamma_B\Gamma_A=2 g_{AB}$,   where the non-vanishing
components of the metric are
$g_{\a\bar\b}=\delta_{\a\bar\b}, g_{+-}=1$. In addition, we define
$\Gamma^B=g^{BA} \Gamma_A$.
The $1$ spinor is a Clifford  vacuum, $\Gamma_{\bar\a}1=\Gamma_+ 1=0$
and  the representation $\Delta_c$
can be constructed by acting on $1$ with the creation operators
$\Gamma^{\bar\a}, \Gamma^+$
or equivalently any spinor can be written as
\be
\eta= \sum_{k=0}^5 {1\over k!}~ \phi_{\bar a_1\dots \bar a_k}~
 \Gamma^{\bar a_1\dots\bar a_k} 1~,~~~~\bar a=\bar\a, +~,
 \la{hbasisa}
\ee
i.e. $\Gamma^{\bar a_1\dots\bar a_k} 1$, for $k=0,\dots,5$, is a basis in the
space of (Dirac) spinors.

Observe that both the timelike basis and the null basis of
$\Delta_c$ are oscillator bases. Because of this, it is
straightforward to adapt the results we have obtained in this paper
for the timelike basis for the Killing spinor equations and for
their integrability conditions to the null basis.

\newsection{N=1 backgrounds}\la{appnone}

In this appendix we summarize the solution of the Killing spinor equations
for backgrounds that admit one Killing spinor with stability subgroup $SU(5)$,
i.e. the spacetime one-form  bilinear is timelike. This case has been
analyzed in \cite{pakis}. The  results, in the form we summarize them below,
 have appeared in \cite{joe}.

The conditions on the geometry are \bea
\Omega_{0,ij}=\Omega_{i,0j}~,~~~~2 \partial_{\bar\alpha} \log
f+\Omega_{0,0\bar\a}&=&0 \nonumber \\ \Omega_{\bar\b,}{}^{\bar
\b}{}_\g -\Omega_{\g,\b}{}^\b-\Omega_{0,0\g}&=&0~. \eea The electric
part of the flux is expressed in terms of the geometry as \bea
G_{\bar\a\b\g}&=&-2i \Omega_{\bar\a,\b\g}+2i g_{\bar\a[\b}
\Omega_{0,0\g]} \nonumber \\ G_{\bar\a_1\bar\a_2\bar\a_3}&=&6i
\Omega_{[\bar\a_1,\bar\a_2\bar\a_3]} \eea and the magnetic part of
the flux is \bea F_{\b_1\dots\b_4}&=&\tfrac{1}{2}
(-\Omega_{0,0\bar\a}+2 \Omega_{\bar\a, \b}{}^\b)
\te^{\bar\a}{}_{\b_1\dots\b_4} \nonumber \\ F_{\b\bar\a\g}{}^\g&=&2i
\Omega_{\bar\a,0\b} +2i g_{\bar\a\b} \Omega_{\bar\g, 0 \d}
g^{\bar\g\d} \nonumber \\ F_{\bar\a\b_1\b_2\b_3}&=&\tfrac{1}{2}
[\Omega_{\bar\a,\bar\g_1\bar\g_2}
\te^{\bar\g_1\bar\g_2}{}_{\b_1\b_2\b_3} + 3 \Omega_{\bar\g_1,
\bar\g_2\bar\g_3} \te^{\bar\g_1\bar\g_2\bar\g_3}{}_{[\b_1\b_2}
g_{\b_3]\bar\a} +12i \Omega_{[\b_1,0\b_2} g_{\b_3]\bar\a}]~.
\la{otF}
 \eea
 The traceless (2,2) part of $F$ is not determined by
the Killing spinor equations.

The conditions on the geometry imply that the one-form $\kappa^f= -f^2 \kappa= f^2 e^0$
is a timelike Killing vector field and the space of orbits of this vector field
has an $SU(5)$ structure with
 $W_5+2df=0$, where
\be
(W_5)_\a=\Omega_{\bar\b,}{}^{\bar\b}{}_\a -\Omega_{\a,\b}{}^\b~,
\ee
is a Gray-Hervella type of class \cite{grayhervella}. We use the above
results to investigate backgrounds
with two supersymmetries.

\newsection {Killing spinor equations in canonical basis}

\label{Killing-list}

To derive the linear system associated with the Killing spinor
equations for the geometry, fluxes and spacetime derivatives of
$f,g, u,v, w,z$ one has to expand ${\cal D}_A\sigma_I$ in the
Hermitian basis (\ref{hermbasis}) and use (\ref{genspcov}). This
computation can be simplified in various ways. First, it is not
necessary to compute both ${\cal D}_\a\sigma_I$ and ${\cal
D}_{\bar\a}\sigma_I$ because since the spinors $\epsilon$ are real
the equations derived from ${\cal D}_\a\epsilon$ are complex
conjugate to those of ${\cal D}_{\bar\a}\epsilon$ and so are not
independent\footnote{Observe though that $\sigma_I$ are complex
spinors and so this complex conjugation relation does not apply for
them.}. In addition, since ${\cal D}_0\epsilon$ is real only  half
of the relations are independent. These are chosen to be along the
basis elements $1$, $\Gamma^{\bar\a}1$ and $\Gamma^{\bar\a\bar\b}1$.
The remaining are related to these by complex conjugation followed
by dualization with the (5,0) form $\epsilon$. So again, we shall
give only the independent conditions. We remark that one can use
these relations between the equations of the linear system to
provide a useful check of the result.

It is intended that the results of this appendix to be used as a
manual to derive the linear system associated with the Killing
spinor equations of any number of spinors. Because of this, we first
state the action of the supercovariant derivative ${\cal
D}_A\sigma_I$ on the appropriate irreducible spinor $\sigma_I$ as a
title of its subsection. Then we expand ${\cal D}_A\sigma_I$ in the
canonical basis. On the left column, we state the basis element of
the oscillator basis (\ref{hermbasis}), and in the right column we give the associated
component.

\subsection{${\cal D}_A 1$}

Evaluating ${\cal D}_01$ and expanding the result in the basis
(\ref{hermbasis}), we find \vskip 0.3cm \leftline{\underline{{${\cal
D}_01$}}} \vskip -0.5cm \bea
1&:&\tfrac{1}{2}\Omega_{0,\g}{}^\g-\tfrac{i}{24}
F_{\g}{}^\g{}_\d{}^\d \cr \Gamma^{\bar \b}1&:& \tfrac{i}{2}
\Omega_{0,0\bar\b}+\tfrac{1}{6} G_{\bar\b\g}{}^\g \cr \Gamma^{\bar
\b_1\bar\b_2}1&:& \tfrac{1}{4} \Omega_{0,\bar\b_1
\bar\b_2}-\tfrac{i}{24}
    F_{\bar\b_1\bar\b_2\g}{}^\g~.
    \la{dzo}
\eea
Similarly, computing ${\cal D}_{\bar\a} 1$ and expanding the result
in the basis (\ref{hermbasis}), we get \vskip 0.3cm
\leftline{\underline{{${\cal D}_{\bar\a} 1$}}} \vskip -0.5cm \bea
1&:& \tfrac{1}{2}\Omega_{\bar \a,\g}{}^\g+\tfrac{i}{12}
G_{\bar\a\g}{}^\g \cr \Gamma^{\bar \b}1 &:& \tfrac{i}{2}
\Omega_{\bar\a,0\bar\b}+\tfrac{1}{12} F_{\bar\a\bar\b\g}{}^\g \cr
\Gamma^{\bar \b_1\bar\b_2}1&:& \tfrac{1}{4} \Omega_{\bar\a,\bar\b_1
\bar\b_2}+\tfrac{i}{24}
    G_{\bar\a\bar\b_1\bar\b_2}
\cr \Gamma^{\bar \b_1\bar\b_2\bar\b_3}1&:& \tfrac{1}{72}
F_{\bar\a\bar\b_1\bar\b_2 \bar \b_3} \cr \Gamma^{\bar
\b_1\bar\b_2\bar\b_3\bar\b_4}1&:& 0 \cr \Gamma^{\bar
\b_1\bar\b_2\bar\b_3\bar\b_4\bar\b_5}1&:& 0~. \la{dabo} \eea As we
have explained the expressions for the remaining basis elements in
(\ref{dzo}) and for ${\cal D}_\a 1$ can be recovered from the above
using complex conjugation.

\subsection{${\cal D}_A e_{12345}$}

The time component of the Killing spinor equations yields \vskip
0.3cm \leftline{\underline{{${\cal D}_0 e_{12345}$}}} \vskip -0.5cm
\bea 1&:& 0 \cr \Gamma^{\bar \b}1&:& \tfrac{i}{144}
F_{\g_1\g_2\g_3\g_4} \te^{\g_1\g_2\g_3\g_4}{}_{\bar\b} \cr
\Gamma^{\bar \b_1\bar\b_2}1&:& -\tfrac{1}{72} G_{\g_1\g_2\g_3}
\te^{\g_1\g_2\g_3}{}_{\bar\b_1\bar\b_2} \la{dabottff} \eea
Similarly the ${\cal D}_{\bar\a} e_{12345}$ yields
\vskip 0.3cm
\leftline{\underline{{${\cal D}_{\bar\a} e_{12345}$}}}
\vskip -0.5cm

\bea 1&:& -\tfrac{1}{72}\te_{\bar\a}{}^{\g_1\g_2\g_3\g_4}
F_{\g_1\g_2\g_3\g_4} \cr \Gamma^{\bar \b}1&:&-\tfrac{i}{36}
\te_{\bar\a\bar\b}{}^{\g_1\g_2\g_3} G_{\g_1\g_2\g_3} \cr
\Gamma^{\bar \b_1\bar\b_2}1&:&-\tfrac{1}{48}
\te_{\bar\a\bar\b_1\bar\b_2}{}^{\g_1\g_2}
 F_{\g_1\g_2\d}{}^\d -\tfrac{1}{48} F_{\bar\a\g_1\g_2\g_3}
 \te^{\g_1\g_2\g_3}{}_{\bar\b_1\bar\b_2}
 \cr
\Gamma^{\bar \b_1\bar\b_2\bar\b_3}1&:&-\tfrac{1}{48}
\Omega_{\bar\a,\g_1\g_2} \te^{\g_1\g_2}{}_{\bar
\b_1\bar\b_2\bar\b_3}-\tfrac{i}{144} \te_{\bar\a\bar
\b_1\bar\b_2\bar\b_3}{}^\g G_{\g\d}{}^\d+\tfrac{i}{96}
G_{\bar\a\g_1\g_2} \te^{\g_1\g_2}{}_{\bar \b_1\bar\b_2\bar\b_3} \cr
\Gamma^{\bar \b_1\bar\b_2\bar\b_3\bar\b_4}1&:&-\tfrac{i}{192}
\Omega_{\bar\a, 0\g} \te^\g{}_{\bar \b_1\bar\b_2\bar\b_3\bar\b_4}-
\tfrac{1}{24^2\cdot 4} F_{\g}{}^\g{}_\d{}^\d \te_{\bar\a\bar
\b_1\bar\b_2\bar\b_3\bar\b_4}- \tfrac{1}{384} F_{\bar\a\g\d}{}^\d
\te^{\g}{}_{\bar \b_1\bar\b_2\bar\b_3\bar\b_4} \cr \Gamma^{\bar
\b_1\bar\b_2\bar\b_3\bar\b_4\bar\b_5}1&:&\tfrac{1}{8\cdot 5!}
[-\tfrac{1}{2}\Omega_{\bar\a,\g}{}^\g+\tfrac{i}{4}
G_{\bar\a\g}{}^\g]
 \te_{\bar\b_1\bar\b_2\bar\b_3\bar\b_4\bar\b_5}~,
\eea
where $\te_{\bar\a_1\cdots\bar\a_5}=\sqrt{2} \epsilon_{\bar\a_1\cdots\bar\a_5}$
and $\epsilon_{\bar 1\cdots \bar5}=1$.

\subsection{$\sqrt{2}\, {\cal D}_A e_k$}

We split up $\alpha$ into\footnote{Note that $k$ is not an index
here but rather a fixed label for a particular spinor $e_k$. The
same holds for the labels of all other spinors $e_{i_1 \cdots i_I}$
in these tables.} $\rho$ and $k$, where $\rho$ are the remaining
four indices: $\rho=(1,\ldots, \hat{k}, \ldots, 5)$. The time
component of the Killing spinor equations yields \vskip 0.3cm
\leftline{\underline{{$\sqrt{2}\,\, {\cal D}_0 e_k$}}} \vskip -0.5cm
\bea 1&:&2 (-\tfrac{i}{2}\Om_{0,0k}+\tfrac{1}{6}G_{k\l}{}^\l) \cr
\Gamma^{\bar \tau}1&:& 2 (\tfrac{1}{2}\Om_{0,\bar \tau
k}+\tfrac{i}{12} F_{\bar \tau k \l}{}^\l) \cr \Gamma^{\bar k}1&:
&\tfrac{1}{2}\Om_{0,\l}{}^\l-\tfrac{1}{2}\Om_{0,k\bar
    k}
    +\tfrac{i}{24}F_\l{}^\l{}_\sigma{}^\sigma- \tfrac{i}{12} F_\l{}^\l{}_{k\bar k}
    \cr
\Gamma^{\bar \tau_1\bar \tau_2}1&:&\tfrac{1}{6}G_{\bar \tau_1\bar
\tau_2k} \cr \Gamma^{\bar \tau\bar k}1&:&-\tfrac{i}{2}\Om_{0,0\bar
\tau}+\tfrac{1}{6} G_{\bar \tau \l}{}^{\l}-\tfrac{1}{6}G_{\bar \tau
k\bar k}~. \la{dzf} \eea The different spatial directions, i.e.
$\bar \rho$ and $\bar k$, yield
\vskip 0.3cm \leftline{\underline{{$\sqrt{2}\,\, {\cal
D}_{\bar\rho}e_k$}}} \vskip -0.5cm \bea 1&:& -i\Om_{\bar
\rho,0k}+\tfrac{1}{6}F_{\bar \rho k\l}{}^\l \cr \Gamma^{\bar
\tau}1&:& \Om_{\bar\rho,\bar\tau k}-\tfrac{i}{6}G_{\bar\rho\bar\tau
k} \cr \Gamma^{\bar k}1&:&
\tfrac{1}{2}\Om_{\bar\rho,\l}{}^\l-\tfrac{1}{2}\Om_{\bar\rho,k\bar
k}
    -\tfrac{i}{12}G_{\bar\rho\l}{}^\l+\tfrac{i}{12}G_{\bar\rho k\bar k}
\cr \Gamma^{\bar \tau_1\bar \tau_2}1&:&
\tfrac{1}{12}F_{\bar\rho\bar\tau_1\bar\tau_2k} \cr \Gamma^{\bar
\tau\bar k}1&:&
    -\tfrac{i}{2}\Om_{\bar\rho,0\bar\tau}+\tfrac{1}{12}F_{\bar\rho\bar\tau\l}{}^\l
    -\tfrac{1}{12}F_{\bar\rho\bar\tau k\bar k}
\cr \Gamma^{\bar \tau_1\bar \tau_2\bar \tau_3}1&:&0 \cr \Gamma^{\bar
\tau_1\bar \tau_2\bar k}1&:&
    (\tfrac{1}{4}\Om_{\bar\rho,\bar\tau_1\bar\tau_2}-\tfrac{i}{24}
    G_{\bar\rho\bar\tau_1\bar\tau_2})
\cr \Gamma^{\bar \tau_1\bar \tau_2\bar \tau_3\bar \tau_4}1&:& 0 \cr
\Gamma^{\bar \tau_1\bar \tau_2\bar \tau_3\bar k}1&:& \tfrac{1}{72}
F_{\bar\rho \bar\tau_1\bar\tau_2\bar\tau_3} \cr \Gamma^{\bar
\tau_1\bar \tau_2\bar \tau_3\bar \tau_4\bar k}1&:& 0~. \la{drbf}
\eea
Next we find that $\sqrt{2}\,\,{\cal D}_{\bar k}e_k$ gives \vskip
0.3cm \leftline{\underline{{$\sqrt{2}\,\, {\cal D}_{\bar k}e_k$}}}
\vskip -0.5cm
\bea
1&:& -(i\Om_{\bar
    k,0k}+\tfrac{1}{12}F_\l{}^\l{}_\m{}^\m
    +\tfrac{1}{3}F_{k\bar k\l}{}^\l)
\cr \Gamma^{\bar \tau}1&:&\Om_{\bar k,\bar \tau
k}-\tfrac{i}{6}G_{\bar \tau\l}{}^\l -\tfrac{i}{3}G_{\bar \tau k\bar
k} \cr \Gamma^{\bar k}1&:& \tfrac{1}{2}\Om_{\bar k,\l}{}^\l
    -\tfrac{1}{2}\Om_{\bar k,k\bar k}-\tfrac{i}{4}G_{\bar k\l}{}^\l
\cr \Gamma^{\bar \tau_1\bar \tau_2}1&:&
-\tfrac{1}{12}F_{\bar\tau_1\bar\tau_2\l}{}^\l
-\tfrac{1}{6}F_{\bar\tau_1\bar\tau_2 k\bar k} \cr \Gamma^{\bar \tau
\bar k}1&:&
    -\tfrac{i}{2}\Om_{\bar k,0\bar\tau}-\tfrac{1}{4}F_{\bar\tau\bar k\l}{}^\l
\cr \Gamma^{\bar \tau_1\bar \tau_2\bar \tau_3}1&:&-\tfrac{i}{36}
G_{\bar\tau_1\bar\tau_2\bar\tau_3} \cr \Gamma^{\bar \tau_1\bar
\tau_2\bar k}1&:& \tfrac{1}{4}\Om_{\bar k,\bar \tau_1 \bar \tau_2}
    -\tfrac{i}{8}G_{\bar k\bar \tau_1 \bar \tau_2}
\cr \Gamma^{\bar \tau_1\bar \tau_2\bar \tau_3\bar \tau_4}1&:&
-\tfrac{1}{144} F_{\bar\tau_1\bar\tau_2\bar\tau_3\bar\tau_4} \cr
\Gamma^{\bar \tau_1\bar \tau_2\bar \tau_3\bar k}1&:&-\tfrac{1}{24}
F_{\bar \tau_1\bar \tau_2\bar \tau_3\bar k} \cr \Gamma^{\bar
\tau_1\bar \tau_2\bar \tau_3\bar \tau_4\bar k}1&:& 0~. \la{dfbf}
\eea

\subsection{$\sqrt{2} {\cal D}_A e_{i_1 \cdots i_4}$}

We split the indices $\alpha$ into $\rho$ and $k$, where $\rho=(i_1,
\ldots, i_4)$ and $k$ is the missing fifth coordinate. In addition
we will use the Levi-Civita symbol $\te_{\bar \rho_1 \cdots \bar
\rho_4}$ which is defined by $\te_{\bar i_1 \cdots \bar i_4} =
\sqrt{2}$. The time component of the Killing spinor equations yields
\vskip 0.3cm \leftline{\underline{{$\sqrt{2}\,\, {\cal D}_{0}e_{i_1
\cdots i_4}$}}} \vskip -0.5cm \bea 1&:&
-\tfrac{i}{72}F_{\l_1\l_2\l_3\l_4}\te^{\l_1\l_2\l_3\l_4} \cr
\Gamma^{\bar \tau}1&:&
-\tfrac{1}{18}G_{\l_1\l_2\l_3}\te^{\l_1\l_2\l_3}{}_{\bar \tau} \cr
\Gamma^{\bar  k}1&:&0 \cr \Gamma^{\bar \tau_1\bar \tau_2}1&:&
\tfrac{1}{2}(-\tfrac{1}{4}\Om_{0,\l_1\l_2}
-\tfrac{i}{24}F_{\l_1\l_2\sigma}{}^\sigma+\tfrac{i}{24}F_{\l_1\l_2
k\bar  k}) \te^{\l_1\l_2}{}_{\bar \tau_1\bar \tau_2} \cr
\Gamma^{\bar \tau\bar  k}1&:& \tfrac{i}{36}F_{\l_1\l_2\l_3\bar  k}\,
\te^{\l_1\l_2\l_3}{}_{\bar \tau}~. \la{dzottf} \eea
The different spatial directions, i.e. $\bar \rho$ and $\bar  k$,
yield
\vskip 0.3cm \leftline{\underline{{$\sqrt{2}\,\, {\cal
D}_{\bar\rho}e_{i_1 \cdots i_4}$}}} \vskip -0.5cm
\bea
1&:&-\tfrac{i}{18}g_{\bar\rho\l_1}G_{\l_2\l_3\l_4}\te^{\l_1\l_2\l_3\l_4}
\cr \Gamma^{\bar \tau}1&:&-(\tfrac{1}{12}F_{\bar\rho\l_1\l_2\l_3}
    +\tfrac{1}{12}g_{\bar\rho\l_1}F_{\l_2\l_3\sigma}{}^\sigma-
    \tfrac{1}{12}g_{\bar\rho\l_1}F_{\l_2\l_3  k\bar  k})\te^{\l_1\l_2\l_3}{}_{\bar\tau}
\cr \Gamma^{\bar  k}1&:&
-\tfrac{1}{18}g_{\bar\rho\l_1}F_{\l_2\l_3\l_4\bar  k}
\te^{\l_1\l_2\l_3\l_4} \cr \Gamma^{\bar \tau_1\bar
\tau_2}1&:&-\tfrac{1}{2}(\tfrac{1}{4}\Om_{\bar\rho,\l_1\l_2}
+\tfrac{i}{8}G_{\bar\rho\l_1\l_2}
    +\tfrac{i}{12}g_{\bar\rho\l_1}G_{\l_2\sigma}{}^\sigma-\tfrac{i}{12}g_{\bar\rho\l_1}
    G_{\l_2  k\bar  k})\te^{\l_1\l_2}{}_{\bar\tau_1\bar\tau_2}
\cr \Gamma^{\bar \tau\bar  k}1&:&
\tfrac{i}{12}\te_{\bar\rho\bar\tau}{}^{\l_1\l_2} G_{\l_1\l_2\bar  k}
\cr \Gamma^{\bar \tau_1\bar \tau_2\bar \tau_3}1&:&
    \tfrac{1}{12}(\tfrac{i}{2}\Om_{\bar\rho,0\l}-\tfrac{1}{4}F_{\bar\rho\l\sigma}{}^\sigma
    +\tfrac{1}{4}F_{\bar\rho\l  k\bar  k}-\tfrac{1}{24}g_{\bar\rho\l}
    F_\sigma{}^\sigma{}_\m{}^\m
    +\tfrac{1}{12}g_{\bar\rho\l}F_{ k\bar k\sigma}{}^\sigma)
    \te^{\l}{}_{\bar\tau_1\bar\tau_2\bar\tau_3}
\cr \Gamma^{\bar \tau_1\bar \tau_2\bar  k}1&:&
-\tfrac{1}{2}(\tfrac{1}{8} F_{\bar\rho \bar  k\l_1\l_2}
    +\tfrac{1}{12}g_{\bar\rho\l_1}F_{\l_2\bar  k\sigma}{}^\sigma)
    \te^{\l_1\l_2}{}_{\bar\tau_1\bar\tau_2}
\cr \Gamma^{\bar \tau_1\bar \tau_2\bar \tau_3\bar \tau_4}1&:&
    \tfrac{1}{96}(-\tfrac{1}{2}\Om_{\bar\rho,\l}{}^\l+\tfrac{1}{2}\Om_{\bar\rho, k\bar  k}
    -\tfrac{i}{4}G_{\bar\rho\l}{}^\l+\tfrac{i}{4}G_{\bar\rho  k\bar  k})
    \te_{\bar\tau_1\bar\tau_2\bar\tau_3\bar\tau_4}
\cr \Gamma^{\bar \tau_1\bar \tau_2\bar \tau_3\bar  k}1&:&
    \tfrac{1}{12}(\tfrac{1}{2}\Om_{\bar\rho,\l\bar  k}+\tfrac{i}{4}G_{\bar\rho\l\bar  k}
    +\tfrac{i}{12}g_{\bar\rho\l}G_{\bar  k\sigma}{}^\sigma)
    \te^\l{}_{\bar\tau_1\bar\tau_2\bar\tau_3}
\cr \Gamma^{\bar \tau_1\bar \tau_2\bar \tau_3\bar \tau_4\bar  k}1&:&
\tfrac{1}{96} (\tfrac{i}{2}\Om_{\bar\rho,0\bar k}
    -\tfrac{1}{4}F_{\bar\rho\bar  k\l}{}^\l)\te_{\bar\tau_1\bar\tau_2\bar\tau_3\bar\tau_4}~.
\la{drbottf}
\eea
Next we turn to $\sqrt{2}\,\, {\cal D}_{\bar  k}e_{i_1 \cdots i_4}$
to find \vskip 0.3cm \leftline{\underline{{$\sqrt{2}\,\, {\cal
D}_{\bar
 k}e_{i_1 \cdots i_4}$}}} \vskip -0.5cm
\bea 1&:& 0 \cr \Gamma^{\bar \tau}1&:&-\tfrac{1}{36}F_{\bar k
\l_1\l_2\l_3}\te^{\l_1\l_2\l_3}{}_{\bar \tau} \cr
    \Gamma^{\bar  k}1&:& 0
\cr \Gamma^{\bar \tau_1\bar
\tau_2}1&:&-\tfrac{1}{2}(\tfrac{1}{4}\Om_{\bar  k,\l_1\l_2}
+\tfrac{i}{24}G_{\bar  k\l_1\l_2})\te^{\l_1\l_2}{}_{\bar \tau_1\bar
\tau_2} \cr \Gamma^{\bar \b\bar  k}1&:& 0 \cr \Gamma^{\bar
\tau_1\bar \tau_2\bar \tau_3}1&:&\tfrac{1}{12}(\tfrac{i}{2}\Om_{\bar
 k,0\l}
    -\tfrac{1}{12}F_{\bar  k\l\sigma}{}^\sigma)
    \te^{\l}{}_{\bar \tau_1\bar \tau_2\bar \tau_3}
\cr \Gamma^{\bar \tau_1\bar \tau_2\bar  k}1&:& 0 \cr \Gamma^{\bar
\tau_1\bar \tau_2\bar \tau_3\bar \tau_4}1&:& \tfrac{1}{96}(
    -\tfrac{1}{2}\Om_{\bar  k,\l}{}^\l+\tfrac{1}{2}\Om_{\bar  k, k\bar  k}
    -\tfrac{i}{12}G_{\bar  k\l}{}^\l)\te_{\bar \tau_1\bar \tau_2\bar \tau_3\bar \tau_4}
\cr
    \Gamma^{\bar \tau_1\bar \tau_2\bar \tau_3\bar  k}1&:&
    \tfrac{1}{24}\Om_{\bar  k,\l\bar  k}\te^\l{}_{\bar
    \tau_1\bar\tau_2\bar\tau_3}
\cr \Gamma^{\bar \tau_1\bar \tau_2\bar \tau_3\bar \tau_4\bar  k}1&:&
    \tfrac{i}{192}\Om_{\bar  k,0\bar  k}\te_{\bar\tau_1\bar\tau_2\bar\tau_3\bar\tau_4}~.
\la{dfbottf}
\eea

\subsection{${\cal D}_A e_{ij} $}

We split the indices $\a$ into $p=(i,j)$ and $a$, which contains the
remaining three indices. We also define $\epsilon_{\bar i \bar j} =
1$. Then the time component of the Killing spinor equations yields
\vskip 0.3cm \leftline{\underline{{${\cal D}_{0}e_{ij}$}}} \vskip
-0.5cm
\bea 1&:&
-2(\tfrac{1}{4}\Om_{0,pq}-\tfrac{i}{24}F_{pqc}{}^c)\ep^{pq} \cr
\Gamma^{\bar a}1&:&-\tfrac{1}{6}G_{\bar a pq}\ep^{pq} \cr
\Gamma^{\bar p}1&:&(\tfrac{i}{2}\Om_{0,0q}-\tfrac{1}{6}G_{qr}{}^r+
\tfrac{1}{6}G_{qa}{}^a)\ep^q{}_{\bar p} \cr \Gamma^{\bar a\bar
b}1&:&\tfrac{i}{24}F_{\bar a\bar b pq}\ep^{pq} \cr \Gamma^{\bar
a\bar p}1&:&(\tfrac{1}{2}\Om_{0,\bar a q} -\tfrac{i}{12}F_{\bar a
qb}{}^b+\tfrac{i}{12}F_{\bar a qr}{}^r)\ep^q{}_{\bar p} \cr
\Gamma^{\bar p\bar
q}1&:&\tfrac{1}{4}(\tfrac{1}{2}\Om_{0,a}{}^a-\tfrac{1}{2}\Om_{0,p}{}^p
    -\tfrac{i}{24}F_a{}^a{}_b{}^b+\tfrac{i}{12}F_a{}^a{}_r{}^r
    -\tfrac{i}{24}F_r{}^r{}_s{}^s)\ep_{\bar p\bar q}
\la{dzott}
\eea
The different spatial directions, i.e. $\bar a$ and $\bar p$, yield
\vskip 0.3cm \leftline{\underline{{${\cal D}_{\bar a}e_{ij}$}}}
\vskip -0.5cm
\bea
    1&:& -2(\tfrac{1}{4}\Om_{\bar a,qr}+\tfrac{i}{24}G_{\bar aqr})\ep^{qr}
    \cr
    \Gamma^{\bar b}1&:& -\tfrac{1}{12}F_{\bar a\bar b qr}\ep^{qr}
    \cr
    \Gamma^{\bar q}1&:& (\tfrac{i}{2}\Om_{\bar a,0r}+\tfrac{1}{12}F_{\bar arb}{}^b-\tfrac{1}{12}F_{\bar ars}{}^s)\ep^r{}_{\bar q}
    \cr
    \Gamma^{\bar b\bar c}1&:& 0
    \cr
    \Gamma^{\bar b\bar q}1&:& (\tfrac{1}{2}\Om_{\bar a,\bar br}+\tfrac{i}{12}G_{\bar a\bar br})\ep^r{}_{\bar q}
    \cr
    \Gamma^{\bar q\bar r}1&:& \tfrac{1}{4}(\tfrac{1}{2}\Om_{\bar a,b}{}^b-\tfrac{1}{2}\Om_{\bar a,s}{}^s
    +\tfrac{i}{12}G_{\bar ab}{}^b-\tfrac{i}{12}G_{\bar as}{}^s)\ep_{\bar q\bar r}
    \cr
    \Gamma^{\bar b\bar c\bar d}1&:& 0
    \cr
    \Gamma^{\bar b\bar c\bar q}1&:& \tfrac{1}{24}F_{\bar a\bar b\bar cr}\ep^r{}_{\bar q}
    \cr
    \Gamma^{\bar b\bar q\bar r}1&:& \tfrac{1}{4}(\tfrac{i}{2}\Om_{\bar a,0\bar b}+\tfrac{1}{12}F_{\bar a\bar b c}{}^c
    -\tfrac{1}{12}F_{\bar a\bar b s}{}^s)\ep_{\bar q\bar r}
    \cr
    \Gamma^{\bar b\bar c\bar d\bar q}1&:& 0
    \cr
    \Gamma^{\bar b\bar c\bar q\bar r}1&:& \tfrac{1}{4}(\tfrac{1}{4}\Om_{\bar a,\bar b\bar c}+\tfrac{i}{24}G_{\bar a\bar b\bar c}
    )\ep_{\bar q\bar r}
    \cr
    \Gamma^{\bar b\bar c\bar d\bar q\bar r}1&:& 0
\eea
and \vskip 0.3cm \leftline{\underline{{${\cal D}_{\bar p}e_{ij}$}}}
\vskip -0.5cm
\bea
    1&:& -2(\tfrac{1}{4}\Om_{\bar p,qr}+\tfrac{i}{12}G_{\bar p qr}-\tfrac{i}{12}g_{\bar p q}G_{ra}{}^a)\ep^{qr}
    \cr
    \Gamma^{\bar b}1&:& (-\tfrac{1}{12}F_{\bar p qr\bar b}+\tfrac{1}{12}g_{\bar p q}F_{r\bar bc}{}^c)\ep^{qr}
    \cr
    \Gamma^{\bar q}1&:& (\tfrac{i}{2}\Om_{\bar p,0r}+\tfrac{1}{4}F_{\bar prb}{}^b-\tfrac{1}{6}F_{\bar prs}{}^s
    -\tfrac{1}{24}g_{\bar p r}F_{b}{}^b{}_c{}^c +\tfrac{1}{12}g_{\bar p r}F_{b}{}^b{}_s{}^s)\ep^r{}_{\bar q}
    \cr
    \Gamma^{\bar b\bar c}1&:& \tfrac{i}{12}G_{r\bar b\bar c}\ep_{\bar
    p}{}^r
    \cr
    \Gamma^{\bar b\bar q}1&:& (\tfrac{1}{2}\Om_{\bar p,\bar b r}+\tfrac{i}{4}G_{\bar p\bar b r}
    +\tfrac{i}{12}g_{\bar p r}G_{\bar bc}{}^c-\tfrac{i}{12}g_{\bar p r}G_{\bar b s}{}^s)\ep^r{}_{\bar q}
    \cr
    \Gamma^{\bar q\bar r}1&:& \tfrac{1}{4}(\tfrac{1}{2}\Om_{\bar p,b}{}^b-\tfrac{1}{2}\Om_{\bar p,s}{}^s
    +\tfrac{i}{4}G_{\bar pb}{}^b-\tfrac{i}{4}G_{\bar ps}{}^s)\ep_{\bar q\bar r}
    \cr
    \Gamma^{\bar b\bar c\bar d}1&:& -\tfrac{1}{36}F_{\bar b\bar c\bar d
    q}\ep_{\bar p}{}^q
    \cr
    \Gamma^{\bar b\bar c\bar q}1&:& (\tfrac{1}{8}F_{\bar pr\bar b\bar c}-\tfrac{1}{24}g_{\bar p r}F_{\bar b\bar c d}{}^d
    +\tfrac{1}{24}g_{\bar p r}F_{\bar b\bar c s}{}^s)\ep^r{}_{\bar q}
    \cr
    \Gamma^{\bar b\bar q\bar r}1&:& \tfrac{1}{4}(\tfrac{i}{2}\Om_{\bar p,0\bar b}+\tfrac{1}{4}F_{\bar p\bar bc}{}^c
    -\tfrac{1}{4}F_{\bar p\bar b s}{}^s)\ep_{\bar q \bar r}
    \cr
    \Gamma^{\bar b\bar c\bar d\bar q}1&:& \tfrac{i}{72}G_{\bar b\bar c\bar
    d}\ep_{\bar p\bar q}
    \cr
    \Gamma^{\bar b\bar c\bar q\bar r}1&:& \tfrac{1}{4}(\tfrac{1}{4}\Om_{\bar p,\bar b\bar c}+\tfrac{i}{8}G_{\bar p\bar b\bar c})\ep_{\bar q\bar r}
    \cr
    \Gamma^{\bar b\bar c\bar d\bar q\bar r}1&:& \tfrac{1}{96}F_{\bar p\bar b\bar c\bar d}\ep_{\bar q\bar r}
\eea
respectively.

\subsection{${\cal D}_A e_{klm} $}

We split the indices $\a$ into $a=(k,l,m)$ and $p$, containing the
remaining two indices. The three-dimensional Levi-Civita symbol
$\te_{\bar a \bar b \bar c}$ is defined by $\te_{\bar k \bar l \bar
m} = \sqrt{2}$. The time component of the Killing spinor equations
yields \vskip 0.3cm \leftline{\underline{{${\cal D}_{0}e_{klm}$}}}
\vskip -0.5cm
\bea
    1&:& -\tfrac{1}{18}G_{abc}\te^{abc}
    \cr
    \Gamma^{\bar a}1&:& -(\tfrac{1}{4}\Om_{0,bc}-\tfrac{i}{24}F_{bcd}{}^d+\tfrac{i}{24}F_{bcp}{}^p)\te^{bc}{}_{\bar a}
    \cr
    \Gamma^{\bar p}1&:&\tfrac{i}{36}F_{abc\bar p}\te^{abc}
    \cr
    \Gamma^{\bar a\bar b}1&:&\tfrac{1}{4}(-\tfrac{i}{2}\Om_{0,0c}-\tfrac{1}{6}G_{cd}{}^d+\tfrac{1}{6}G_{cp}{}^p)\te^c{}_{\bar a\bar b}
    \cr
    \Gamma^{\bar a\bar p}1&:& \tfrac{1}{12}G_{bc\bar p}\te^{bc}{}_{\bar a}
    \cr
    \Gamma^{\bar p\bar q}1&:& 0
\eea
The different spatial directions, i.e. $\bar a$ and $\bar p$, yield
\vskip 0.3cm \leftline{\underline{{${\cal D}_{\bar a}e_{klm}$}}}
\vskip -0.5cm
\bea
    1&:& -2(\tfrac{1}{36}F_{\bar abcd}-\tfrac{1}{24}g_{\bar
    ab}F_{cdp}{}^p)\te^{bcd}
    \cr
    \Gamma^{\bar b}1&:& - (\tfrac{1}{4}\Om_{\bar a,cd}-\tfrac{i}{8}G_{\bar acd}-\tfrac{i}{12}g_{\bar a c}G_{de}{}^e
    +\tfrac{i}{12}g_{\bar a c}G_{dp}{}^p)\te^{cd}{}_{\bar b}
    \cr
    \Gamma^{\bar q}1&:& \tfrac{i}{12}G_{bc\bar q}\te^{bc}{}_{\bar a}
    \cr
    \Gamma^{\bar b\bar c}1&:& \tfrac{1}{4}(-\tfrac{i}{2}\Om_{\bar a,0d}-\tfrac{1}{4}F_{\bar ade}{}^e
    +\tfrac{1}{4}F_{\bar adq}{}^q- \tfrac{1}{24}g_{\bar a d} (F_{e}{}^e{}_f{}^f - 2 F \cont{e} \cont{p}
    + F_{p}{}^p{}_q{}^q)) \te^d{}_{\bar b\bar c}
    \cr
    \Gamma^{\bar b\bar q}1&:& (\tfrac{1}{8}F_{\bar acd\bar q}+\tfrac{1}{12}g_{\bar a c}F_{d\bar qe}{}^e
    -\tfrac{1}{12}g_{\bar a c}F_{d\bar qr}{}^r)\te^{cd}{}_{\bar b}
    \cr
    \Gamma^{\bar q\bar r}1&:& \tfrac{1}{24}F_{bc\bar q\bar r}\te^{bc}{}_{\bar a}
    \cr
    \Gamma^{\bar b\bar c\bar d}1&:& \tfrac{1}{24}(-\tfrac{1}{2}\Om_{\bar a,e}{}^e+\tfrac{1}{2}\Om_{\bar a,q}{}^q
    +\tfrac{i}{4}G_{\bar ae}{}^e-\tfrac{i}{4}G_{\bar aq}{}^q)\te_{\bar b\bar c\bar d}
    \cr
    \Gamma^{\bar b\bar c\bar q}1&:& -\tfrac{1}{4}(\tfrac{1}{2}\Om_{\bar a,d\bar q}-\tfrac{i}{4}G_{\bar ad\bar q}
    -\tfrac{i}{12}g_{\bar a d}G_{\bar qe}{}^e+\tfrac{i}{12}g_{\bar a d}G_{\bar qr}{}^r)\te^d{}_{\bar b\bar c}
    \cr
    \Gamma^{\bar b\bar q\bar r}1&:& \tfrac{i}{24}G_{c\bar q\bar
    r}\te^{c}{}_{\bar a\bar b}
    \cr
    \Gamma^{\bar b\bar c\bar d\bar q}1&:& \tfrac{1}{24}(\tfrac{i}{2}\Om_{\bar a,0\bar q}+\tfrac{1}{4}F_{\bar a\bar q e}{}^e
    -\tfrac{1}{4}F_{\bar a\bar q r}{}^r)\te_{\bar b\bar c\bar d}
    \cr
    \Gamma^{\bar b\bar c\bar q\bar r}1&:& \tfrac{1}{4}(\tfrac{1}{8}F_{\bar ad \bar q\bar r}
    +\tfrac{1}{24}g_{\bar a d}F_{\bar q\bar re}{}^e)\te^{d}{}_{\bar b\bar c}
    \cr
    \Gamma^{\bar b\bar c\bar d\bar q\bar r}1&:& \tfrac{1}{24}(\tfrac{1}{4}\Om_{\bar a,\bar q\bar r} -\tfrac{i}{8}G_{\bar a\bar q\bar r})\te_{\bar b\bar c\bar d}
\eea
and \vskip 0.3cm \leftline{\underline{{${\cal D}_{\bar p}e_{klm}$}}}
\vskip -0.5cm
\bea
    1&:& -\tfrac{1}{36}F_{\bar pbcd}\te^{bcd}
    \cr
    \Gamma^{\bar b}1&:& -(\tfrac{1}{4}\Om_{\bar p,cd}-\tfrac{i}{24}G_{\bar pcd})\te^{cd}{}_{\bar b}
    \cr
    \Gamma^{\bar q}1&:& 0
    \cr
    \Gamma^{\bar b\bar c}1&:& \tfrac{1}{4}(-\tfrac{i}{2}\Om_{\bar p,0d}-\tfrac{1}{12}F_{\bar pde}{}^e+\tfrac{1}{12}F_{\bar pdq}{}^q)\te^d{}_{\bar b\bar c}
    \cr
    \Gamma^{\bar b\bar q}1&:& \tfrac{1}{24}F_{\bar p\bar q cd}\te^{cd}{}_{\bar b}
    \cr
    \Gamma^{\bar q\bar r}1&:& 0
    \cr
    \Gamma^{\bar b\bar c\bar d}1&:& \tfrac{1}{24}(-\tfrac{1}{2}\Om_{\bar p,e}{}^e+\tfrac{1}{2}\Om_{\bar p,q}{}^q
    +\tfrac{i}{12}G_{\bar pe}{}^e-\tfrac{i}{12}G_{\bar pq}{}^q)\te_{\bar a\bar b\bar c}
    \cr
    \Gamma^{\bar b\bar c\bar q}1&:& -\tfrac{1}{4}(\tfrac{1}{2}\Om_{\bar p,d\bar q}+\tfrac{i}{12}G_{\bar p\bar qd})\te^d{}_{\bar b\bar c}
    \cr
    \Gamma^{\bar b\bar q\bar r}1&:& 0
    \cr
    \Gamma^{\bar b\bar c\bar d\bar q}1&:& -\tfrac{1}{24}(-\tfrac{i}{2}\Om_{\bar p,0\bar q}-\tfrac{1}{12}F_{\bar p\bar qe}{}^e)\te_{\bar b\bar c\bar d}
    \cr
    \Gamma^{\bar b\bar c\bar q\bar r}1&:& 0
    \cr
    \Gamma^{\bar b\bar c\bar d\bar q\bar r}1&:& \tfrac{1}{96}\Om_{\bar p,\bar q\bar r}\te_{\bar a\bar b\bar c}
\eea
respectively.

\newsection{Integrability conditions in canonical basis} \label{integrability-list}

\subsection{${{\cal I}}_A 1$}

Inserting $1$ in (\ref{intcond}) and expanding in the different
$\Gamma$-structures, one finds that the integrability conditions
with $A = 0$ give rise to
\vskip 0.3cm \leftline{\underline{{${{\cal I}}_{0} 1$}}} \vskip
-0.5cm
 \bea
   1 &:& -i E_{00} - 12 L_{0 \alpha}{}^{\alpha} - 120 B_{0
  \alpha}{}^{\alpha}{}_{\beta}{}^{\beta} \cr
   \Gamma^{\bar \a} 1&:&  E_{0 \bar \a} - 6 i L_{\bar \a
\beta}{}^{\beta} - 60 i
  B_{\bar \a \beta}{}^{\beta}{}_{\gamma}{}^{\gamma}  \cr
   \Gamma^{\bar \a \bar{\b}}1 &:& -6 L_{0 \bar \a \bar{\b}} - 120
B_{0 \bar \a \bar{\b}}
  \cont{\gamma}
 \eea
For $A = \bar \a$ we find
\vskip 0.3cm \leftline{\underline{{${{\cal I}}_{\bar \a} 1$}}}
\vskip -0.5cm
 \bea
   1 &:& - i E_{0 \bar \a} - 6 L_{\bar \a} \cont{\beta} - 60 B_{\bar
\a} \cont{\beta} \cont{\gamma}  \cr
   \Gamma^{\bar \b}1 &:& E_{\bar \a \bar{\b}} + 6 i L_{0 \bar \a
\bar{\b}} + 120 i
  B_{0 \bar \a \bar{\b}} \cont{\gamma}
  \cr
   \Gamma^{\bar \b \bar{\g}}1 &:& - 3 L_{\bar \a \bar{\b} \bar{\g}} -
60 B_{\bar \a \bar{\b}
  \bar{\g}} \cont{\delta}
  \cr
   \Gamma^{\bar \b \bar{\g} \bar{\d}} 1&:& 20 i B_{0 \bar \a \bar{\b}
\bar{\g} \bar{\d}} \cr
   \Gamma^{\bar \b \bar{\g} \bar{\d} \bar{\e}}1 &:& -5 B_{\bar \a
\bar{\b} \bar{\g} \bar{\d} \bar{\e}} \cr
   \Gamma^{\bar{\b}_1 \cdots \bar{\b}_5}1 &:&  0
 \eea

\subsection{${{\cal I}}_A e_{12345}$}

For the basis element $e_{12345}$ we find the following
integrability conditions for $A =0$:
\vskip 0.3cm \leftline{\underline{{${{\cal I}}_{0} e_{12345}$}}}
\vskip -0.5cm
\bea
   1 &:&  4 i B_{ \a {\b} {\g} {\d} {\e}} \te^{ \a {\b} {\g} {\d}
{\e}} \cr
   \Gamma^{\bar \a}1 &:& -20 B_{0  \b {\g} {\d} {\e}} \te_{\bar
\a}{}^{\b \g \d {\e}} \cr
   \Gamma^{\bar \a \bar{\b}}1 &:& \tfrac{1}{2} (-i L_{ \g {\d} {\e}} +
20 i B_{ \g {\d}
  {\e}} \cont{\phi}) \te_{\bar \a \bar{\b} }{}^{{\g} {\d} {\e}}
\eea For $A = \bar \alpha$ we find
\vskip 0.3cm \leftline{\underline{{${{\cal I}}_{\bar \a}
e_{12345}$}}} \vskip -0.5cm
 \bea
   1 &:& 20 i g_{\bar \alpha {\b}_1} B_{0 {\b}_2 \cdots {\b}_5
  } \te^{{\b}_1 \cdots {\b}_5} \cr
   \Gamma^{\bar \b}1 &:& 2 (g_{\bar \alpha {\g}} L_{{\d} {\e} {\phi}
} + 20
  g_{\bar \alpha {\g}} B_{{\d} {\e} {\phi}}
  \cont{\kappa} - 15 B_{\bar \alpha {\g} {\d} {\e} {\phi}}) \te_{\bar
\b}{}^{\g \d \e \phi} \cr
   \Gamma^{\bar \b \bar{\g}} 1&:& - \tfrac{1}{2} (3 i g_{\bar \alpha
{\d}} L_{0 {\e} {\phi}} - 60 i
  g_{\bar \alpha {\d}} B_{0 {\e} {\phi} } \cont{\kappa} -
  60 i  B_{0 \bar \alpha {\d} {\e} {\phi}} ) \te_{\bar \b \bar \g}{}^{\d \e
\phi} \cr
   \Gamma^{\bar \b \bar{\g} \bar \d} 1&:& - \tfrac{1}{12} (-6 g_{\bar
\alpha \e} L_{\phi } \cont{\kappa} - 9
  L_{\bar \alpha \e \phi} + 60 g_{\bar \alpha \e} B_{\phi}
  \cont{\kappa} \cont{\lambda} + 180 B_{\bar \alpha \e \phi}
  \cont{\kappa}) \te_{\bar \b \bar \g \bar \d}{}^{\e \phi}
  \cr
   \Gamma^{\bar \b \bar{\g} \bar \d \bar \e}1 &:& \tfrac{1}{96}(E_{\bar \alpha \phi} - 6 i g_{\bar \alpha \phi} L_{0}
  \cont{\kappa} - 18 i L_{0 \bar \alpha \phi} + 60 i g_{\bar \alpha \phi}
  B_{0} \cont{\kappa} \cont{\lambda} + 360 i B_{0 \bar \alpha \phi}
  \cont{\kappa}) \te_{\bar \b \bar \g \bar \d \bar \e}{}^{\phi}
  \cr
   \Gamma^{\bar \b_1 \cdots \bar \b_5}1 &:& \tfrac{1}{960}(i E_{0 \bar
\alpha} + 18 L_{\bar \alpha} \cont{\kappa}- 180
  B_{\bar \alpha} \cont{\kappa} \cont{\lambda}) \te_{\bar \b_1 \cdots \bar
\b_5}
 \eea
where $\te_{\bar 1 \cdots \bar 5} = \sqrt{2}$.

\subsection{$\sqrt{2}\, {{\cal I}}_A e_k$}

Next we consider the contributions from $ \sqrt{2} e_ k$. We split
up $\alpha$ into\footnote{Note that $k$ is not an index here but
rather a fixed label for a particular spinor $e_k$. The same holds
for the labels of all other spinors $e_{i_1 \cdots i_I}$ in these
tables.} $\rho$ and $k$, where $\rho$ are the remaining four
indices: $\rho=(1,\ldots, \hat{k}, \ldots, 5)$. The $A = 0$
integrability conditions amount to
\vskip 0.3cm \leftline{\underline{{$\sqrt{2} \, {{\cal I}}_{0} e_{
k}$}}} \vskip -0.5cm
 \bea
   1 &:& -2 (-E_{0 k} - 6 i L \cont{\l} {}_k - 60 i B \cont{\l}
  \cont{\m} {}_k) \cr
   \Gamma^{\bar \lambda}1 &:&  -2(12 L_{0 \bar \lambda k} + 240 B_{0
\bar \lambda}
  \cont{\mu} {}_{k})
  \cr
   \Gamma^{\bar \lambda \bar{\mu}}1 &:& -2 (- 3 i L_{\bar \lambda
\bar \mu k} - 60 i
  B_{\bar \lambda \bar \mu} \cont{\nu} {}_k)
  \cr
   \Gamma^{\bar k}1 &:&  i E_{00} - 12 L_0 \cont{\lambda} + 12 L_{0 k
\bar{k}}
   - 120 B_0 \cont{\lambda} \cont{\mu} + 240 B_{0} \cont \lambda {}_{k
\bar{k}} \cr
   \Gamma^{\bar{\lambda} \bar k } 1&:&  E_{0 \bar \lambda} + 6 i
L_{\bar \lambda} \cont{\mu} -
   6 i L_{\bar \lambda k \bar{k}} + 60 i B_{\bar \lambda} \cont{\mu}
   \cont{\nu} - 120 i B_{\bar \lambda} \cont{\mu} {}_{k \bar{k}}
 \eea
The $A = \bar \lambda$ integrability conditions on $\sqrt{2} e_k$
yield
\vskip 0.3cm \leftline{\underline{{$\sqrt{2} \, {{\cal I}}_{\bar
\lambda} e_{k}$}}} \vskip -0.5cm
 \bea
  1 &:& 2 E_{\bar \lambda k} - 12 i L_{0 \bar \lambda k} - 240 i
 B_{0 \bar \lambda} \cont{\mu} {}_{k} \cr
  \Gamma^{\bar \mu} 1&:&  -12 L_{\bar \lambda \bar \mu k} - 240 B_{
\bar \lambda \bar
 \mu} \cont{\nu} {}_k \cr
  \Gamma^{\bar \mu \bar \nu}1 &:&  -120 i B_{0 \bar \lambda \bar \mu
\bar \nu
 k} \cr
  \Gamma^{\bar \mu \bar \nu \bar \rho} 1&:&   -40 B_{\bar \lambda
\bar \mu \bar \nu \bar
 \rho k} \cr
  \Gamma^{\bar \mu \bar \nu \bar \rho \bar \sigma}1 &:&
 0 \cr
  \Gamma^{\bar k} 1&:&  i E_{0 \bar \lambda} - 6 L_{\bar \lambda}
\cont{\mu} +
 6 L_{\bar \lambda k \bar k} -60 B_{\bar \lambda} \cont{\mu} \cont{\nu} +
120
 B_{\bar \lambda} \cont{\mu} {}_{k \bar k} \cr
  \Gamma^{\bar \mu \bar k}1 &:&  E_{\bar \lambda \bar \mu} - 6 i L_{0
\bar \lambda \bar
 \mu} - 120 i B_{0 \bar \lambda \bar \mu} \cont{\nu} +120 i B_{0 \bar
\lambda \bar
 \mu k \bar k} \cr
  \Gamma^{\bar \mu \bar \nu \bar k} 1&:&  -3 L_{\bar \lambda \bar \mu
\bar
 \nu} - 60 B_{\bar \lambda \bar \mu \bar \nu} \cont \rho + 60 B_{\bar
\lambda \bar \mu \bar
 \nu k \bar k} \cr
  \Gamma^{\bar \mu \bar \nu \bar \rho \bar k} 1&:&  - 20 i B_{0 \bar
\lambda \bar \mu \bar \nu \bar \rho} \cr
  \Gamma^{\bar \mu \bar \nu \bar \rho \bar \sigma \bar k} 1&:& 0
 \eea
Finally, the $A = \bar k$ integrability conditions give the
following contributions:
\vskip 0.3cm \leftline{\underline{{$\sqrt{2} \, {{\cal I}}_{\bar k}
e_{k}$}}} \vskip -0.5cm
 \bea
  1 &:& 2 E_{k \bar k} + 12 i L_0 \cont \lambda + 24 i L_{0 k \bar
 k} + 120 i B_0 \cont \lambda \cont \mu + 480 i B_0 \cont \lambda {}_{k
\bar
 k} \cr
  \Gamma^{\bar \lambda} 1&:& - 12 L_{\bar \lambda} \cont{\mu} - 24
L_{\bar \lambda k \bar k}
 -120 B_{\bar \lambda} \cont \mu \cont \nu - 480 B_{\bar \lambda}
\cont{\mu}
 {}_{k \bar k} \cr
  \Gamma^{\bar \lambda \bar \mu} 1&:& 6 i L_{0 \bar \lambda \bar \mu}
+ 120 i
 B_{0 \bar \lambda \bar \mu} \cont \nu + 240 i B_{0 \bar \lambda \bar \mu
k
 \bar k} \cr
  \Gamma^{\bar \lambda \bar \mu \bar \nu} 1&:& - 2 L_{\bar \lambda
\bar \mu \bar
 \nu} - 40 B_{\bar \lambda \bar \mu \bar \nu} \cont \rho - 80 B_{\bar
\lambda \bar \mu \bar \nu k \bar
 k} \cr
  \Gamma^{\bar \lambda \bar \mu \bar \nu \bar \rho}1 &:& 10 i B_{0
\bar \lambda \bar \mu \bar
 \nu \bar \rho} \cr
  \Gamma^{\bar k}1 &:& i E_{0 \bar k} - 18 L \cont \lambda {}_{\bar
k} -
 180 B \cont \lambda \cont \mu {}_{\bar k} \cr
  \Gamma^{\bar \lambda \bar k}1 &:& E_{\bar \lambda \bar k} + 18 i
L_{0 \bar \lambda
 \bar k} + 360 i B_{0 \bar \lambda} \cont{\mu} {}_{\bar k} \cr
  \Gamma^{\bar \lambda \bar \mu \bar k}1 &:& - 9 L_{\bar \lambda \bar
\mu \bar
 k} - 180 B_{\bar \lambda \bar \mu} \cont{\nu} {}_{\bar k} \cr
  \Gamma^{\bar \lambda \bar \mu
 \bar \nu\bar k}1 &:& 60 i B_{0 \bar \lambda \bar \mu
 \bar \nu \bar k} \cr
  \Gamma^{\bar \lambda \bar \mu \bar \nu \bar \rho \bar k}1 &:& -15 B_{\bar \lambda \bar \mu \bar \nu \bar \rho \bar
 k}
 \eea

\subsection{$\sqrt{2} {{\cal I}}_A e_{i_1 \cdots i_4}$}

Next we consider $\sqrt{2} e_{i_1 \cdots i_4}$. We split the indices
$\alpha$ into $\rho$ and $k$, where $\rho=(i_1, \ldots, i_4)$ and
$k$ is the missing fifth coordinate. In addition we will use the
Levi-Civita symbol $\te_{\bar \rho_1 \cdots \bar \rho_4}$ which is
defined by $\te_{\bar i_1 \cdots \bar i_4} = \sqrt{2}$. The $A =0$
integrability conditions are
\vskip 0.3cm \leftline{\underline{{$\sqrt{2} \, {{\cal I}}_{0}
e_{i_1 \cdots i_4}$}}} \vskip -0.5cm
\bea
   1 &:& -40 B_{0  \lambda  \mu  \nu  \rho} \te^{ \lambda  \mu  \nu
\rho} \cr
   \Gamma^{\bar{\lambda}}1 &:& -2 (i L_{\mu \nu \rho} - 20 i B_{\mu
\nu \rho} \cont{\sigma} + 20 i
  B_{\mu \nu \rho k \bar k}) \te_{\bar \lambda}{}^{\mu  \nu  \rho} \cr
   \Gamma^{\bar{\lambda} \bar \mu}1 &:&  \tfrac{1}{2} (6 L_{0 \nu
\rho} - 120
  B_{0 \nu \rho} \cont{\sigma} + 120 B_{0 \nu \rho k \bar k}) \te_{\bar
\lambda \bar \mu }{}^{\nu  \rho} \cr
   \Gamma^{\bar k} 1&:&  -20 i B_{ \lambda  \mu  \nu  \rho  \bar k}
\te^{ \lambda  \mu  \nu  \rho} \cr
   \Gamma^{\bar \lambda \bar{k}} 1&:&  -80 B_{0 \mu \nu \rho \bar k}
\te_{\bar \lambda}{}^{\mu  \nu  \rho} \eea The $A = \bar \lambda$
integrability conditions on $\sqrt{2} e_{i_1 \cdots i_4}$ give rise
to
\vskip 0.3cm \leftline{\underline{{$\sqrt{2} \, {{\cal I}}_{\bar
\lambda} e_{i_1 \cdots i_4}$}}} \vskip -0.5cm
 \bea
   1 &:& 4 (g_{\bar \lambda  \mu} L_{ \nu  \rho \sigma}- 15 B_{\bar
\lambda  \mu  \nu  \rho \sigma} - 20 g_{\bar         \lambda  \mu}
(B_{ \nu  \rho \sigma} \cont{\tau} - B_{ \nu  \rho  \sigma k \bar
k})) \te^{\mu \nu \rho \sigma} \cr
   \Gamma^{\bar \mu}1 &:&  2 (- 3 i g_{\bar \lambda \nu} L_{0 \rho
\sigma} + 60 i B_{0 \bar \lambda \nu \rho \sigma} + 60 i g_{\bar
\lambda \nu} (B_{0 \rho \sigma} \cont{\tau} - B_{0 \rho \sigma k
\bar k})) \te_{\bar \mu}{}^{\nu \rho \sigma} \cr
   \Gamma^{\bar \mu \bar \nu}1 &:&  - \tfrac{1}{2} (- 9 L_{\bar
\lambda \rho \sigma} - 6 g_{\bar \lambda \rho} (L_{\sigma}
\cont{\tau} - L_{\sigma k \bar k}) + 180 B_{\bar \lambda \rho
\sigma} \cont{\tau} - 180 B_{\bar \lambda \rho \sigma k \bar k} +
\cr
 && + 60 g_{\bar \lambda \rho} (
B_{\sigma} \cont{\tau}
  \cont{\omega} - 2 B_{\sigma} \cont{\tau} {}_{k \bar k})) \te_{\bar \mu
\bar \nu}{}^{\rho \sigma} \cr
   \Gamma^{\bar \mu \bar \nu \bar \rho}1 &:&  - \tfrac{1}{12} (E_{\bar
\lambda \sigma} + 18 i L_{0 \bar \lambda
  \sigma} + 6 i g_{\bar \lambda \sigma} (L_0 \cont{\tau} - L_{0 k \bar k})
  - 360 i B_{0 \bar \lambda
\sigma} \cont{\tau} + 360 i B_{0 \bar \lambda \sigma k \bar k} + \cr
 && - 60 i g_{\bar \lambda
\sigma} (B_0 \cont{\tau} \cont{\omega} - 2 B_0 \cont{\tau} {}_{k
\bar
  k})) \te_{\bar \mu \bar \nu \bar \rho}{}^{
\sigma} \cr
   \Gamma^{\bar \mu_1 \cdots \bar \mu_4} 1&:&  \tfrac{1}{96} (- i E_{0
\bar \lambda} + 18 L_{\bar \lambda} \cont{\nu} - 18 L_{\bar \lambda
k \bar k} - 180 B_{\bar \lambda} \cont{\nu} \cont{\rho} + 360
B_{\bar \lambda} \cont{\nu} {}_{k \bar
  k}) \te_{\bar \mu_1 \cdots \bar \mu_4} \cr
   \Gamma^{\bar k} 1&:& - 80 i g_{\bar \lambda \mu} B_{0  \nu
    \rho
\sigma \bar k} \te^{\mu \nu \rho \sigma} \cr
   \Gamma^{\bar \mu \bar k}1 &:&  2 (3 g_{\bar \lambda \nu} L_{\rho
\sigma
  \bar k}- 60 B_{\bar \lambda \nu \rho \sigma \bar k} - 60 g_{\bar \lambda
\nu} B_{\rho
\sigma} \cont{\tau}
  {}_{\bar k}) \te_{\bar \mu}{}^{\nu \rho
\sigma} \cr
   \Gamma^{\bar \mu \bar \nu \bar k}1 &:&  - \tfrac{1}{2} (- 6 i
g_{\bar \lambda \rho} L_{0
\sigma
  \bar k} + 180 i B_{0 \bar \lambda \rho
\sigma \bar k} + 120 i g_{\bar \lambda \rho} B_{0 \sigma}
\cont{\tau} {}_{
  \bar k}) \te_{\bar \mu \bar \nu}{}^{\rho
\sigma} \cr
   \Gamma^{\bar \mu \bar \nu \bar \rho \bar k}1 &:&  -\tfrac{1}{12}
(-18 L_{\bar \lambda \sigma \bar k} - 6 g_{\bar \lambda \sigma} L
\cont{\tau}
  {}_{\bar k}+ 360 B_{\bar \lambda
\sigma} \cont{\tau} {}_{\bar k} + 60 g_{\bar \lambda \sigma} B
\cont{\tau} \cont{\omega}
  {}_{\bar k}) \te_{\bar \mu \bar \nu \bar \rho}{}^{
\sigma} \cr
   \Gamma^{\bar \mu_1 \cdots \bar \mu_4 \bar k} 1&:&  \tfrac{1}{96}
(E_{\bar \lambda \bar k} + 18 i L_{0 \bar \lambda \bar k} - 360 i
B_{0 \bar \lambda} \cont{\nu} {}_{\bar k}) \te_{\bar \mu_1 \cdots
\bar \mu_4}
 \eea
The $A = \bar k$ integrability conditions on $\sqrt{2} e_{i_1 \cdots
i_4}$ lead to
\vskip 0.3cm \leftline{\underline{{$\sqrt{2} \, {{\cal I}}_{\bar k}
e_{i_1 \cdots i_4}$}}} \vskip -0.5cm
 \bea
   1 &:& - 20 B_{ \lambda  \mu
  \nu  \rho \bar k} \te^{\lambda \mu \nu \rho} \cr
   \Gamma^{\bar \lambda}1 &:&  - 40 i B_{0  \mu  \nu \rho \bar k}
\te_{\bar \lambda}{}^{\mu \nu \rho} \cr
   \Gamma^{\bar \lambda \bar \mu} 1&:&  - \tfrac{1}{2}(- 3 L_{ \nu
\rho \bar k} + 60
 B_{ \nu \rho} \cont{
\sigma} {}_{\bar k}) \te_{\bar \lambda \bar \mu}{}^{\nu \rho} \cr
   \Gamma^{\bar \lambda \bar \mu \bar \nu} 1&:&  -\tfrac{1}{12}
(E_{\rho \bar k} - 6 i L_{0 \rho \bar k}
 + 120 i B_{0  \rho} \cont{
\sigma} {}_{\bar k}) \te_{\bar \lambda \bar \mu \bar \nu}{}^{\rho}
\cr
   \Gamma^{\bar \lambda_1 \cdots \bar \lambda_4}1 &:&  \tfrac{1}{96} (
-i E_{0 \bar k} + 6 L \cont{\mu} {}_{\bar k} - 60 B \cont{\mu}
\cont{\nu} {}_{\bar k}) \te_{\bar \lambda_1 \cdots \bar \lambda_4}
\cr
   \Gamma^{\bar k} 1&:& 0 \cr
   \Gamma^{\bar \mu \bar k} 1&:& 0 \cr
   \Gamma^{\bar \mu \bar \nu \bar k} 1&:& 0 \cr
   \Gamma^{\bar \mu \bar \nu \bar \rho \bar k}1 &:& 0 \cr
   \Gamma^{\bar \lambda_1 \cdots \bar \lambda_4 \bar k}1 &:&
\tfrac{1}{96} E_{\bar k \bar k} \te_{\bar \lambda_1 \cdots \bar
\lambda_4} \eea

\subsection{${{\cal I}}_A e_{ij} $}

We now turn to the contributions from $e_{ij}$. We split the indices
$\a$ into $p=(i,j)$ and $a$, which contains the remaining three
indices. We also define $\epsilon_{\bar i \bar j} = 1$.  The $A = 0$
integrability conditions on $e_{ij}$ give rise to
\vskip 0.3cm \leftline{\underline{{${{\cal I}}_{0} e_{ij}$}}} \vskip
-0.5cm
 \bea
   1 &:& - (- 12 L_{0pq} - 240 B_0 \cont a {}_{pq}) \epsilon^{pq} \cr
   \Gamma^{\bar a}1 &:& -(-6 i L_{\bar a pq} -120 i B_{\bar a} \cont
  b {}_{pq}) \epsilon^{pq} \cr
   \Gamma^{\bar a \bar b}1 &:&  120 B_{0 \bar a \bar b pq}
\epsilon^{pq} \cr
   \Gamma^{\bar q}1 &:& -(- E_{0 p} + 6 i L \cont a {}_p - 6 i L_p
\cont r +
  60 i B \cont a \cont b {}_p - 120 i B \cont a {}_p \cont r)
\epsilon^p{}_{\bar q} \cr
   \Gamma^{\bar a \bar q}1 &:& -(12 L_{0 \bar a p} + 240 B_{0 \bar a}
\cont
  b {}_p - 240 B_{0 \bar a p} \cont r) \epsilon^p{}_{\bar q} \cr
   \Gamma^{\bar p \bar q}1 &:& -\tfrac{1}{2} \epsilon_{\bar p \bar
q}(\tfrac{1}{2} i E_{00} + 6 L_0 \cont a -
  6 L_0 \cont r + 60 B_0 \cont a \cont b - 120 B_0 \cont a
  \cont r + 60 B_0 \cont r \cont s)
 \eea
The $A = \bar a$ integrability conditions on $e_{ij}$ read
\vskip 0.3cm \leftline{\underline{{${{\cal I}}_{\bar a} e_{ij}$}}}
\vskip -0.5cm
 \bea
   1 &:& \tfrac{1}{2} \epsilon^{pq} (12 L_{\bar a pq} + 240 B_{\bar a}
\cont b
  {}_{pq}) \cr
   \Gamma^{\bar b} 1&:& - 120 i \epsilon^{pq} B_{0 \bar a \bar b pq} \cr
   \Gamma^{\bar b \bar c} 1&:&  60 \epsilon^{pq} B_{\bar a \bar b
\bar c
  pq} \cr
   \Gamma^{\bar b \bar c \bar d}1 &:&  0 \cr
   \Gamma^{\bar q}1 &:&  (E_{\bar a p} + 6 i
  L_{0 \bar a p} + 120 i B_{0 \bar a} \cont b {}_p - 120 i B_{0
  \bar a p} \cont r) \epsilon^p{}_{\bar q} \cr
   \Gamma^{\bar b \bar q}1 &:&  (- 6 L_{\bar a
  \bar b p} - 120 B_{\bar a \bar b} \cont c {}_p + 120 B_{\bar
  a \bar b p} \cont r) \epsilon^p{}_{\bar q} \cr
   \Gamma^{\bar b \bar c \bar q} 1&:& 60 i B_{0 \bar a \bar b \bar c
p} \epsilon^p{}_{\bar q} \cr
   \Gamma^{\bar b \bar c \bar d \bar q} 1&:&  0  \cr
   \Gamma^{\bar p \bar q}1 &:&  \tfrac{1}{2} (- \tfrac{1}{2} i E_{0
\bar a} - 3
  L_{\bar a} \cont b + 3 L_{\bar a} \cont r - 30 B_{\bar a}
  \cont b \cont c + 60 B_{\bar a} \cont b \cont r - 30 B_{\bar
  a} \cont r \cont s) \epsilon_{\bar p \bar q} \cr
   \Gamma^{\bar b \bar p \bar q}1 &:&  \tfrac{1}{2} (\tfrac{1}{2}
E_{\bar a
  \bar b} + 3 i L_{0 \bar a \bar b} + 60 i B_{0 \bar a \bar
  b} \cont c - 60 i B_{0 \bar a \bar b} \cont r) \epsilon_{\bar p \bar q} \cr
   \Gamma^{\bar b \bar c \bar p \bar q}1 &:&  \tfrac{1}{2} (-
\tfrac{3}{2}
  L_{\bar a \bar b \bar c} + 30 B_{\bar a \bar b \bar c}
  \cont r) \epsilon_{\bar p \bar q} \cr
   \Gamma^{\bar b \bar c \bar d \bar p \bar q} 1&:&   0
 \eea
Finally, the $A = \bar p$ integrability conditions are given by
\vskip 0.3cm \leftline{\underline{{${{\cal I}}_{\bar p} e_{ij}$}}}
\vskip -0.5cm
 \bea
   1 &:& \tfrac{1}{2} \epsilon^{qr}(- 24 g_{\bar p q} (L \cont a
{}_{r}- L_{r}
  \cont s) + 36 L_{\bar p qr} - 240 g_{\bar p q} (B \cont a \cont b
  {}_{r}  - 2 B \cont a {}_{r} \cont s) + 720 B \cont a {}_{\bar p
  qr}) \cr
   \Gamma^{\bar a}1 &:& \tfrac{1}{2} \epsilon^{qr}(- 24 i g_{\bar p q}
L_{0 \bar a r} -
  480 i g_{\bar p q} (B_{0 \bar a} \cont b {}_{r} - B_{0 \bar a
  r} \cont s) + 720 i B_{0 \bar a \bar p qr}) \cr
   \Gamma^{\bar a \bar b}1 &:& \tfrac{1}{2} \epsilon^{qr}(- 12 g_{\bar
p q} L_{\bar a
  \bar b r} - 240 g_{\bar p q} (B_{\bar a \bar b} \cont c {}_{r} - B_{\bar
a \bar b r} \cont
  s) + 360 B_{\bar a \bar b \bar p qr}) \cr
   \Gamma^{\bar a \bar b \bar
  c}1 &:& - 40 i \epsilon^{qr} g_{\bar p q} B_{0 \bar a \bar b \bar
  c r} \cr
   \Gamma^{\bar r} 1&:& (E_{\bar p q} - 6 i
  g_{\bar p q} (L_{0} \cont a - L_0 \cont s) + 18 i L_{0 \bar p
  q} - 60 i g_{\bar p q} (B_0 \cont a \cont b - 2 B_0 \cont a \cont s +
B_0 \cont s \cont
  t) + \cr
 && + 360 i B_{0} \cont a {}_{\bar p q} - 360 i B_{0 \bar p q}
  \cont s) \epsilon^q{}_{\bar r} \cr
   \Gamma^{\bar a \bar r} 1&:& (- 6 g_{\bar p
  q} (L_{\bar a} \cont b - L_{\bar a} \cont s) + 18 L_{\bar a
  \bar p q} - 60 g_{\bar p q} (B_{\bar a} \cont b \cont c - 2
  B_{\bar a} \cont b \cont s + B_{\bar a} \cont s \cont t) +
  360 B_{\bar a} \cont b {}_{\bar p q} + \cr
&& - 360  B_{\bar a \bar p
  q} \cont s) \epsilon^q{}_{\bar r} \cr
   \Gamma^{\bar a \bar b \bar r} 1&:& (- 3
  i g_{\bar p q} L_{0 \bar a \bar b} - 60 i g_{\bar p q} (B_{0
  \bar a \bar b} \cont c - B_{0 \bar a \bar b} \cont s) + 180
  i B_{0 \bar a \bar b \bar p q}) \epsilon^q{}_{\bar r} \cr
   \Gamma^{\bar a \bar b \bar c \bar r}1 &:& (- g_{\bar p q} L_{\bar
a \bar b \bar c} + 20 g_{\bar p q}
  B_{\bar a \bar b \bar c} \cont s + 60 B_{\bar a \bar b \bar
  c \bar p q}) \epsilon^q{}_{\bar r} \cr
   \Gamma^{\bar q \bar r} 1&:& \tfrac{1}{2} (- \tfrac{1}{2} i E_{0 \bar
p} - 9
  L \cont a {}_{\bar p} + 9 L_{\bar p} \cont s - 90 B \cont a
  \cont b {}_{\bar p} + 180 B \cont a {}_{\bar p} \cont s) \epsilon_{\bar
q \bar r} \cr
   \Gamma^{\bar a \bar q \bar r} 1&:&  \tfrac{1}{2}(\tfrac{1}{2}
E_{\bar a
  \bar p} - 9 i L_{0 \bar a \bar p} - 180 i B_{0 \bar a} \cont
  b {}_{\bar p} + 180 i B_{0 \bar a \bar p} \cont s) \epsilon_{\bar q \bar
r} \cr
   \Gamma^{\bar a \bar b \bar q \bar r}1 &:& \tfrac{1}{2} (-
\tfrac{9}{2}
  L_{\bar a \bar b \bar p} - 90 B_{\bar a \bar b} \cont c
  {}_{\bar p}  + 90 B_{\bar a \bar b \bar p} \cont s) \epsilon_{\bar q
\bar r} \cr
   \Gamma^{\bar a \bar b \bar c \bar q \bar r}1 &:&  - 15 i B_{0
  \bar a \bar b \bar c \bar p} \epsilon_{\bar q \bar r}
 \eea

\subsection{${{\cal I}}_A e_{klm} $}

We split the indices $\a$ into $a=(k,l,m)$ and $p$, containing the
remaining two indices. The three-dimensional Levi-Civita symbol
$\te_{\bar a \bar b \bar c}$ is defined by $\te_{\bar k \bar l \bar
m} = \sqrt{2}$. The $A =0$ integrability conditions for $e_{klm}$
read
\vskip 0.3cm \leftline{\underline{{${{\cal I}}_{0} e_{klm}$}}}
\vskip -0.5cm
 \bea
   1 &:& 2(- i L_{ a
  b  c} - 20 i B_{ a  b  c} \cont p) \te^{a b c} \cr
   \Gamma^{\bar a}1 &:& (6 L_{0  b c} -120 B_{0
   b  c} \cont d +120 B_{0 b c} \cont p) \te_{\bar a}{}^{b c} \cr
   \Gamma^{\bar a \bar b}1 &:& -\tfrac{1}{4}(- E_{0
  c} + 6 i L_{c} \cont d - 6 i L_{c} \cont p - 60 i
  B_{c} \cont d \cont e + 120 i B_{c} \cont d \cont
  p - 60 i B_{c} \cont p \cont q) \te_{\bar a \bar b}{}^{c} \cr
   \Gamma^{\bar p}1 &:& - 80 B_{0  a  b
  c \bar p} \te^{a b c} \cr
   \Gamma^{\bar a \bar p}1 &:& -(-3 i L_{ b  c \bar p} + 60 i
  B_{ b  c} \cont d {}_{\bar p} - 60 i B_{ b c \bar p}
  \cont q) \te_{\bar a}{}^{b c} \cr
   \Gamma^{\bar p \bar q}1 &:& - 20 i B_{ a  b
  c \bar p \bar q} \te^{ a  b  c}
 \eea
Similarly, for $A = \bar a$ we find
\vskip 0.3cm \leftline{\underline{{${{\cal I}}_{\bar a} e_{klm}$}}}
\vskip -0.5cm
 \bea
   1 &:& -2(3
  i g_{\bar a  b} L_{0 c  d} - 60 i g_{\bar a  b}
  (B_{0  c  d} \cont e - B_{0  c  d} \cont p)
  - 60 i B_{0 \bar a  b  c  d}) \te^{b c d} \cr
   \Gamma^{\bar b}1 &:& -(-6 g_{\bar a  c}
  (L_{ d} \cont e - L_{\bar d} \cont p) - 9 L_{\bar a
   c  d} + 180 B_{\bar a  c  d} \cont e - 180 B_{\bar a  c d}
  \cont p + \cr
  && + 60 g_{\bar a  c} (B_{d} \cont e \cont f
  -2 B_{d} \cont e \cont p + B_{c} \cont p \cont q) ) \te_{\bar b}{}^{c
d} \cr
   \Gamma^{\bar b \bar c}1 &:& \tfrac{1}{4}( E_{\bar a  d} -
  6 i g_{\bar a  d} (L_0 \cont e - L_0 \cont p) - 18 i L_{0 \bar a
   d} + 60 i g_{\bar a  d} (B_0 \cont e \cont f - 2 B_0
  \cont e \cont p + B_0 \cont p \cont q) + \cr
 && + 360 i B_{0 \bar a  d}
  \cont e - 360 i B_{0 \bar a  d} \cont  p) \te_{\bar b \bar c}{}^{d} \cr
   \Gamma^{\bar b \bar c \bar d}1 &:& \tfrac{1}{24}(i E_{0 \bar a} +
18 L_{\bar a}
  \cont e - 18 L_{\bar a} \cont p - 180 B_{\bar a} \cont e \cont f + 360
  B_{\bar a} \cont e \cont p - 180 B_{\bar a} \cont p \cont q) \te_{\bar b
\bar c \bar d} \cr
   \Gamma^{\bar p}1 &:& 2 (
  3 g_{\bar a   b} L_{ c  d \bar p} - 60 g_{\bar a   b}
  (B_{ c  d} \cont e {}_{\bar p} + 60 B_{ c  d \bar p}
  \cont q) - 60 B_{\bar a  b  c  d \bar p}) \te^{b c d} \cr
   \Gamma^{\bar b \bar p} 1&:& ( 6 i
  g_{\bar a   c} L_{0  d \bar  p} - 120 i g_{\bar a   c} (B_{0  d} \cont e
{}_{\bar p} - B_{0  d \bar p} \cont
  q) - 180 i B_{0 \bar a   c   d \bar p}) \te_{\bar b}{}^{c d} \cr
   \Gamma^{\bar b \bar c \bar p} 1&:& -\tfrac{1}{4}(-6 g_{\bar a  d}
  (L \cont e {}_{\bar p} - L_{\bar p} \cont q) - 18 L_{\bar a  d \bar p} +
60 g_{\bar a
   d} (B \cont e \cont f {}_{\bar p} - 2 B \cont e {}_{\bar p} \cont q)
   + 360 B_{\bar a  d} \cont e {}_{\bar p} + \cr
 && - 360 B_{\bar a  d \bar p}
\cont
  q) \te_{\bar b \bar c}{}^{d} \cr
   \Gamma^{\bar b \bar c \bar d \bar p}1 &:& - \tfrac{1}{24}(E_{\bar a
\bar p} - 18 i L_{0 \bar a
  \bar p} + 360 i B_{0 \bar a} \cont b {}_{\bar p} - 360 i B_{0 \bar a
\bar p} \cont q) \te_{\bar b \bar c \bar d} \cr
   \Gamma^{\bar p \bar q}1 &:& - 60 i g_{\bar a b} B_{0
  c d \bar  p \bar q} \te^{b c d} \cr
   \Gamma^{\bar b \bar p \bar q}1 &:& -(3 g_{\bar a  c} L_{ d \bar p
\bar q} -60 g_{\bar a
   c} B_{ d } \cont e {}_{\bar p \bar q} -90 B_{\bar a  c  d \bar p \bar
q}) \te_{\bar b}{}^{c d} \cr
   \Gamma^{\bar b \bar c \bar p \bar q} 1&:& \tfrac{1}{4}(3 i g_{\bar
a  d} L_{0 \bar p \bar q} -60 i g_{\bar a  d} B_0 \cont
  e {}_{\bar p \bar q} -180 i B_{0 \bar a  d \bar p \bar q}) \te_{\bar b
\bar c}{}^{d} \cr
   \Gamma^{\bar b \bar c \bar d \bar p \bar q}1 &:& \tfrac{1}{24} (-9
L_{\bar a \bar p \bar q} +  180 B_{\bar a} \cont b {}_{\bar p \bar q})
\te_{\bar b \bar c \bar d}
 \eea
For $A = \bar p$ the integrability conditions on $e_{klm}$ lead to
\vskip 0.3cm \leftline{\underline{{${{\cal I}}_{\bar p} e_{klm}$}}}
\vskip -0.5cm
 \bea
   1 &:& - 40 i B_{0
  a  b  c \bar p} \te^{a b c} \cr
   \Gamma^{\bar a}1 &:& -(- 3 L_{b  c \bar p} + 60 B_{ b  c} \cont d
{}_{\bar p} - 60
  B_{ b  c \bar p} \cont q) \te_{\bar a}{}^{b c} \cr
   \Gamma^{\bar a \bar b} 1&:& \tfrac{1}{4} (E_{c \bar p} + 6 i
  L_{0 c \bar p} - 120 i B_{0 c} \cont d {}_{\bar p} + 120 i B_{0
  c \bar p} \cont q) \te_{\bar a \bar b}{}^{c} \cr
   \Gamma^{\bar a \bar b \bar c}1 &:& \tfrac{1}{24}(i E_{0 \bar p} + 6
L \cont d
  {}_{\bar p} - 6 L_{\bar p} \cont q - 60 B \cont d \cont e {}_{\bar p} +
120 B \cont
  d {}_{\bar p} \cont q ) \te_{\bar a \bar b \bar c} \cr
   \Gamma^{\bar q} 1&:& 40 B_{ a  b  c \bar p \bar q} \te^{a b c} \cr
   \Gamma^{\bar a \bar q}1 &:& - 60 i B_{0
   b  c \bar p \bar q} \te_{\bar a}{}^{b c} \cr
   \Gamma^{\bar a \bar b \bar q} 1&:& -\tfrac{1}{4}(6 L_{ c \bar p
\bar q} -
  120 B_{c} \cont d {}_{\bar p \bar q}) \te_{\bar a \bar b}{}^{c} \cr
   \Gamma^{\bar a \bar b \bar c \bar q} 1&:& - \tfrac{1}{24}(E_{\bar p
\bar q} - 6 i L_{0 \bar p \bar q} +
  120 i B_{0} \cont d {}_{\bar p \bar q}) \te_{\bar a \bar b \bar c} \cr
   \Gamma^{\bar q \bar r} 1&:& 0 \cr
   \Gamma^{\bar a \bar q \bar r} 1&:& 0 \cr
   \Gamma^{\bar a \bar b \bar q \bar r} 1&:& 0 \cr
   \Gamma^{\bar a \bar b \bar c \bar q \bar r} 1&:& 0
 \eea

\newsection{The linear system for $SU(4)$ invariant spinors}

\subsection{The conditions}

To solve (\ref{sufkspin}), we collect from the appendices above the
terms associated with ${\cal D}_A (e_5+e_{1234})$  and  $i{\cal D}_A
(1-e_{12345})$. In addition, we decompose the expressions that arise
in $i{\cal D}_A (1-e_{12345})$ in terms of $SU(4)$ representations.
In practice this means of splitting the holomorphic  index
$\alpha=(\rho, 5)$, where   $\rho=1,2,3,4$. Using this, the
conditions arising from Killing spinor equations for $\eta_2$
involving derivatives along the time direction are
 \bea
  0 & = & g_3^{-1}
\partial_0(g_1+ig_2)-i\Omega_{0,05}+\tfrac{1}{3} G_{5\rho}{}^\rho
-\tfrac{i}{72} F_{\rho_1\rho_2\rho_3\rho_4}
\te^{\rho_1\rho_2\rho_3\rho_4} \la{original-tfone} \\
  0 & = & i g_3^{-1}
g_2 [i\Omega_{0,0\bar\rho}+\tfrac{1}{3} (G_{\bar\rho
5}{}^5+G_{\bar\rho \s}{}^\s)]+ \Omega_{0,\bar\rho 5}+\tfrac{i}{6}
F_{\bar\rho 5 \s}{}^\s -\tfrac{1}{18} G_{\s_1\s_2\s_3}
\te^{\s_1\s_2\s_3}{}_{\bar\rho} \la{original-tftwo} \\ 0 & = &
\partial_0\log g_3 +i g_3^{-1} g_2 [i\Omega_{0,0\bar 5}+\tfrac{1}{3} G_{\bar 5 \s}{}^\s]+
\tfrac{1}{2} \Omega_{0,\s}{}^\s-\tfrac{1}{2} \Omega_{0,5\bar5} +\tfrac{i}{24} F_{\rho}{}^\rho{}_\s{}^\s \cr && -\tfrac{i}{12}
F_\rho{}^\rho{}_{5\bar 5} \la{original-tfthree} \\ 0 & = &  i g_3^{-1} g_2
[\tfrac{1}{2} \Omega_{0, \bar\rho\bar\s}- \tfrac{i}{12}
(F_{\bar\rho\bar\s 5}{}^5+F_{\bar\rho\bar\s \l}{}^\l)] +\tfrac{1}{6}
G_{\bar\rho\bar\s 5}+ [-\tfrac{1}{8} \Omega_{0,\l_1\l_2}
~~~~~~~~~\\ && -\tfrac{i}{48} F_{\l_1\l_2\tau}{}^\tau +\tfrac{i}{48}
F_{\l_1\l_2 5\bar5}] \te^{\l_1\l_2}{}_{\bar\rho\bar\s}
\la{original-tffour} \\ 0 & = &  i g_3^{-1} g_2 [ \Omega_{0, \bar\rho \bar
5}-\tfrac{i}{6} F_{\bar\rho\bar5\l}{}^\l] + \tfrac{i}{36}
F_{\s_1\s_2\s_3 \bar 5} \te^{\s_1\s_2\s_3}{}_{\bar \rho} - \tfrac{i}{2} \Omega_{0,0\bar\rho}+\tfrac{1}{6}
G_{\bar\rho\l}{}^\l-\tfrac{1}{6} G_{\bar\rho5\bar5} \la{original-tffive}
\eea We have used the conditions on the geometry and fluxes arising
from
 the Killing spinor equations of the
first spinor to simply somewhat the above expression. Similarly the
conditions arising from
 Killing spinor equations for $\eta_2$ involving derivatives
 along the spatial directions are
 \bea
0 & = &
g_3^{-1} (\partial_{\bar\rho} g_1-g_1 \partial_{\bar\rho}\log f
+i\partial_{\bar\rho} g_2) +i g_3^{-1} g_2 [ \partial_{\bar
\rho}\log f +
(\Omega_{\bar\rho,\s}{}^\s+\Omega_{\bar\rho,5\bar5})+\tfrac{i}{6}
(G_{\bar\rho \s}{}^\s +G_{\bar\rho 5\bar 5})] \cr && -i\Om_{\bar
\rho,05}+\smallfrac{1}{6}F_{\bar \rho5\l}{}^\l
-\smallfrac{i}{18}g_{\bar\rho\l_1}G_{\l_2\l_3\l_4}\te^{\l_1\l_2\l_3\l_4}
\la{original-afone} \\ 0 & = &   ig_3^{-1} g_2 [i\Omega_{\bar\rho,0\bar
\s}+\tfrac{1}{6} (F_{\bar \rho\bar \s \l}{}^\l +F_{\bar \rho\bar
\s 5}{}^5)] + \Om_{\bar\rho,\bar\s 5}-\smallfrac{i}{6}G_{\bar\rho\bar\s
5} \cr &&
 -(\smallfrac{1}{12}F_{\bar\rho\l_1\l_2\l_3}
    +\smallfrac{1}{12}g_{\bar\rho\l_1}F_{\l_2\l_3\tau}{}^\tau
 -\smallfrac{1}{12}g_{\bar\rho\l_1}F_{\l_2\l_3 5\bar 5})\te^{\l_1\l_2\l_3}{}_{\bar\s}
 \la{original-aftwo} \\
0 & = &
\partial_{\bar \rho} \log g_3+ig_3^{-1} g_2 [i\Omega_{\bar\rho, 0\bar 5}
+\tfrac{1}{6} F_{\bar \rho \bar 5 \l}{}^\l]
+\smallfrac{1}{2}\Om_{\bar\rho,\l}{}^\l-\smallfrac{1}{2}\Om_{\bar\rho,5\bar 5}
    -\smallfrac{i}{12}G_{\bar\rho\l}{}^\l+\smallfrac{i}{12}G_{\bar\rho 5\bar 5})
    \cr
    && -\smallfrac{1}{18}g_{\bar\rho\l_1}F_{\l_2\l_3\l_4\bar 5}\te^{\l_1\l_2\l_3\l_4}
\la{original-afthree} \\
0 & = &   ig_3^{-1} g_2 [\tfrac{1}{2}\Omega_{\bar\rho,
\bar\s_1\bar\s_2} +\tfrac{i}{12} G_{\bar\rho\bar\s_1\bar\s_2}]
+\smallfrac{1}{12}F_{\bar\rho\bar\s_1\bar\s_25}-\smallfrac{1}{2}(\smallfrac{1}{4}\Om_{\bar\rho,\l_1\l_2}
+\smallfrac{i}{8}G_{\bar\rho\l_1\l_2} \cr
  &&  +\smallfrac{i}{12}g_{\bar\rho\l_1}G_{\l_2\tau}{}^\tau-\smallfrac{i}{12}g_{\bar\rho\l_1}G_{\l_2 5\bar 5})\te^{\l_1\l_2}{}_{\bar\s_1\bar\s_2} \la{original-affour} \\ 0 & = & ig_3^{-1} g_2 [\Omega_{\bar\rho,\bar\s\bar
5}+\tfrac{i}{6} G_{\bar\rho\bar\s\bar
5}]-\smallfrac{i}{2}\Om_{\bar\rho,0\bar\s}
+\smallfrac{1}{12}F_{\bar\rho\bar\s\l}{}^\l
    -\smallfrac{1}{12}F_{\bar\rho\bar\s 5\bar 5}+\smallfrac{i}{12}\te_{\bar\rho\bar\s}{}^{\l_1\l_2}G_{\l_1\l_2\bar 5} \la{original-affive} \\
0 & = &  ig_3^{-1} g_2 [\tfrac{1}{36}
F_{\bar\rho\bar\s_1\bar\s_2\bar\s_3}]
+\smallfrac{1}{12}(\smallfrac{i}{2}\Om_{\bar\rho,0\l}-\smallfrac{1}{4}F_{\bar\rho\l\tau}{}^\tau
    +\smallfrac{1}{4}F_{\bar\rho\l 5\bar 5}-\smallfrac{1}{24}g_{\bar\rho\l}F_\tau{}^\tau{}_\s{}^\s
    \cr
&&    +\smallfrac{1}{12}g_{\bar\rho\l}F_{5\bar5\tau}{}^\tau)\te^{\l}{}_{\bar\s_1\bar\s_2\bar\s_3}
\la{original-afsix} \\
0 & = & ig_3^{-1} g_2 [\tfrac{1}{12}
F_{\bar\rho\bar\s_1\bar\s_2\bar 5}]+
\smallfrac{1}{4}\Om_{\bar\rho,\bar\s_1\bar\s_2}
-\smallfrac{i}{24}G_{\bar\rho\bar\s_1\bar\s_2}-\smallfrac{1}{2}(\smallfrac{1}{8}F_{\bar\rho
\bar 5\l_1\l_2}
    +\smallfrac{1}{12}g_{\bar\rho\l_1}F_{\l_2\bar 5\tau}{}^\tau)\te^{\l_1\l_2}{}_{\bar\s_1\bar\s_2}
\la{original-afseven} \\
0 & = &
\partial_{\bar\rho} \log g_3-\smallfrac{1}{2}\Om_{\bar\rho,\l}{}^\l+\smallfrac{1}{2}\Om_{\bar\rho,5\bar 5}
    -\smallfrac{i}{4}G_{\bar\rho\l}{}^\l+\smallfrac{i}{4}G_{\bar\rho 5\bar 5}
\la{original-afeight} \\
0 & = &
 \smallfrac{1}{72}F_{\bar\rho
\bar\s_1\bar\s_2\bar\s_3}+\smallfrac{1}{12}(\smallfrac{1}{2}\Om_{\bar\rho,\l\bar
5}+\smallfrac{i}{4}G_{\bar\rho\l\bar 5}
    +\smallfrac{i}{12}g_{\bar\rho\l}G_{\bar 5\tau}{}^\tau)\te^\l{}_{\bar\s_1\bar\s_2\bar\s_3}
\la{original-afnine} \\
0 & = &
g_3^{-1}(\partial_{\bar\rho}g_1-g_1\partial_{\bar\rho}\log
f-i\partial_{\bar \rho} g_2) + ig_3^{-1} g_2 \partial_{\bar\rho}\log
f
 +2(\smallfrac{i}{2}\Om_{\bar\rho,0\bar 5}
    -\smallfrac{1}{4}F_{\bar\rho\bar 5\l}{}^\l)
 \la{original-aften}
\eea
and along the fifth direction we have
 \bea
  0 & = & g_3^{-1}
(\partial_{\bar5} g_1-g_1 \partial_{\bar5}\log f+i\partial_{\bar5}
g_2) +i g_3^{-1} g_2 [ \partial_{\bar 5}\log f +
(\Omega_{\bar5,\s}{}^\s+\Omega_{\bar5,5\bar5})+\tfrac{i}{6} G_{\bar5
\s}{}^\s] \cr && -(i\Om_{\bar
    5,05}+\smallfrac{1}{12}F_\l{}^\l{}_\tau{}^\tau
    +\smallfrac{1}{3}F_{5\bar 5\l}{}^\l)
    \la{original-fffone} \\
0 & = & ig_3^{-1} g_2 [i\Omega_{\bar 5,0\bar \s}+\tfrac{1}{6}
F_{\bar 5\bar \s \l}{}^\l] +(\Om_{\bar 5,\bar \s
5}-\smallfrac{i}{6}G_{\bar \s\l}{}^\l-\smallfrac{i}{3}G_{\bar \s 5\bar 5})
-\tfrac{1}{36} F_{\bar5\l_1\l_2\l_3} \te^{\l_1\l_2\l_3}{}_{\bar\s}
\la{original-ffftwo} \\
0 & = &
\partial_{\bar 5} \log g_3+\smallfrac{1}{2}\Om_{\bar 5,\l}{}^\l
    -\smallfrac{1}{2}\Om_{\bar 5,5\bar 5}-\smallfrac{i}{4}G_{\bar 5\l}{}^\l
\la{original-fffthree} \\
0 & = & ig_3^{-1} g_2 [\tfrac{1}{2}\Omega_{\bar 5,
\bar\s_1\bar\s_2}+\tfrac{i}{12} G_{\bar 5\bar\s_1\bar\s_2}]
-(\smallfrac{1}{12}F_{\bar\s_1\bar\s_2\l}{}^\l+\smallfrac{1}{6}F_{\bar\s_1\bar\s_2
5\bar 5}) \cr && -\smallfrac{1}{2}(\smallfrac{1}{4}\Om_{\bar
5,\l_1\l_2}+\smallfrac{i}{24}G_{\bar 5\l_1\l_2})\te^{\l_1\l_2}{}_{\bar
\s_1\bar \s_2} \la{original-ffffour} \\
0 & = & ig_3^{-1} g_2
[\Omega_{\bar5,\bar\s\bar 5}]-\smallfrac{i}{2}\Om_{\bar
5,0\bar\s}-\smallfrac{1}{4}F_{\bar\s\bar 5\l}{}^\l
 \la{original-fffive}
\\
0 & = &  ig_3^{-1} g_2 [\tfrac{1}{36} F_{\bar
5\bar\s_1\bar\s_2\bar\s_3}]-\smallfrac{i}{36}G_{\bar\s_1\bar\s_2\bar\s_3}+
\smallfrac{1}{12}(\smallfrac{i}{2}\Om_{\bar 5,0\l}
    -\smallfrac{1}{12}F_{\bar 5\l\tau}{}^\tau)\te^{\l}{}_{\bar \s_1\bar \s_2\bar \s_3}
 \la{original-fffsix}
\\
0 & = &  \Om_{\bar 5,\bar \s_1\bar \s_2}
    -\smallfrac{i}{2}G_{\bar 5\bar \s_1\bar \s_2}
\la{original-fffseven} \\
0 & = &
-\smallfrac{1}{144}F_{\bar\s_1\bar\s_2\bar\s_3\bar\s_4}+\smallfrac{1}{96}(\partial_{\bar
5} \log g_3
    -\smallfrac{1}{2}\Om_{\bar 5,\l}{}^\l+\smallfrac{1}{2}\Om_{\bar 5,5\bar 5}
    -\smallfrac{i}{12}G_{\bar 5\l}{}^\l)\te_{\bar \s_1\bar \s_2\bar \s_3\bar \s_4}
\la{original-fffeight} \\
0 & = &  -F_{\bar\s_1\bar\s_2\bar\s_3\bar
5}+\Omega_{\bar 5,\l\bar 5} \te^\l{}_{\bar\s_1\bar\s_2\bar\s_3}
\la{original-fffnine} \\
0 & = &  g_3^{-1}(\partial_{\bar
5}g_1-g_1\partial_{\bar 5}\log f-i\partial_{\bar 5} g_2) + ig_3^{-1}
g_2 \partial_{\bar 5}\log f \la{original-ffften} \eea

Using the conditions arising from the $SU(5)$ invariant spinor which
has been collected in appendix B, we can substitute for the  fluxes
${\cal F}$ and rewrite the above equations   in terms of the
connection.
  The conditions arising
from Killing spinor equations for $\eta_2$ involving derivatives
along the time direction then become \bea
0 & = & g_3^{-1}\partial_0 g_1-i \Om_{0,05}+i \Om_{0,0\bar 5}~, \la{tfone-1} \\
0 & = & i g_3^{-1}\partial_0 g_2-\smallfrac{2i}{3}(-\Om_{5,\l}{}^\l+\Om_{\bar
5,\l}{}^\l-\Om_{5,5\bar5}+\Om_{\bar5,5\bar5}+\Om_{0,05}+\Om_{0,0\bar5})~,
\la{tfone-2} \\
0 & = & -\smallfrac{1}{3}g_3^{-1}g_2(\Om_{0,0\bar\rho}-2\Om_{\bar\rho,\l}{}^\l-2\Om_{\bar\rho,5\bar5})
+\Omega_{0,\bar\rho 5}+\smallfrac{1}{3}\Om_{\bar\rho,05}+\tfrac{i}{3}
\Om_{\l_1,\l_2\l_3} \te^{\l_1\l_2\l_3}{}_{\bar\rho}~, \la{tftwo} \\
0 & = &
\partial_0\log
g_3-\smallfrac{1}{6}g_3^{-1}g_2(\Om_{0,05}+\Om_{0,0\bar5}-2(-\Om_{5,\l}{}^\l+\Om_{\bar
5,\l}{}^\l-\Om_{5,5\bar5}+\Om_{\bar5,5\bar5}))~, \la{tfthreeone}\\
0 & = & -\smallfrac{1}{3}g_3^{-1}g_2[\Om_{0,05}-\Om_{0,0\bar5}+2(\Om_{5,\l}{}^\l+\Om_{\bar
5,\l}{}^\l+\Om_{5,5\bar5}+\Om_{\bar5,5\bar5})]\nonumber\\
&&+\tfrac{4}{3}
\Omega_{0,5\bar 5} -\tfrac{4}{3} \Omega_{0,\l}{}^\l~, \la{tfthree} \\
0 & = & \smallfrac{1}{6}g_3^{-1}g_2(2\Om_{\l_1,\l_2
5}\te^{\l_1\l_2}{}_{\bar\s_1\bar\s_2}+\Om_{5,\l_1\l_2}\te^{\l_1\l_2}{}_{\bar\s_1\bar\s_2})\nn \\
&& +\smallfrac{1}{3}(i \Omega_{5, \bar\s_1\bar\s_2}+ i \Omega_{[\bar\s_1,
\bar\s_2] \bar5} -\tfrac{1}{2} \Omega_{0, \l_1\l_2}
\te^{\l_1\l_2}{}_{\bar\s_1\bar\s_2})~, \la{tffour} \\
 0 & = &
\smallfrac{1}{3}g_3^{-1}g_2
\Om_{\l_1,\l_2\l_3}\te^{\l_1\l_2\l_3}{}_{\bar\rho}+
\smallfrac{i}{6}(2\Omega_{\bar 5, \bar
5\bar\rho}+2\Omega_{5,\bar\rho\bar5}-5\Omega_{0,0\bar\rho}-2\Omega_{\s,\bar\rho}{}^{\s})~,
\la{tffive} \eea where the grouped equations constitute the split
into real and imaginary parts.

Similarly, the conditions arising from
 Killing spinor equations for $\eta_2$ involving derivatives
 along the spatial $\bar\rho$ directions become
 \bea
 0 & = &
g_3^{-1}(\partial_{\bar\rho}g_1-g_1\partial_{\bar\rho}\log
f+i\partial_{\bar\rho}g_2)
+\smallfrac{i}{3}g_3^{-1}g_2(-\smallfrac{1}{2}\Om_{0,0\bar\rho}+4\Om_{\bar\rho,\l}{}^\l+4\Om_{\bar\rho,5\bar5})\nn\\
&& -\smallfrac{4i}{3}\Om_{\bar
\rho,05}+\smallfrac{1}{3}\Om_{\l_1,\l_2\l_3}\te^{\l_1\l_2\l_3}{}_{\bar\rho}~,
\la{afone} \cr
 0 & = &
\smallfrac{i}{3}g_3^{-1}g_2(\Om_{5,\l_1\l_2}\te^{\l_1\l_2}{}_{\bar\rho\bar\s}+2\Om_{\l_1,\l_2
5}\te^{\l_1\l_2}{}_{\bar\rho\bar\s})+\Omega_{\bar\rho,
\bar\s5}-\Omega_{\bar\rho, \bar\s\bar5}+\tfrac{1}{3}
\Omega_{5,\bar\rho\bar\s}-\Omega_{\bar 5,\bar\rho\bar\s}\nn\\
 && -\tfrac{2}{3} \Omega_{[\bar\rho,\bar\s]\bar5} -\tfrac{i}{3}
\Omega_{0,\l_1\l_2} \te^{\l_1\l_2}{}_{\bar\rho\bar\s} ~,
 \la{aftwo} \\
 0 & = &
\smallfrac{i}{3}g_3^{-1}g_2\Om_{\l_1,\l_2\l_3}\te^{\l_1\l_2\l_3}{}_{\bar\rho}+
\partial_{\bar\rho}\log g_3+\tfrac{1}{2} \Omega_{\bar\rho,\l}{}^\l-\tfrac{1}{2}\Omega_{\bar\rho,5\bar5}
-\tfrac{1}{6}
\Omega_{\s,\bar\rho}{}^\s\nn\\
&&
+\tfrac{1}{6}\Omega_{5,\bar\rho\bar5}+\tfrac{2}{3}\Omega_{\bar5,
\bar5\bar\rho}- \tfrac{1}{6}\Omega_{0,0\bar\rho} ~, \la{afthree} \\
 0 & = &
-\smallfrac{4i}{3}g_3^{-1}g_2(\Om_{\bar\rho,\bar\s_1\bar\s_2}-\Om_{[\bar\s_1,\bar\s_2]\bar\rho})
+\Omega_{\bar\rho,\l_1\l_2}\te^{\l_1\l_2}{}_{\bar\s_1\bar\s_2}
\nn\\
&& +[\tfrac{1}{3}\Omega_{5,5\tau}-\tfrac{1}{3}\Omega_{\bar\s,\tau}{}^{\bar\s}+\tfrac{1}{3}\Omega_{\bar5,\tau
5}
+\tfrac{1}{6}\Omega_{0,0\tau}]\te^\tau{}_{\bar\rho\bar\s_1\bar\s_2}~, \la{affour} \\
 0 & = &
\smallfrac{i}{3}g_3^{-1}g_2(2\Om_{\bar\rho,\bar\s\bar5}-\Om_{\bar5,\bar\rho\bar\s}
+\Om_{\bar\s,\bar\rho\bar5}) -\smallfrac{2i}{3}\Om_{0,\bar\rho
\bar\s}+\smallfrac{1}{6}\Om_{\l_1,\l_25}\te^{\l_1\l_2}{}_{\bar\rho\bar\s}\nonumber\\
&&+\smallfrac{1}{6}\Om_{\bar5,\l_1\l_2}\te^{\l_1\l_2}{}_{\bar\rho\bar\s}
~, \la{affive} \\
 0 & = &
-\smallfrac{i}{6}g_3^{-1}g_2(\Om_{0,05}+2\Om_{5,\s}{}^\s+2\Om_{5,5\bar5})g_{\bar\rho\l}+
i \Omega_{\bar\rho, 0\l}+\tfrac{1}{2}
F_{\bar\rho\l5\bar5}-\tfrac{i}{3}\Omega_{0,\s}{}^\s g_{\bar\rho\l} \cr && -
\tfrac{2i}{3} \Omega_{0,5\bar5} g_{\bar\rho\l} ~, \la{afsix} \\
 0 & = &
\smallfrac{i}{6}g_3^{-1}g_2(\Om_{0,0[\l_1}+2\Om_{[\l_1|,\tau}{}^\tau+2\Om_{[\l_1|,5\bar
5})g_{\l_2]\bar\rho}+\tfrac{1}{4} F_{\bar\rho\bar5\l_1\l_2}\nn\\
&&
-\tfrac{1}{8} (\Omega_{\bar\rho,\bar\s_1\bar\s_2}
+\Omega_{[\bar\rho,\bar\s_1\bar\s_2]})
\te^{\bar\s_1\bar\s_2}{}_{\l_1\l_2} +\tfrac{i}{3} g_{\bar\rho[\l_1}
\Omega_{\bar5, 0\l_2]} ~,
\la{afseven} \\
 0 & = &
\partial_{\bar\rho}\log g_3-\tfrac{1}{2} \Omega_{\bar\rho,\l}{}^\l-\tfrac{1}{2}
\Omega_{\l,\bar\rho}{}^\l +\tfrac{1}{2} \Omega_{\bar\rho,5\bar5}
+\tfrac{1}{2}\Omega_{5,\bar\rho\bar5}-\tfrac{1}{2}\Omega_{0,0\bar\rho}~,
\la{afeight} \\
 0 & = &
\tfrac{1}{3}[-\tfrac{1}{2}\Omega_{0,05}-\tfrac{1}{2}\Omega_{0,0\bar5}-\Omega_{5,\l}{}^\l
+\Omega_{\bar5,\l}{}^\l-\Omega_{5,5\bar5}+\Omega_{\bar5,5\bar5}]
g_{\bar\rho\s}
+\Omega_{\bar\rho,\s\bar5}+\Omega_{\s,\bar\rho\bar5}~, \la{afnine}
\\
 0 & = &
 g_3^{-1}(\partial_{\bar\rho}g_1-g_1\partial_{\bar\rho}\log
f-i\partial_{\bar\rho} g_2)+ig_3^{-1}g_2\partial_{\bar\rho}\log f\nonumber\\
&&+
4i\Omega_{\bar\rho,0\bar5}-\Omega_{\l_1,\l_2\l_3}
\te^{\l_1\l_2\l_3}{}_{\bar\rho}~.
 \la{aften}
\eea The conditions arising from
 Killing spinor equations for $\eta_2$ involving derivatives
 along the spatial $\bar5$ direction become
\bea
 0 & = & g_3^{-1}(\partial_{\bar5}g_1-g_1\partial_{\bar5}\log
f+i\partial_{\bar5}
g_2)+\smallfrac{i}{3}g_3^{-1}g_2[-\smallfrac{1}{2}\Om_{0,0\bar5}+4\Om_{\bar5,\l}{}^\l
+4\Om_{\bar5,5\bar5}] \cr && +\tfrac{8i}{3}\Om_{0,5\bar5}+\tfrac{4i}{3}
\Omega_{0,\l}{}^\l ~,
    \la{fffone} \\
 0 & = &
-\smallfrac{i}{3}g_3^{-1}g_2\Om_{\l_1,\l_2\l_3}\te^{\l_1\l_2\l_3}{}_{\bar\s}
+ \Omega_{\bar5,\bar\s5}-\tfrac{2}{3}
\Omega_{5,\bar\s\bar5}+\tfrac{1}{3} \Omega_{\bar5,\bar5\bar\s}
-\tfrac{1}{3}
\Omega_{\rho,\bar\s}{}^\rho-\tfrac{5}{6}\Omega_{0,0\bar\s} ~,
\la{ffftwo} \\
 0 & = &
\partial_{\bar5} \log g_3-\Omega_{\bar5,5\bar5}-\tfrac{1}{2}\Omega_{0,0\bar5}~,
\la{fffthree} \\
 0 & = &
-\smallfrac{2i}{3}g_3^{-1}g_2(\Om_{\bar5,\bar\s_1\bar\s_2}-\Om_{[\bar\s_1,\bar\s_2]\bar5})+
\tfrac{1}{2}
\Omega_{5,\l_1\l_2}\te^{\l_1\l_2}{}_{\bar\s_1\bar\s_2}+\tfrac{5}{6}\Omega_{\l_1,\l_25}
\te^{\l_1\l_2}{}_{\bar\s_1\bar\s_2}\nn \\
 && +\tfrac{1}{3}
\Omega_{\bar5, \l_1\l_2} \te^{\l_1\l_2}{}_{\bar\s_1\bar\s_2}
-\tfrac{4i}{3} \Omega_{0,\bar\s_1\bar\s_2} ~, \la{ffffour} \\
 0 & = &
ig_3^{-1}g_2\Om_{\bar5,\bar\s\bar5}+2i\Om_{0,\bar\s\bar5}-\smallfrac{1}{2}
\Om_{\l_1,\l_2\l_3}\te^{\l_1\l_2\l_3}{}_{\bar\s}
~,
 \la{fffive}
\\
 0 & = &
\smallfrac{i}{72}g_3^{-1}g_2(\Om_{0,0\l}+2\Om_{\l,\tau}{}^\tau+2
\Om_{\l,5\bar5})\te^{\l}{}_{\bar\s_1\bar\s_2\bar\s_3}+\smallfrac{1}{6}
\Om_{[\bar\s_1,\bar\s_2\bar\s_3]}+\smallfrac{i}{18}\Om_{\bar5,0\l}
\te^{\l}{}_{\bar\s_1\bar\s_2\bar\s_3} ~,
 \la{fffsix}
\\
0 & = & \Omega_{5, \s_1\s_2}+\Omega_{[\s_1,\s_2]5}=0~, \la{fffeight} \\
 0 & = &
\partial_{\bar5}\log g_3-\tfrac{2}{3} (\Omega_{\bar5,\l}{}^\l-\Omega_{5,\l}{}^\l)
+\tfrac{1}{3} \Omega_{\bar 5, 5\bar5}+\tfrac{2}{3} \Omega_{5, 5\bar5}
-\tfrac{1}{6} \Omega_{0,0\bar5}+\tfrac{1}{3} \Omega_{0,05}~.
\la{fffnine}
\\
 0 & = &
 \Omega_{\lambda,} \cont{\rho} + \Omega_{\lambda , 5 \bar 5} + \Omega_{\bar 5, \lambda \bar 5}
 + \tfrac{1}{2} \Omega_{0,0 \lambda} ~,
 \la{fffseven}
\\
 0 & = &
g_3^{-1}(\partial_{\bar5}g_1-g_1\partial_{\bar5}\log
f-i\partial_{\bar5}g_2)+ig_3^{-1}g_2\partial_{\bar5}\log f~.
\la{ffften}
\eea

\subsection{The solution to the Killing spinor equations with $g_2 \neq 0$}

Here we shall investigate the case $g_2\neq0$. Taking
the trace of (\ref{afsix}), we get
\be
-\tfrac{2i}{3}g_3^{-1}g_2(2\Om_{5,\l}{}^\l+2\Om_{5,5\bar5}+\Om_{0,05})
-\tfrac{7i}{3}\Om_{0,\l}{}^\l-\tfrac{8i}{3}\Om_{0,5\bar5}-\tfrac{1}{2}F_{5\bar5\l}{}^\l=0~.
\ee
Substituting
\be
F_{5\bar5\l}{}^\l=-2i(\Om_{0,\l}{}^\l+2\Om_{0,5\bar5})
\ee
from the
$N=1$ results given in appendix \ref{appnone}, we find
\bea
-\tfrac{1}{3} g_3^{-1} g_2 [\Omega_{0,05}+2\Omega_{5,\r}{}^\r+
2\Omega_{5,5\bar 5}] -\tfrac{1}{3} \Omega_{0,5\bar5}-\tfrac{2}{3}
\Omega_{0,\r}{}^\r=0~. \la{R1}
\eea
Using the above formulae in (\ref{tfthreeone}), we find
\be
\partial_0 \log g_3=0
\ee
and in (\ref{tfthree})
\be
\Om_{0,5\bar5}=0~.
\ee
{}From (\ref{tfone-2}) and (\ref{R1}), we find
\be
 g_3^{-1}\partial_0 g_2- (\Om_{0,05}+\Om_{0,0\bar5})=0
\ee and (\ref{tfone-1}) gives \be
g_3^{-1}\partial_0 g_1 - i(\Om_{0,05}-\Om_{0,0\bar5})=0~. \ee

Using (\ref{R1}) in (\ref{afnine}) yields
\be
\Om_{(\bar\rho,\s)5}=0~.
\ee
Taking the difference between
(\ref{fffthree}) and (\ref{fffnine}) gives
\be
2(\Om_{5,\l}{}^\l-\Om_{\bar5,\l}{}^\l)+4\Om_{\bar5,5\bar5}+2\Om_{5,5\bar5}
+\Om_{0,05}+\Om_{0,0\bar5}=0~,
\ee
which together with (\ref{R1})
yields \be \Om_{5,5\bar5}=0~. \ee
 By instead adding the equations
(\ref{fffthree}) and (\ref{fffnine}), and using (\ref{R1}), we get
\be
\partial_{\bar5}\log g_3-\tfrac{1}{2}\Om_{0,0\bar5}=\partial_{\bar5}\log (g_3 f)=0~.
\ee Next use (\ref{R1}) in (\ref{fffone}) and (\ref{ffften}) to find
\bea
\partial_{\bar5}\log(g_2 f^{-1})=\partial_{\bar5}\log(g_1 f^{-1})=0~.
\eea If we combine (\ref{afsix}) and (\ref{R1}) we can solve for one
of the components of the $F$ flux in terms of the geometry \be
i\Om_{0,\bar\rho\l}+\tfrac{1}{2}F_{\bar\rho\l5\bar5}=0 ~. \ee
The symmetric part of (\ref{aftwo}) implies that
\be
\Om_{(\bar\rho,\bar\s)5}=\Om_{(\bar\rho,\bar\s)\bar5}~.
\ee
Similarly, the symmetric part of (\ref{affive}) yields \be
\Om_{(\bar\rho,\bar\s)\bar5}=0 \ee and thus \be
\Om_{(\bar\rho,\bar\s)5}=\Om_{(\bar\rho,\bar\s)\bar5}=0~. \ee Using
(\ref{fffeight}) the antisymmetric part of (\ref{affive}) yields
\be
-\tfrac{2i}{3}g_3^{-1}g_2\Om_{\bar5,\bar\rho\bar\s}-\tfrac{2i}{3}\Om_{0,\bar\rho\bar\s}
-\tfrac{1}{6}\Om_{5,\l_1\l_2}\te^{\l_1\l_2}{}_{\bar\rho\bar\s}
+\tfrac{1}{6}\Om_{\bar5,\l_1\l_2}\te^{\l_1\l_2}{}_{\bar\rho\bar\s}=0~,
\ee
which coincides with (\ref{tffour}) and (\ref{ffffour}). The dual of
(\ref{fffsix}) coincides with (\ref{tftwo}), and taking the trace of
(\ref{afseven}) and using the result
\be
F_{\l\bar5\tau}{}^\tau=-2i\Om_{0,\l\bar5}
\ee
from the $N=1$ solution, we find
\be
i g_3^{-1}g_2(\Om_{0,0\r}+2\Om_{\r,\s}{}^\s+2\Om_{\r,5\bar5})
+\Om_{\bar\l_1,\bar\l_2\bar\l_3}\te^{\bar\l_1,\bar\l_2\bar\l_3}{}_
\r~,
\la{K5}
\ee
which combined with (\ref{tftwo}) yields
\be
\Om_{\bar\l_1,\bar\l_2\bar\l_3}\te^{\bar\l_1,\bar\l_2\bar\l_3}{}_
\r+2i \Om_{0,\r\bar5}=0~.
\la{epsrel}
\ee
Dualizing (\ref{affour}) with $\te^{\bar\s_1\bar\s_2\bar\r}{}_{\l}$
yields
\be
\Om_{\bar\r,\l}{}^{\bar\r}+\Om_{5,5\l}+\Om_{\bar5,\l5}+\tfrac{1}{2}\Om_{0,0\l}=0~.
\la{K1}
\ee
Taking the sum and difference of (\ref{tffive}) and (\ref{ffftwo})
we find
\be
\Om_{\bar5,\bar\r 5}=\Om_{5,\bar\r\bar5}
\ee
and
\be
\tfrac{2i}{3}g_3^{-1}g_2
\Om_{\l_1\l_2\l_3}\te^{\l_1\l_2\l_3}{}_{\bar\r}+\tfrac{2}{3}\Om_{\s,\bar\r}{}^\s-\tfrac{2}{3}\Om_{5,\bar\r\bar5}
-\tfrac{2}{3}\Om_{\bar5,\bar5\bar\r}+\tfrac{5}{3}\Om_{0,0\bar\r}=0~.
\la{K2}
\ee
In the same way the sum and difference between (\ref{afthree}) and
(\ref{afeight}), and using (\ref{K2}), yield
\be
-\Om_{\bar\r,\l}{}^\l+\tfrac{1}{2}\Om_{0,0\bar\r}+\Om_{\bar\r,5\bar5}-\Om_{\bar5,\bar5\bar\r}=0
\la{K3}
\ee
and
\be
2\partial_{\bar\r}\log
g_3-\Om_{\s,\bar\r}{}^\s-\tfrac{3}{2}\Om_{0,0\bar\r}+\Om_{5,\bar\r\bar5}+\Om_{\bar5,\bar5\bar\r}=0~.
\la{K4}
\ee
Equation (\ref{K1}) can be simplified using the $N=1$ result
\be
-\Om_{\bar\rho,\l}{}^{\bar\rho}-\Om_{\bar5,\l5}-\Om_{\l,\tau}{}^\tau-\Om_{\l,5\bar5}
-\Om_{0,0\l}=0
\la{K7}
\ee
and (\ref{fffseven}) yielding
\be
\Om_{5,5\l}=-\Om_{\bar5,\l \bar5}~.
\ee
Combining (\ref{K3}) and (\ref{fffseven}) we find
\be
\Om_{\r,5\bar5}=0~.
\ee
The equations (\ref{K5}) and (\ref{fffive}) can be simplified, using
(\ref{fffseven}) and (\ref{epsrel}), to
\be
g_3^{-1}g_2 \Om_{\bar5,\l\bar5}=-\Om_{0,\l\bar5}
\ee
and
\be
\Om_{0,\s\bar5}=\Om_{0,\s5}~.
\ee
Using (\ref{epsrel}) and (\ref{K7}) the equations (\ref{K2}) and
(\ref{K4}) can be rewritten as
\be
-g_3^{-1}g_2 \Om_{0,\r\bar5}-\Om_{\bar5,\r
5}-\Om_{5,5\r}+\Om_{0,0\r}=0
\ee
and
\be
\partial_{\bar\r}\log
g_3+\Om_{5,\bar\r\bar5}+\Om_{\bar5,\bar5\bar\r}-\tfrac{1}{2}\Om_{0,0\bar\r}=0~.
\ee
By combining (\ref{afone}) and (\ref{aften}), using (\ref{epsrel}),
we find
\be
\partial_{\bar\r}\log(g_1/f)=0
\ee
and
\be
\partial_{\bar\r}\log(g_2/f)-2g_3 g_2^{-1}\Om_{0,\bar\r\bar5}=0~.
\ee
Using the above results (\ref{afseven}) can be solved for one
component of the $F$ flux
\be
F_{\bar\rho\bar5\l_1\l_2}-\tfrac{8i}{3}\Om_{0,\bar5[\l_1}g_{\l_2]\bar\r}
-\tfrac{1}{2}(\Om_{\bar\rho,\bar\s_1\bar\s_2}+\Om_{[\bar\r,\bar\s_1\bar\s_2]})\te^{\bar\s_1\bar\s_2}{}_{\l_1\l_2}=0~.
\ee
Taking the sum of (\ref{tffour}) and (\ref{aftwo}), and using (\ref{fffeight}), yields
\be
\Om_{[\bar\rho,\bar\s]5}+\Om_{5,\bar\rho\bar\s}=0~.
\ee
and by substituting the above results back into (\ref{affour}) we
find
\be \Om_{\bar\rho,\l_1\l_2}+\tfrac{2}{3}(\Om_{5,5[\l_1}-\Om_{\bar5,5[\l_1}+\tfrac{1}{2}\Om_{0,0[\l_1})g_{\l_2]\bar\rho}
-\tfrac{i}{6}g_3^{-1}g_2(\Om_{\bar\rho,\bar\s_1\bar\s_2}-\Om_{\bar\s_1,\bar\s_2
\bar\r}) \te^{\bar\s_1\bar\s_2}{}_{\l_1\l_2}=0~.
\ee

\end{document}